\documentclass{jfm}

\usepackage{graphicx}
\usepackage{caption}
\usepackage{subcaption}
\usepackage{subcaption}
\usepackage{newtxtext}
\usepackage{newtxmath}
\usepackage{natbib}
\usepackage{hyperref}
\usepackage[normalem]{ulem}

\hypersetup{
    colorlinks = true,
    urlcolor   = blue,
    citecolor  = black,
}

\newcommand{\RomanNumeralCaps}[1]
\linenumbers

\usepackage{cleveref}

\crefformat{section}{\S#2#1#3}

 \title{Evolution of the rotating Rayleigh-Taylor instability under the influence of magnetic fields.} 

\author{Narinder Singh\aff{1}
 and Anikesh Pal\aff{1}
  \corresp{\email{pala@iitk.ac.in}}}

\affiliation{\aff{1}Department of Mechanical Engineering, Indian Institute of Technology Kanpur, Kanpur 208016, U.P., India}

\begin{document}
\maketitle

\begin{abstract}
 The combined effects of the imposed vertical mean magnetic field ($B_0$, scaled as the Alfv\`{e}n velocity) and rotation on the heat transfer phenomenon driven by the Rayleigh-Taylor (RT) instability are investigated using direct numerical simulations. In the hydrodynamic (HD) case ($B_0=0$), as the rotation rate ($f$) increases from $4$ to $8$, the Coriolis force suppresses the growth of the mixing layer height ($h$) and the vertical velocity fluctuations ($u_3^{\prime}$), leading to a reduction in the heat transport, characterized by the Nusselt number ($Nu$). In non-rotating magnetohydrodynamic (MHD) cases, we find a significant delay in the onset of RT instability with increasing $B_0(=0.1,0.15,0.3)$, consistent with the linear theory in the literature. The imposed $B_0$ forms vertically elongated thermal plumes that exhibit larger $u_3^{\prime}$ and efficiently transport heat between the bottom hot fluid and the upper cold fluid with limited horizontal mixing. Therefore, due to higher $u_3^{\prime}$, we observe an enhancement in heat transfer in the initial regime of unbroken elongated plumes in non-rotating MHD cases compared to the corresponding HD case. In the mixing regime of broken small-scale structures, the flow is collimated along the vertical magnetic field lines due to the imposed $B_0$, resulting in a decrease in $u_3^{\prime}$ and an increase in the growth of $h$ compared to the non-rotating HD case. This increase in $h$ enhances heat transfer in the mixing regime of non-rotating MHD over the corresponding HD case. When rotation is added along with the imposed $B_0$, the growth and breakdown of vertically elongated plumes are inhibited by the Coriolis force, reducing $h$ and $u_3'$. Consequently, heat transfer is also reduced in the rotating MHD cases compared to the corresponding non-rotating MHD cases. Interestingly, the heat transfer in the rotating MHD cases remains higher than in corresponding rotating HD cases due to the vertical stretching and collimation of flow structures along the vertical magnetic field lines. This also suggests that the mean magnetic field mitigates the instability-suppressing effect of the Coriolis force. The presence of the ultimate state regime $Nu\simeq Ra^{1/2}Pr^{1/2}$, where $Ra$ is the Rayleigh number, and $Pr$ is the Prandtl number, in the non-rotation HD and MHD cases for $B_0=0.1,0.15$ is observed. However, HD and MHD cases ($B_0=0.1,0.15$) at $f=4,8$ show a departure from this ultimate state scaling. Further, the dynamic balance between different forces is analyzed to understand the behavior of the thermal plumes. The turbulent kinetic energy budget reveals the conversion of the turbulent kinetic energy, generated by the buoyancy flux, into turbulent magnetic energy.
\end{abstract}

\begin{keywords}

\end{keywords}

\section{Introduction} \label{sec: intro} 
 Turbulent thermal convection is an important phenomenon that plays a key role in understanding the heat transport and dynamics of several natural and engineering flows \citep{siggia1994high}. One important mechanism that induces turbulent thermal convection is the Rayleigh-Taylor (RT) instability \citep{boffetta2017incompressible}. The RT instability is a buoyancy-induced hydrodynamic instability that occurs at the interface between colder (denser) and hotter (lighter) fluids when the hotter fluid supports the colder one in the presence of a gravitational field or when the hotter fluid is accelerated into the colder one \citep{rayleigh1882investigation,taylor1950instability}. In the presence of a magnetic field, the RT instability is known as magnetic RT instability \citep{bucciantini2004magnetic}, which plays an important role in engineering applications and astrophysical phenomena. One of its engineering applications is in the inertial confinement fusion (ICF) implosions in which the interactions between the self-generated \citep{zhang2022self,walsh2023nonlinear} or applied \citep{walsh2022magnetized} magnetic fields and RT instability can affect the hot spot's temperature and the target's thermal losses. Therefore, magnetohydrodynamic effects should be considered while designing the ICF devices \citep{walsh2021biermann}. 
 For efficient fusion in ICF devices \citep{ZHOU20171a}, a controlled mixing of the hot and the cold fluid is desirable. Therefore, suppression of the RT instability may be achieved by rotating the spherical target as suggested by \cite{tao2013nonlinear}. In astrophysical systems, the magnetic RT instability is crucial in the filamentary structures of the Crab Nebula (remnant of a supernova explosion) \citep{hester1996wfpc2}, type-Ia supernovae \citep{cabot2006reynolds,hristov2018magnetohydrodynamical}, young supernova remnants \citep{jun1996origin,abarzhi2019supernova}, pulsar wind nebulae in expanding supernova remnants \citep{bucciantini2004magnetic}, and quiescent \citep{ryutova2010observation} and solar \citep{hillier2017magnetic} prominences. Along with the magnetic fields, rotational motions (such as steller, solar, and planetary rotations) are also present in many astrophysical settings that can simultaneously influence RT instability and turbulence. For example, accretion through RT instability at the disc-magnetosphere interface depends on the stellar rotation rates and magnetic fields \citep{kulkarni2008accretion}. The rotation of the Sun causes prominences related to magnetic RT instability to rotate around and be present on the solar disk \citep{ryutova2010observation,hillier2017magnetic}. \\

The RT instability arises at the slightly perturbed interface between two unequal-density fluids subjected to an acceleration in a direction opposite to the mean density gradient such that the denser (or colder) fluid is pushed and accelerated by the lighter (or hotter) one \citep{rayleigh1882investigation,taylor1950instability}. 
The instability is triggered when there is a misalignment of pressure ($p$) and density ($\rho$) gradients such that $\nabla p \cdot \nabla \rho <0$, where pressure gradient results from the acceleration. This misalignment results in the generation of baroclinic torque ($\sim \nabla p \times \nabla \rho$) at the perturbed interface which in turn creates vorticity ($\boldsymbol{\omega}$) and induces velocity field ($\boldsymbol{u}$), as described by the two-dimensional vorticity equation \citep{kundu2015fluid, roberts_jacobs_2016}:
\begin{equation}
    \label{vorticity eq}
    \frac{D \boldsymbol{\omega}}{Dt}=\underbrace{\frac{1}{\rho^2} \left( \nabla p \times \nabla \rho \right)}_{\displaystyle \mathrm{baroclinic\, torque}},
\end{equation}
where $\boldsymbol{\omega} = \nabla \times \boldsymbol{u}$. This results in increased misalignment of the gradient vectors and baroclinic torque. Note that for the thermal convection problem, the density differences reflect temperature fluctuations in RT configuration.
At the initial linear stage, for initial multi-mode perturbations, the short-wavelength ($\lambda$) perturbations (or larger wavenumber $k(=2\pi/\lambda)$) grow more rapidly (exponentially) than the long-wavelength perturbations (smaller $k$) according to the growth rate 
\begin{equation}
    \label{growth rate}
        \sigma=\sqrt{kg\frac{\rho_2-\rho_1}{\rho_2+\rho_1}}
\end{equation}
as predicted by the linear stability analysis \citep{taylor1950instability,chandrasekhar1968hydrodynamic,sharp1984overview,youngs1984numerical} in the absence
 of viscous effects, magnetic field, and rotation. Here $\rho_1$ is the density of the light fluid, and $\rho_2$ is the density of the heavy fluid. As the perturbation amplitude reaches a size of the order of $\sim0.5\lambda$, the linear phase breaks down, and the non-linear effects emerge in the next stage. The exponential growth rate of short-wavelength perturbations slows down. In contrast, the long-wavelength perturbations grow more rapidly, resulting in the appearance of larger fluid structures of ‘spikes’ of heavy fluid growing into light fluid and ‘bubbles’ of light fluid growing into heavy fluid. As the spikes grow, the small structures develop due to the relative motion between two fluids, which triggers secondary Kelvin-Helmholtz (KH) instability along the side of the spike and causes it to become mushroom-shaped. As a result, the drag force on the spike increases, slowing its growth \citep{sharp1984overview}. In the fully nonlinear phase, the bubbles and spikes interact non-linearly, resulting in bubble merging, where the smaller bubbles and spikes merge to produce larger ones, and bubble competition, where the larger bubbles and spikes envelop the smaller ones. Eventually, the fluid structures break down due to various mechanisms, such as shear and interpenetration, leading to the turbulent mixing of two fluids \citep{YOUNGS1989270,Youngs_1994,COOK_DIMOTAKIS_2001,ramaprabhu2005numerical,ZHOU20171a,ZHOU20171b}. In the late non-linear regime, the mixing layer evolves self-similarly with  quadratic time evolution, which is described as \citep{ristorcelli2004rayleigh,cook2004mixing,dimonte2004comparative,cabot2006reynolds}: 
 \begin{equation}
 \label{h2}
    h_{RT}(t) = \alpha gt^2 \frac{\rho_2-\rho_1}{\rho_2+\rho_1},
 \end{equation}
 where $h_{RT}$ is the half-height of the mixing region, $\alpha$ is the nonlinear growth rate of the mixing layer and $t$ is the time. \\
 
In the case of thermal convection driven by RT turbulence, the heat transfer efficiency is quantified by the non-dimensional Nusselt number ($Nu$), which is defined as the ratio of total heat transfer (convective and conductive) to conductive heat transfer, while the turbulent intensity is measured by the Reynolds number ($Re$) \citep{boffetta2010statistics,boffetta2017incompressible}. The $Nu$ and $Re$ depend on two parameters: Rayleigh number ($Ra$) representing a dimensionless measure of the temperature difference between the fluid that forces the system and Prandtl number ($Pr$) representing the ratio of kinematic viscosity and thermal diffusivity. Due to the irrelevance of boundaries, the ultimate state of thermal convection develops in RT turbulence, according to which $Nu \simeq Ra^{1/2} Pr^{1/2}$ and $Re \simeq Ra^{1/2} Pr^{-1/2}$, as reported by \cite{boffetta2009kolmogorov,boffetta2010nonlinear,boffetta2012ultimate} using 3-D numerical simulations for incompressible, miscible fluids with a small Atwood number $\mathcal{A}= \left(\rho_2-\rho_1\right)/\left(\rho_2+\rho_1\right)=\beta\Theta/2$. 
Here $\beta$ is the thermal expansion coefficient, and $\Theta$ is the initial temperature jump across the fluid layer. \\
 
Understanding the physical mechanisms that can significantly influence and control the growth of RT instability and turbulence has attracted the attention of researchers to study the influence of rotation on RT instability. \cite{chandrasekhar1968hydrodynamic} performed a linear stability analysis to study the effect of uniform rotation on the RT instability in inviscid fluids and concluded that it could reduce the growth of RT instability. \cite{carnevale2002rotational} conducted 3-D numerical simulations for low Atwood number ($\mathcal{A}$) flows 
and demonstrated that the rotation, along with viscosity and diffusion, can greatly retard the growth of RT instability and stabilize it indefinitely for high enough rotation rates. \cite{tao2013nonlinear} theoretically studied the rotating system with a rotation axis perpendicular to the direction of acceleration of the interface in a cylindrical fluid domain, in contrast to the studies by \cite{chandrasekhar1968hydrodynamic} and \cite{carnevale2002rotational}, which considered the rotation axis parallel to the interface acceleration, i.e., the gravity vector. \cite{tao2013nonlinear} demonstrated the stabilization effect of rotation on the non-linear stage of RT instability in the cylindrical system and suggested that this effect can apply to the equatorial region of a rotating sphere with potential applications in the ICF. The stabilizing effect of the Coriolis force on the RT instability in a cylindrical system rotating about an axis parallel to the direction of acceleration was investigated experimentally by \cite{baldwin2015inhibition}, theoretically by \cite{scase2017rotating}, and both experimentally and theoretically by \cite{scase2020magnetically}. \cite{boffetta2016rotating} conducted direct numerical simulations (DNS) to study the effects of rotation on the fully developed RT turbulence of incompressible and miscible fluids subjected to the Boussinesq approximation. They reported that rotation reduces the intensity of turbulent fluctuations and slows down the growth rate of the mixing layer. They also found that an increase in rotation rates reduces the heat transfer efficiency (measured by Nusselt number, $Nu$) as a function of time and at a given Rayleigh number, $Ra$, (i.e., the height of the mixing layer) compared to the non-rotating case. Moreover, the violation of the ultimate state scaling (where $Nu \propto Ra^{1/2}$ and $Re \propto Ra^{1/2}$ for $Pr=1$) at high rotation rates was also observed. Recently, \cite{wei2022small} investigated the effects of rotation on the small-scale characteristics, heat transfer performance, and scaling law in the mixing zone of RT turbulence. \\

The presence of a magnetic field adds complexity to the evolution of RT instability and its analysis. The effect of the imposed, uniform magnetic field tangential and normal to the fluid interface between the idealized two inviscid, electrically conducting fluids was studied by \cite{chandrasekhar1968hydrodynamic} using linear stability analysis. Later \cite{jun1995numerical} used two-dimensional (2-D) numerical ideal magnetohydrodynamics (MHD) simulations for incompressible magnetic fluids to demonstrate the influence of the imposed magnetic fields. They reported that the growth of multiple-wavelength initial perturbations in the linear regime decreases with an increase in the strength of the normal magnetic field, which is consistent with the linear theory \citep{chandrasekhar1968hydrodynamic}. In the non-linear regime, increasing the normal magnetic field up to a certain strength enhances the growth of the bubbles due to the collimation of flow along the magnetic field lines. However, beyond this field strength, the growth is significantly suppressed. 
Using 3-D numerical MHD simulations with uniform tangential magnetic field, \cite{jun1995numerical} revealed that as the instability grows, a stronger magnetic field (and energy) component is produced across the magnetic field lines, i.e., along the gravity vector than along the field direction, resulting in the formation of elongated sheetlike fingers. The non-linear phase of the RT instability in the presence of a uniform tangential magnetic field in two inviscid, perfectly conducting fluids was investigated numerically by \cite{stone2007nonlinear}. They solved the 3-D ideal MHD equations of compressible gas dynamics close to the incompressible limit. They reported that the tension forces produced by even a small tangential field on small scales are sufficient to suppress the shear between the rising and descending fluid structures. This reduces the mixing and enhances the non-linear growth of the RT instability. \cite{stone2007magnetic} used the same numerical setup of \cite{stone2007nonlinear} to study the structures and dynamics of the nonlinear evolution of the RT instability for a range of shear angles (in the horizontal plane) between the upper and lower magnetic fields relevant to the optical filaments in the Crab Nebula. They demonstrated that
the change in the magnetic field direction (i.e., non-zero angle) can delay the onset of instability and produce isolated, large-scale isotropic bubbles with smooth surfaces and bulbous tips. Using the same numerical configuration of \cite{stone2007magnetic}, \cite{carlyle2017non} examined the effect of tangential magnetic fields that were stronger than those imposed by \cite{stone2007magnetic,stone2007nonlinear}. They observed that increasing the strength of the magnetic field reduces the nonlinear growth rate of the rising bubbles of the RT instability. Apart from these numerical investigations, theoretical studies were carried out to examine the RT instability with sheared magnetic fields at the interface contact discontinuity by \cite{ruderman2014rayleigh,hillier2016nature,ruderman2018rayleigh} and with an oblique magnetic field by \cite{vickers2020magnetic}\\ 

From the above discussion, it is evident that the majority of the studies associated with the understanding of the effect of imposed uniform magnetic fields on RT turbulence were focused on the tangential and sheared magnetic fields only. However, rotation also significantly affects the evolution of RT instability. To the best of our knowledge, the evolution of the RT instability and the subsequent turbulence under the simultaneous influence of magnetic fields and rotation is absent from the literature. Therefore, we perform DNS of RT turbulence in the presence of both the vertical magnetic fields normal to the interface and rotation in an electrically conducting, incompressible, hot-cold miscible fluid system in an unstably stratified configuration. Our primary goal is to investigate how the simultaneous inclusion of vertical magnetic field and rotation affects the onset of RT instability, turbulence, heat transfer efficiency, and fluid structures. To achieve these goals, the paper is organized as follows. The governing equations with boundary conditions, numerical methodology, solver validation, and the set-up for the simulations are detailed in \cref{sec: numerical details}. We present the numerical results in \cref{{sec: results}} and conclude in \cref{sec: conclusions}. \\

\section{Methods}\label{sec: numerical details}
\subsection{Governing equations}
\begin{figure}
	\centering
  		\includegraphics[width=0.75\textwidth,trim={4.0cm 0.0cm 4.0cm 0cm},clip]{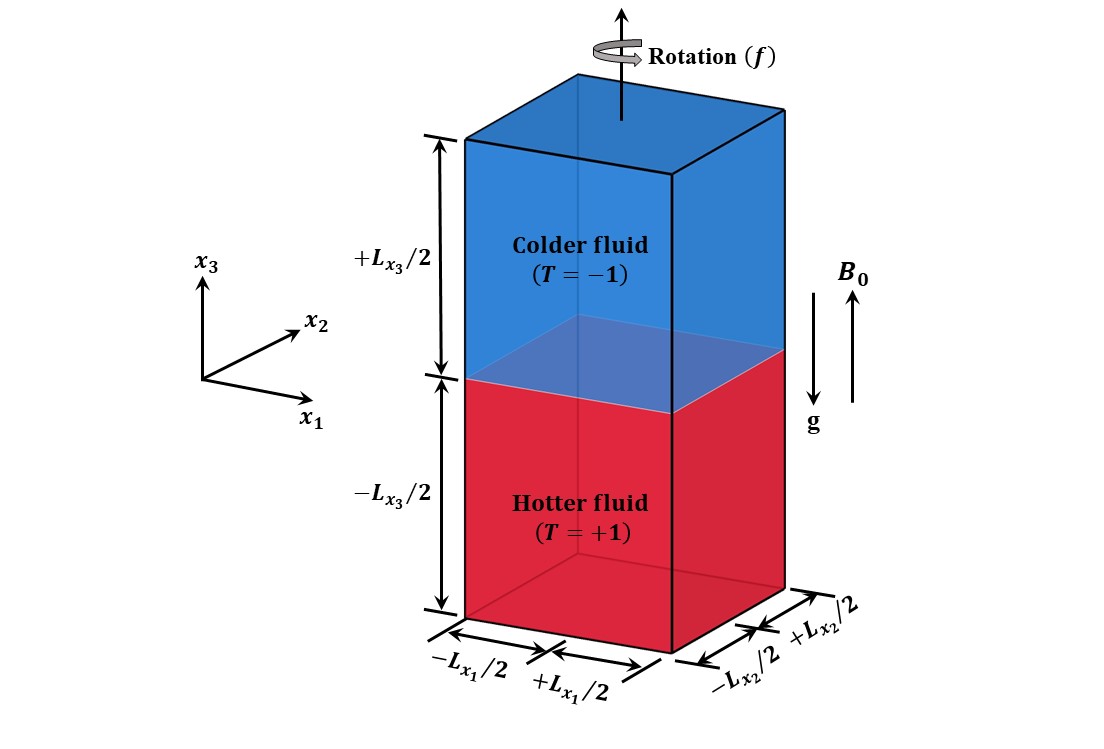} 		
  	\caption{Sketch of the computational domain of RT configuration for electrically conducting miscible fluids, where the colder fluid of temperature $T=-1$ is placed above the hotter fluid of temperature $T=+1$ under gravity $g$. This configuration is rotated about the vertical axis $x_3$ with uniform Coriolis's frequency $f$, and uniform mean magnetic field $B_0$ is imposed normal to the interface between colder and hotter fluid (or anti-parallel to $g$).}
 \label{fig: domain}
\end{figure}

We consider unstable Rayleigh-Taylor configuration of electrically conducting miscible fluids, where the colder fluid (temperature $T=-1$) is supported by the hotter fluid ($T=1$) against gravitational acceleration ($g$) in a rectangular domain as shown in figure \ref{fig: domain}. This configuration is subject to rotation about the vertical axis ($x_3$) with a constant Coriolis frequency $f$. We impose a uniform mean magnetic field ($B_0$), scaled as the Alfv\`{e}n velocity \citep{davidson2013turbulence}, normal to the initial interface, and anti-parallel to $g$. Here, we scale the magnetic field $\boldsymbol{B}$ $(= (B_1, B_2, B_3))$ with a Alfv\`{e}n velocity $\boldsymbol{V}_A$, such that $\boldsymbol{B}=\boldsymbol{V}_A= \boldsymbol{H}/\sqrt{\rho_0\mu_0}$, where $\boldsymbol{H}$ is the true magnetic field, $\mu_0$ is the magnetic permeability and $\rho_0$ is the reference density \citep{davidson2013turbulence}. 
Under the Boussinesq and magnetohydrodynamic (MHD) approximations, the governing equations for an unsteady, incompressible, electrically conducting flow are given as \citep{naskar_pal_2022a,naskar_pal_2022b}: \\
 \begin{subequations}
    \begin{equation}
        \label{continuity1}
        \frac{\partial u_j}{\partial x_j}=\frac{\partial B_j}{\partial x_j}=0,
    \end{equation}
    \begin{equation}
        \label{momentum1}
         \underbrace{\frac{\partial u_i}{\partial t}}_{\displaystyle \boldsymbol{F}_I} +  \underbrace{u_j \frac{\partial u_i}{\partial x_j}}_{\displaystyle \boldsymbol{F}_A} =  \underbrace{- \frac{\partial P}{\partial x_i}}_{\displaystyle \boldsymbol{F}_P}  + \underbrace{f\epsilon_{ij3} u_j\hat{e}_3}_{\displaystyle \boldsymbol{F}_C} + \underbrace{B_j \frac{\partial B_i}{\partial x_j}}_{\displaystyle \boldsymbol{F}_L} + \underbrace{\beta g T\delta_{i3}}_{\displaystyle \boldsymbol{F}_B} + \underbrace{\nu \frac{\partial^2 u_i}{\partial x_j \partial x_j}}_{\displaystyle \boldsymbol{F}_V},
    \end{equation}
    \begin{equation}
        \label{temp eq1}
         \frac{\partial{T}}{\partial t} + u_j \frac{\partial T}{\partial x_j}= \kappa \frac{\partial^2 u_i}{\partial x_j \partial x_j},
    \end{equation}
    \begin{equation}
        \label{mag eq1}
        \frac{\partial{B_i}}{\partial t} + u_j \frac{\partial B_i}{\partial x_j}= B_j \frac{\partial u_i}{\partial x_j}+\mathcal{D}\frac{\partial^2 B_i}{\partial x_j \partial x_j}.
    \end{equation}
 \end{subequations}
Here, $\boldsymbol{u} = (u_1, u_2, u_3)$ is the velocity field vector,  
${P}$ is the reduced pressure, and the Cartesian coordinate system is denoted as $(x_1,x_2,x_3)$. The different forces in the momentum equation \ref{momentum1} are identified as the inertial $F_I$, the advection $F_A$, the pressure gradient $F_P$, the Coriolis $F_C$, the Lorentz $F_L$, the buoyancy $F_B$, and the viscous $F_V$ forces. The temperature field $T$ is related to density $\rho$ via
the thermal expansion coefficient $\beta$ as $\rho=\rho_0\left[1-\beta(T-T_0)\right]$, where $\rho_0$ and $T_0$ are the reference value of density and temperature, respectively. The fluid has kinematic viscosity $\nu$, thermal diffusivity $\kappa$, and magnetic diffusivity $\mathcal{D}$. Initially, at time $t=0$, both the upper colder fluid (heavier of density $\rho_2$) layer separated by the lower hotter fluid (lighter of density $\rho_1$) layer are at rest, i.e., $T(\boldsymbol{x},0)=-({\Theta}/{2}) sgn(x_3)$ and $\boldsymbol{u}(\boldsymbol{x},t)=0$, where $\Theta$ denotes the initial temperature jump across the layers that fix the Atwood number $\mathcal{A}= \left(\rho_2-\rho_1\right)/\left(\rho_2+\rho_1\right)=\beta\Theta/2$ \citep{boffetta2017incompressible}. The mean magnetic field is imposed vertically ($B_0$), such that $B_3(\boldsymbol{x},0)=B_0$, while $B_2(\boldsymbol{x},0)=B_3(\boldsymbol{x},0)=0$. Note that the vertical mean magnetic field ($B_0$) is imposed at each time step. We seed initial perturbations by adding a $10\%$ of white noise to $T(\boldsymbol{x},0)$ in a small region around the interface at $x_3=0$ (center of the vertical domain height). \\

 \captionsetup[subfigure]{textfont=normalfont,singlelinecheck=off,justification=raggedright, labelfont=bf, textfont=bf,font=large}


We apply impenetrable and stress-free (free-slip) velocity boundary conditions at the top and bottom boundaries, written as 
\begin{equation}
    u_3=\frac{\partial u_1}{\partial x_3}=\frac{\partial u_2}{\partial x_3}=0, \quad \mathrm{at}\ x_3=\pm {L_{x_3}}/{2}, 
\end{equation}
where $L_{x_3}$ is the vertical domain height. We consider periodic boundary conditions in horizontal directions $x_1$ and $x_2$ for all the variables. The thermal boundary conditions are isothermal at the top and bottom boundaries, such that 
\begin{equation}
    T=1 \ \mathrm{(hot)}\ \mathrm{at}\  x_3= -{L_{x_3}}/{2}, \quad \mathrm{and} \quad T=-1 \ \mathrm{(cold)} \ \mathrm{at}\  x_3= +{L_{x_3}}/{2}.
\end{equation}

We use perfectly conducting boundary conditions for the magnetic field at the top and bottom boundaries as \citep{naskar_pal_2022b}:
\begin{equation}
    \frac{\partial B_1}{\partial x_3}=\frac{\partial B_2}{\partial x_3}=0, \ B_3=B_0,  \quad \mathrm{at}\ x_3=\pm {L_{x_3}}/{2}. 
\end{equation}

Since the horizontal directions are homogeneous, we perform Reynolds decomposition of all the variables $\mathcal{\phi}= \left\{u_i, P, T, B_i\right\}$ into horizontally averaged mean component $\overline{ \mathcal{\phi}}$ and fluctuating component $\mathcal{\phi}'$, such that
\begin{subequations}
\begin{equation}
\label{reynolds decomposition p3}
 \mathcal{\phi}({x_1,x_2,x_3},t)=\overline{ \mathcal{\phi} }({x_3},t) + \mathcal{\phi}'({x_1,x_2,x_3},t),
 \end{equation}
 \begin{equation}
\label{average p3}
 \quad \overline{\mathcal{\phi}}({x_3},t)= \frac{1}{L_{x_1} L_{x_2}} \int_{-L_{x_2}/2}^{{+L_{x_2}/2}} \int_{-L_{x_1}/2}^{{+L_{x_1}/2}} \mathcal{\phi} ({x_1,x_2,x_3},t)\mathrm{d}x_1\mathrm{d}x_2.
\end{equation}
\end{subequations}
Here, the overbar denotes the average in the horizontal directions, and $L_{x_1}$ and $L_{x_2}$ represent the domain lengths in $x_1$ and $x_2$, respectively. We calculate root mean square ($r.m.s.$) of $\phi$ as
\begin{equation}
    \label{rms}
    \phi_{r.m.s.}(x_3,t)=\left( \overline{\phi'^2({x_1,x_2,x_3},t)}\right)^{1/2}, 
\end{equation}
where, $\phi'=\phi -\overline{\phi}$ from equation \ref{reynolds decomposition p3}. 

The spatial average of $\phi$, denoted by the angled brackets ($\langle \phi \rangle$), inside the mixing layer of height $h$ (defined later) is computed as
 \begin{equation}
    \label{average p31}
 \langle \phi \rangle(t)= \frac{1}{h} \int_{-h/2}^{{+h/2}} \overline{\phi} ({x_3},t)\mathrm{d}x_3. \\
\end{equation} 

\subsection {Numerical methodology and solver validation}
The governing equations \ref{continuity1} - \ref{mag eq1} are solved in a Cartesian coordinate system on a staggered grid arrangement, where the vector quantities ($\boldsymbol{u}$ and $\boldsymbol{B}$) are stored at the cell faces, and scalar quantities ($T$ and $P$) are stored at the cell centers, using the finite-difference method. All the spatial derivatives are computed using a second-order central finite-difference scheme, and the time advancement is performed using an explicit third-order Runge–Kutta method except for the diffusion terms, which are solved implicitly using the Crank–Nicolson method \citep{brucker2010comparative,pal2020deep,naskar_pal_2022a,naskar_pal_2022b,singh_pal_2023}. The projection method is employed to obtain the divergence-free velocity field where the pressure Poisson equation is solved using a parallel multigrid iterative solver to calculate the dynamic pressure. Similarly, to keep the magnetic field solenoidal, an elliptic divergence cleaning algorithm is employed \citep{brackbill1980effect}. This numerical solver has been extensively validated and used for studies of turbulent mixing driven by Faraday instability \citep{singh_pal_2023,singh_pal_2024}, three-phase flows \citep{singh_singh_pal_2023}, rapidly rotating convection-driven dynamos \citep{naskar_pal_2022a,naskar_pal_2022b}, rotating convection \citep{pal2020evolution}, and several stratified free-shear and wall-bounded turbulent flows \citep{brucker2010comparative, pal2013spatial, pal2015effect, pal2020deep}. \\

\begin{figure}
 	\centering
  	\begin{subfigure}{0.495\textwidth}
  		\centering
        \caption{}  \label{subfig: h val}
  		\includegraphics[width=1.0\textwidth,trim={0cm 0cm 0cm 0cm},clip]{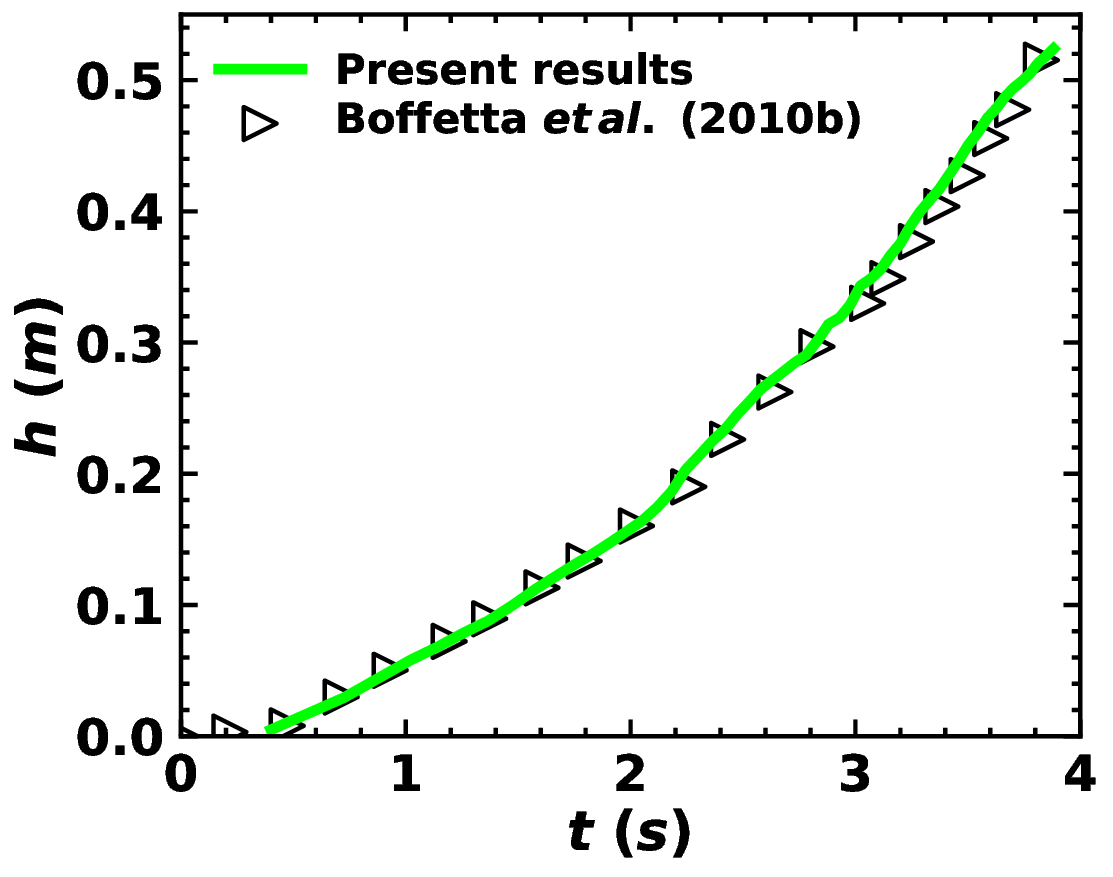}   		
  	\end{subfigure}
 	\hfill
   	\begin{subfigure}{0.495\textwidth}
   		\centering
        \caption{}  \label{subfig: Nu val}
   		\includegraphics[width=1.0\textwidth,trim={0cm 0cm 0cm 0cm},clip]{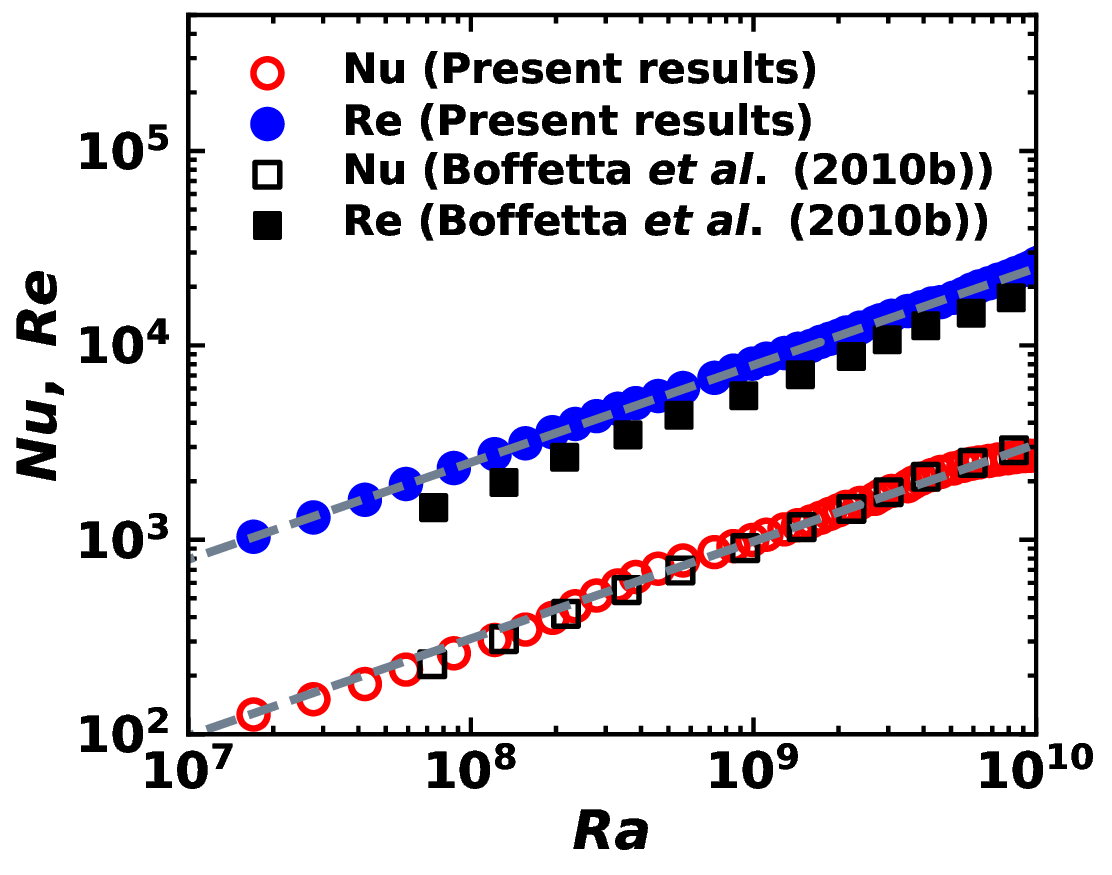}   		
   	\end{subfigure}
  	\caption{Comparison between present simulations and results obtained by \cite{boffetta2010statistics} for a (\textit{a}) temporal evolution of the mixing layer of height $h$ and (\textit{b}) spatially averaged Nusselt number ($Nu=1+{\langle u_3' T' \rangle h}/{(\kappa \Theta)}$) and Reynolds number ($Re=\sqrt{(u_{1,r.m.s.}^2+u_{2,r.m.s.}^2+u_{3,r.m.s.}^2)}\ h/\nu$) as a function of Rayleigh number ($Ra=\beta g \Theta h^3/ (\nu \kappa)$). Dashed gray lines represent the ultimate state predictions, i.e., $Nu \simeq Ra^{1/2}Pr^{1/2}$ and $Re \simeq Ra^{1/2}Pr^{-1/2}$ scaling for $Pr=1$ \citep{boffetta2010statistics}.  }
 \label{fig:validation}
\end{figure}

We further validate our numerical solver by 
replicating the results of \cite{boffetta2010statistics} for Rayleigh–Taylor turbulence without rotation and magnetic field. The simulation is performed in a horizontally periodic domain of size $L_{x_1}=L_{x_2}$ and $L_{x_3}=4L_{x_1}$ with uniform grids $N_{x_1}=N_{x_2}=512$ and $N_{x_3}=2048$. Other parameters are $\beta g=2.0$, $\Theta=1$, $\nu=\kappa=4.8 \times 10^{-6}$. The comparison between the results obtained in the present simulations and the results of \cite{boffetta2010statistics} for the temporal evolution of the mixing layer height ($h$) is shown in figures \ref{subfig: h val} and for the Nusselt number ($Nu=1+{\langle u_3' T' \rangle h}/{(\kappa \Theta)}$) and Reynolds number ($Re=\sqrt{(u_{1,r.m.s.}^2+u_{2,r.m.s.}^2+u_{3,r.m.s.}^2)}\ h/\nu$) as a function of Rayleigh number ($Ra=\beta g \Theta h^3/ (\nu \kappa)$) is shown in figure \ref{subfig: Nu val}. Here, $u_3'$ is the fluctuating vertical velocity, $T'$ is fluctuating temperature, $\langle \ \rangle$ is the spatial average inside $h$ (defined in equation \ref{average p31}), the $r.m.s.$ (defined in equation \ref{rms}) horizontal $u_{1,r.m.s.}$, $u_{2,r.m.s.}$, and vertical $u_{3,r.m.s.}$ velocities are taken at mid-plane $x_3=0$. The mixing layer height $h$ is measured based on the threshold on the horizontally averaged mean temperature profile, such that $\bar{T} (\pm h/2)  = \mp 0.8 \Theta/2$. We can observe that our results match the results of \cite{boffetta2010statistics}. \\

We also show the validation of our solver by reproducing the results of non-magnetic rotating convection (RC) and rotating dynamo convection (DC) of \cite{stellmach2004cartesian} in a rotating plane layer of an electrically conducting fluid confined between two parallel plates, heated from the bottom and cooled from the top (see \cite{naskar_pal_2022a} for details) in appendix \ref{appA}. \\

 \captionsetup[subfigure]{textfont=normalfont,singlelinecheck=off,justification=raggedright, labelfont=bf, textfont=bf,font=large}

\subsection {Problem set-up and simulation parameters}
\begin{figure*}
    \centering
    \hspace{-2.8cm} (\textbf{a})\includegraphics[width=0.38\textwidth,trim={0.6cm 0.0cm 0.0cm 0.0cm},clip]{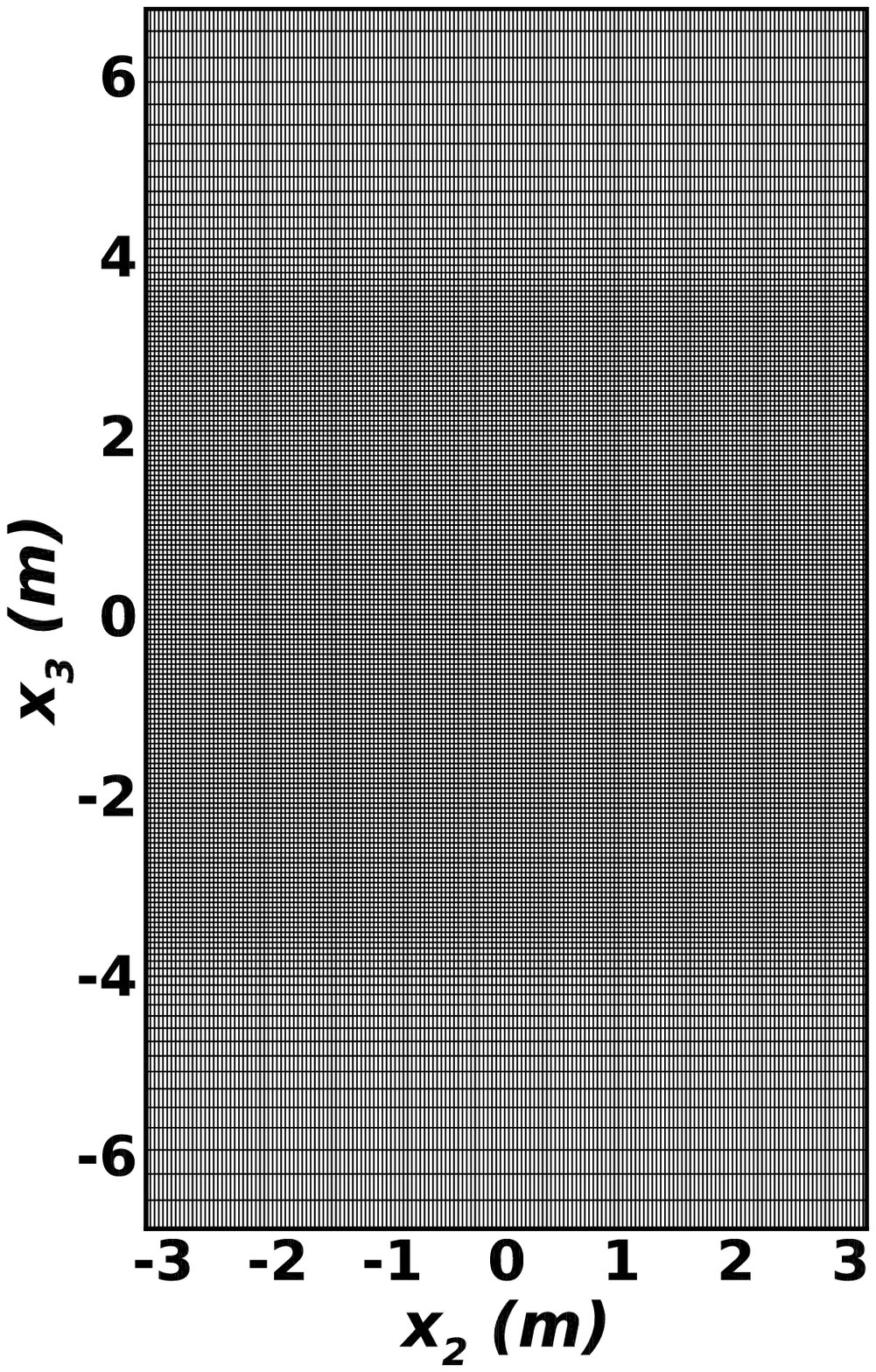} 
    \hspace{0.00cm} (\textbf{b})\includegraphics[width=0.75\textwidth,trim={0.4cm 0.0cm 0.0cm 0.0cm},clip]{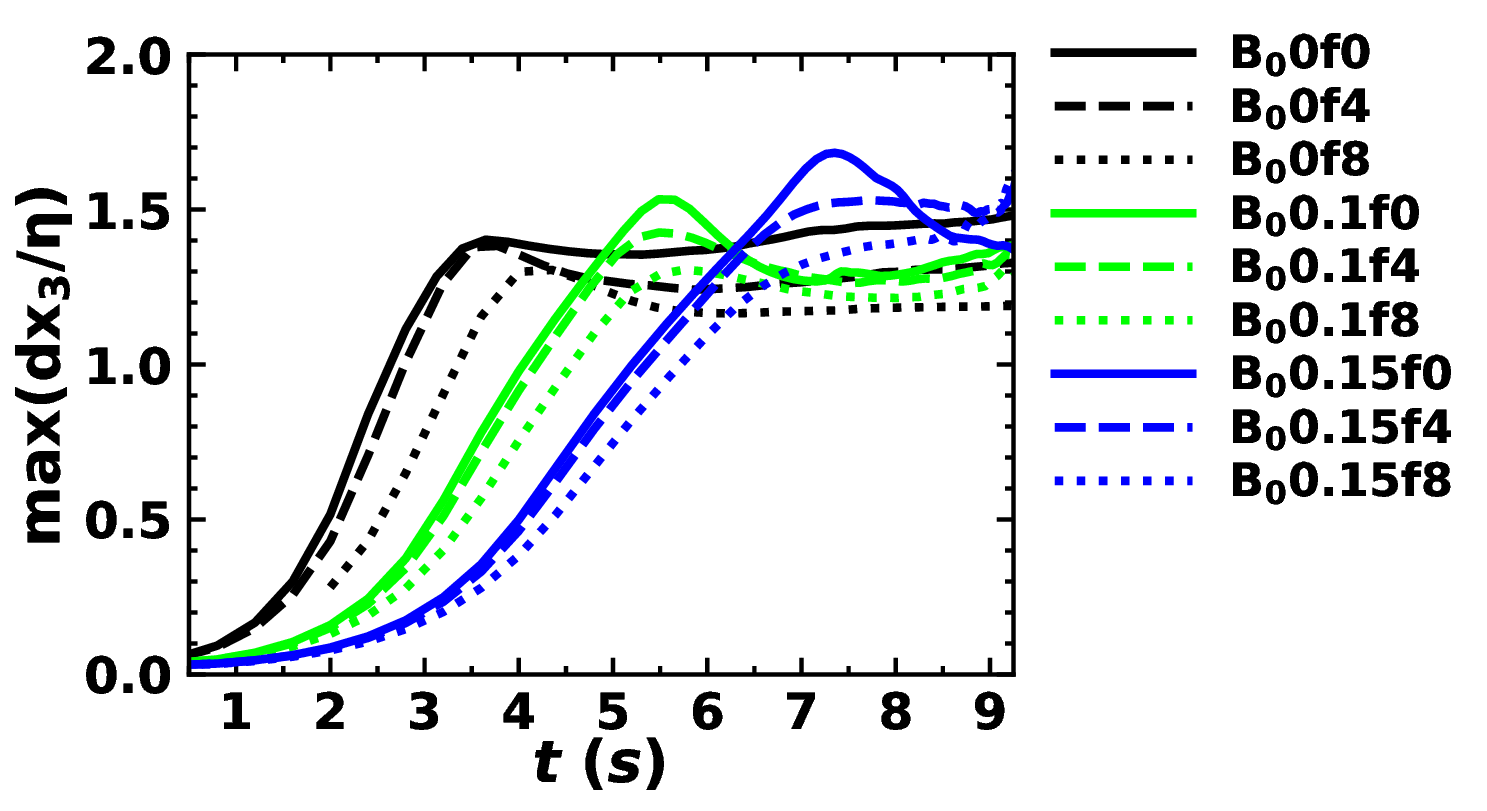} 
    \hspace{-3.0cm}
    \caption{(\textit{a}) Grid clustering at the vertical center region for resolving all the length scales in the mixing layer shown in the vertical $x_2-x_3$ center plane ($x_1=0$). (\textit{b}) Temporal evolution of the maximum vertical grid spacing $dx_3$ normalized with the Kolmogorov length scale $\eta=\left(\nu^3/\epsilon_v \right)^{1/4}$ for non-rotation and rotation cases with $B_0=0,0.1,0.15$. Here $\nu$ is the kinematic viscosity and $\epsilon_v$ is the horizontally averaged viscous dissipation rate (defined later in equation \ref{diss}).}
 \label{fig: resolution}
\end{figure*}

 We perform all simulations in the computational domain of size $L_{x_1}=L_{x_2}=2\pi \:m$ in the horizontal directions and $L_{x_3}=4.32\pi \:m$ which includes sponge region of thickness $L_s\simeq0.2\pi \:m$ utilizing $20$ grid points at the top and bottom boundaries ($x_3$). These sponge regions are employed near the top and bottom boundaries to control the spurious reflections from the disturbances propagating out of the domain, where damping functions gradually relax the values of the velocities to their corresponding values at the boundary \citep{singh_pal_2023}. This is accomplished by adding the damping functions to the right-hand side of the momentum equation \ref{momentum1}, as explained in \cite{brucker2010comparative}. The sponge region is always kept sufficiently far away from the mixing zone and, therefore, does not affect the dynamics of the mixing of fluids. We use uniform grids $N_{x_1}=N_{x_2}=1024$ in the horizontal directions, while non-uniform grids $N_{x_3}=1536$ are used in the vertical direction $x_3$. The grids are clustered at the vertical center region of thickness $\simeq2.23\pi \:\mathrm{m}$ with $\Delta x_{3_{min}}=\Delta x_1=\Delta x_2$ as illustrated in figure \ref{fig: resolution}a. This grid arrangement maintains a $max(dx_3/\eta) < 2$ \citep{brucker2010comparative}, where $\eta$ is the Kolmogorov scale $\left(\eta=\left(\nu^3/\epsilon_v \right)^{1/4}\right)$, as shown in figure \ref{fig: resolution}b which is fine enough to resolve all the length scales in the mixing region. Here $\nu$ is the kinematic viscosity and $\epsilon_v$ is the horizontally averaged viscous dissipation rate, which is defined later in equation \ref{diss}. We summarize all the parameters for different simulation cases in table \ref{tab: parameters}. Hereafter, we refer to non-magnetic ($B_0 = 0$) cases as hydrodynamic (HD) cases and magnetic ($B_0 \neq 0$) cases as MHD cases. We refer to each simulation case with a unique name; for example, the B$_0$0.1f4 case represents $B_0=0.1$ and $f=4$. \\  


 \begin{table}
  \begin{center}
  \def~{\hphantom{0}}
  \setlength{\tabcolsep}{15pt} 
  \renewcommand{\arraystretch}{1.2} 
  \begin{tabular}{lccc}
     Case          & $B_0\:(\mathrm{ms^{-1}})$   & $f\:(\mathrm{s^{-1}})$ 
     \\[3pt]
       B$_0$0f0    & 0  & 0  
       \\
       B$_0$0f4    & 0  & $ 4$
       \\
       B$_0$0f8    & 0  & $8$
       \\       
       B$_0$0.1f0  & 0.1 & 0  
       \\
       B$_0$0.1f4  & 0.1 & $ 4$    
       \\
       B$_0$0.1f8  & 0.1 & $ 8$    
       \\
       B$_0$0.15f0  & 0.15 & 0            
       \\
       B$_0$0.15f4  & 0.15 & $ 4$  
       \\
       B$_0$0.15f8  & 0.15 & $8$  
       \\
       B$_0$0.3f0  & 0   & $0$  
       \\
       B$_0$0.3f4  & 0.3 & $4$   
       \\
       B$_0$0.3f8  & 0.3 & $ 8$   
       \\
  \end{tabular}
  \caption{Simulation parameters: For all cases we use Atwood number $\mathcal{A}g=1.0$, $\beta g = 1.0$, $g=9.81\:\mathrm{m\:s^{-2}}$, $\Theta=2$, kinematic viscosity $\nu$,  thermal diffusivity $\kappa$, and magnetic diffusivity $\mathcal{D}$ respectively are $\nu=\kappa=\mathcal{D}=3\times10^{-4}\:\mathrm{m^2\:s^{-1}}$, such that Pradntl number $Pr=\nu / \kappa=1$ and magnetic Pradntl number $Pr_m=\nu/\mathcal{D}=1$. }  \label{tab: parameters}
  \end{center}
\end{table}

\section{Results}\label{sec: results}
\begin{figure}
 	\centering
  	\begin{subfigure}{0.245\textwidth}
  		\centering
        \caption{}  \label{Temp B00f0}
  		\includegraphics[width=1.0\textwidth,trim={0cm 0cm 1cm 0cm},clip]{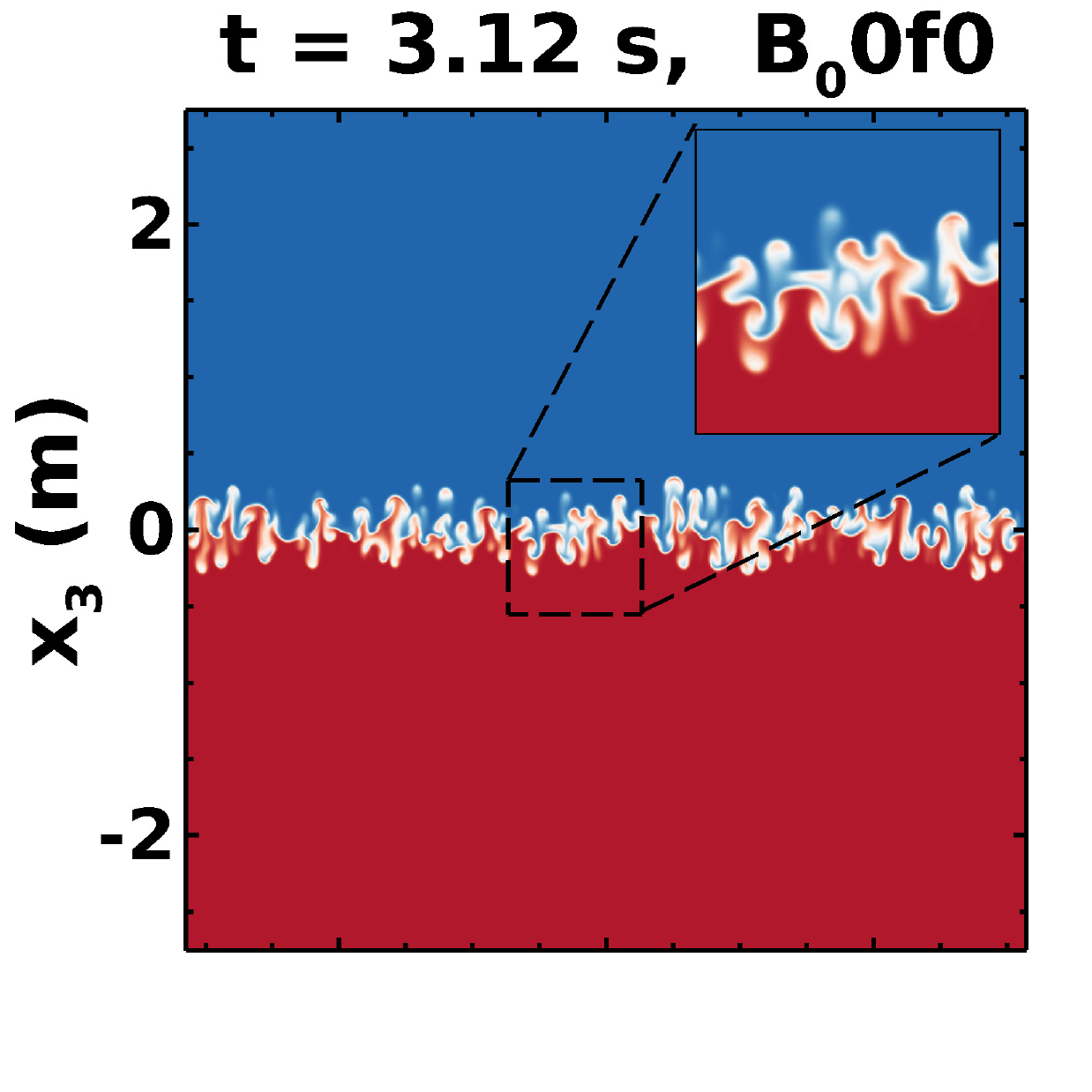}   	 
  	\end{subfigure}
  	\begin{subfigure}{0.245\textwidth}
  		\centering
         \caption{}  \label{Temp B00f8}
  		\includegraphics[width=1.0\textwidth,trim={0cm 0cm 1cm 0cm},clip]{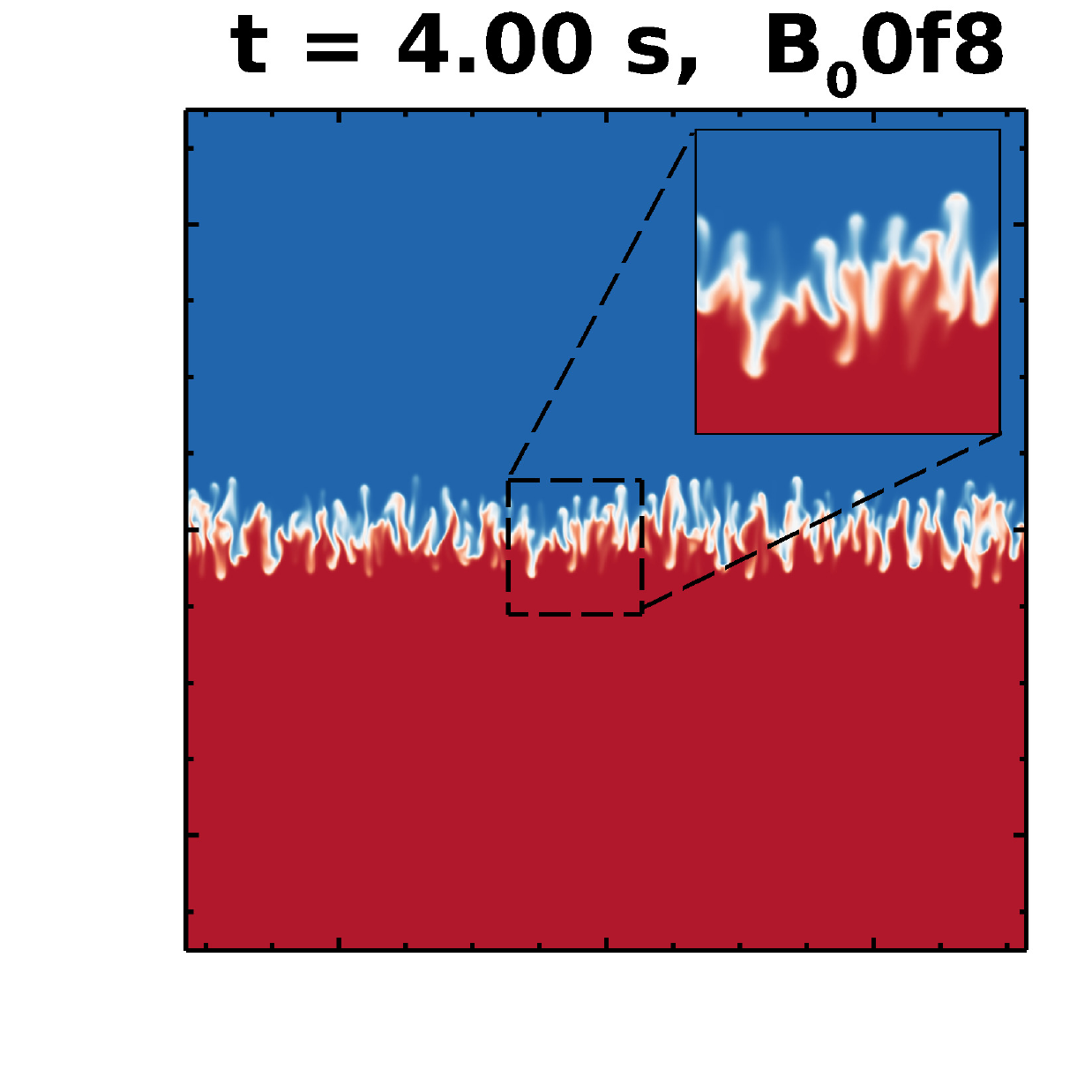}
   	\end{subfigure}
  	\begin{subfigure}{0.245\textwidth}
  		\centering
        \caption{}  \label{Temp B015f0}
  		\includegraphics[width=1.0\textwidth,trim={0cm 0cm 1cm 0cm},clip]{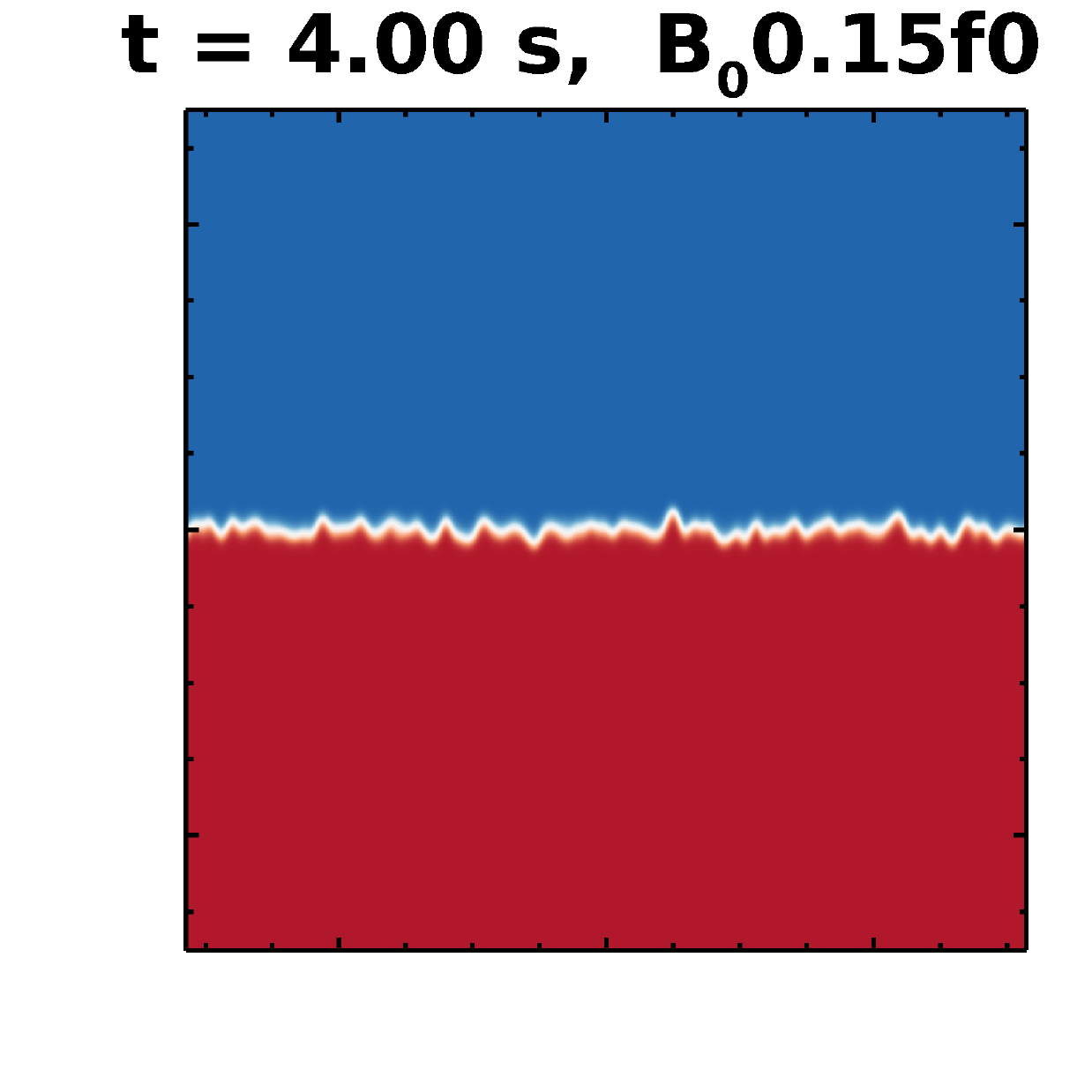}
  	\end{subfigure}
  	\begin{subfigure}{0.245\textwidth}
  		\centering
        \caption{}  \label{Temp B015f8}
        \includegraphics[width=1.0\textwidth,trim={0cm 0cm 1cm 0cm},clip]{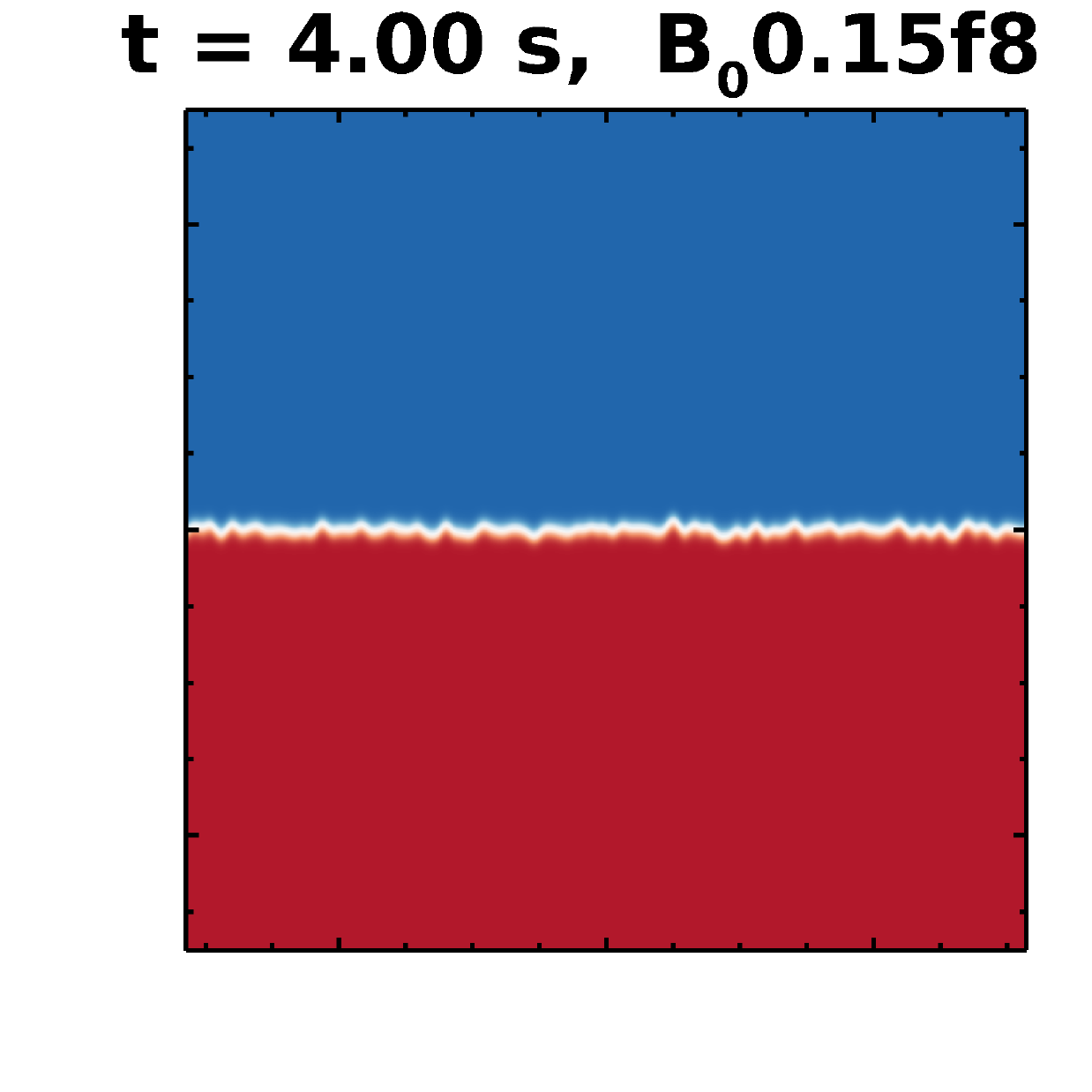}
  	\end{subfigure}   \\
  	\begin{subfigure}{0.245\textwidth}
  		\centering
  		\includegraphics[width=1.0\textwidth,trim={0cm 0cm 1cm 0cm},clip]{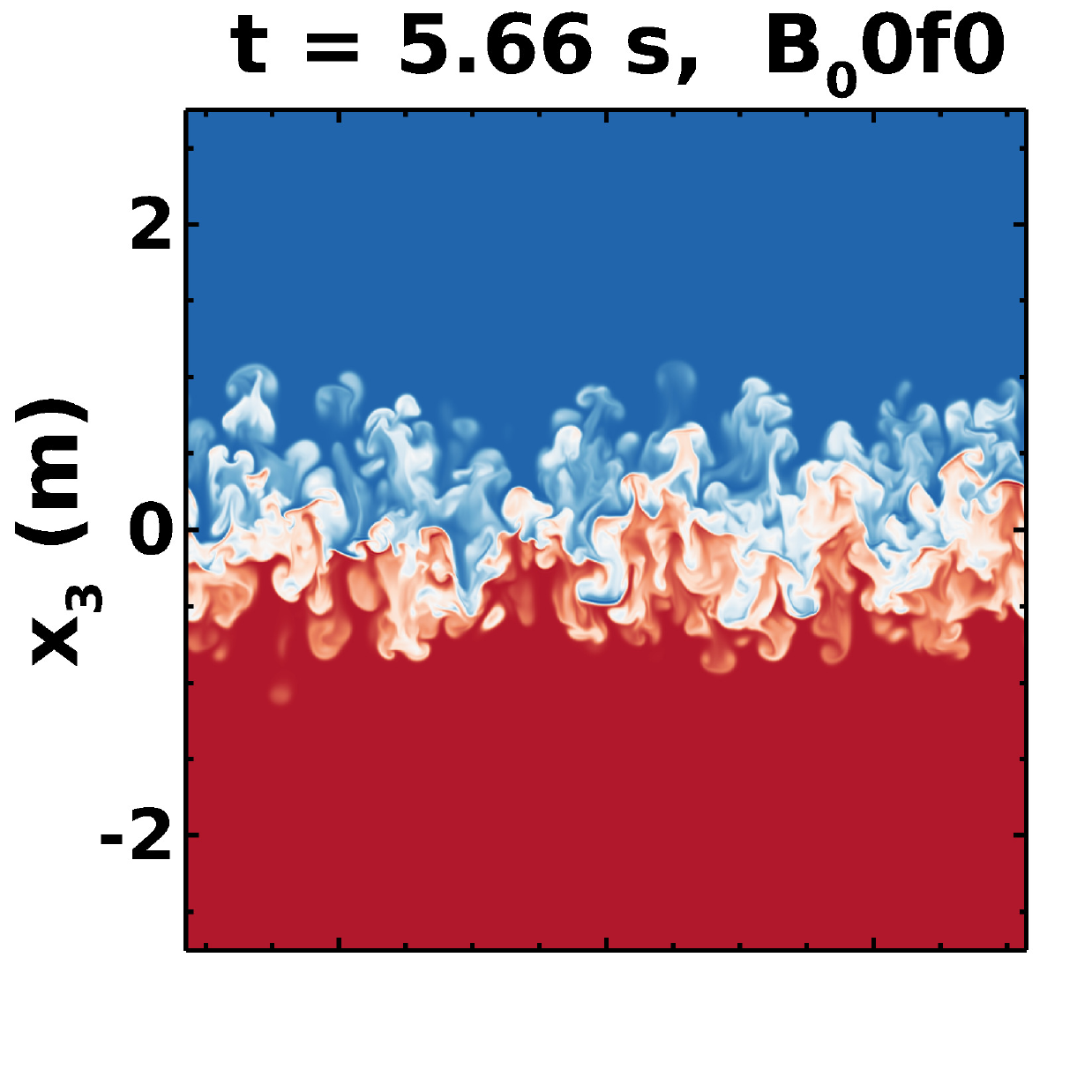}
  	\end{subfigure}
  	\begin{subfigure}{0.245\textwidth}
  		\centering
  		\includegraphics[width=1.0\textwidth,trim={0cm 0cm 1cm 0cm},clip]{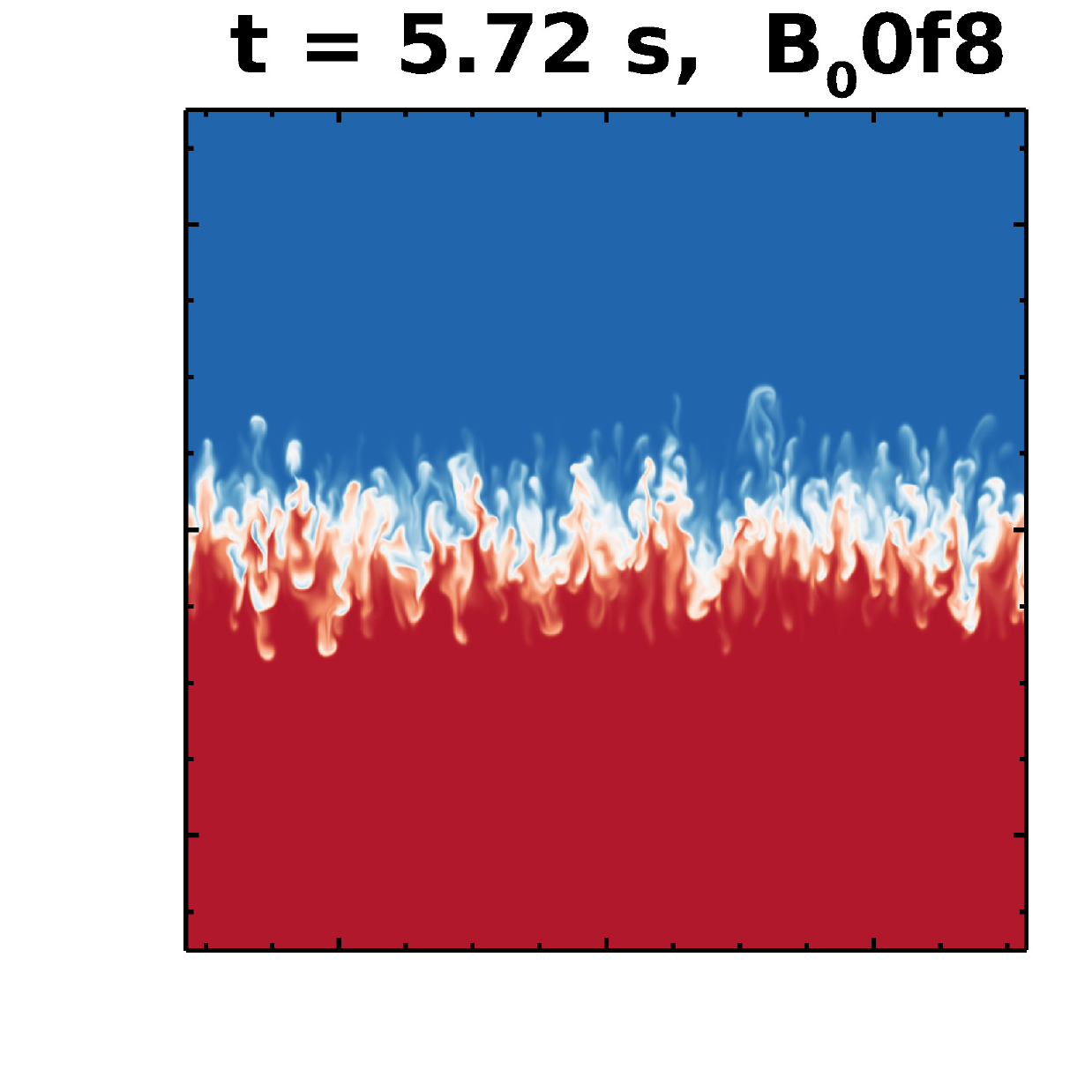}
  	\end{subfigure}
  	\begin{subfigure}{0.245\textwidth}
  		\centering
  		\includegraphics[width=1.0\textwidth,trim={0cm 0cm 1cm 0cm},clip]{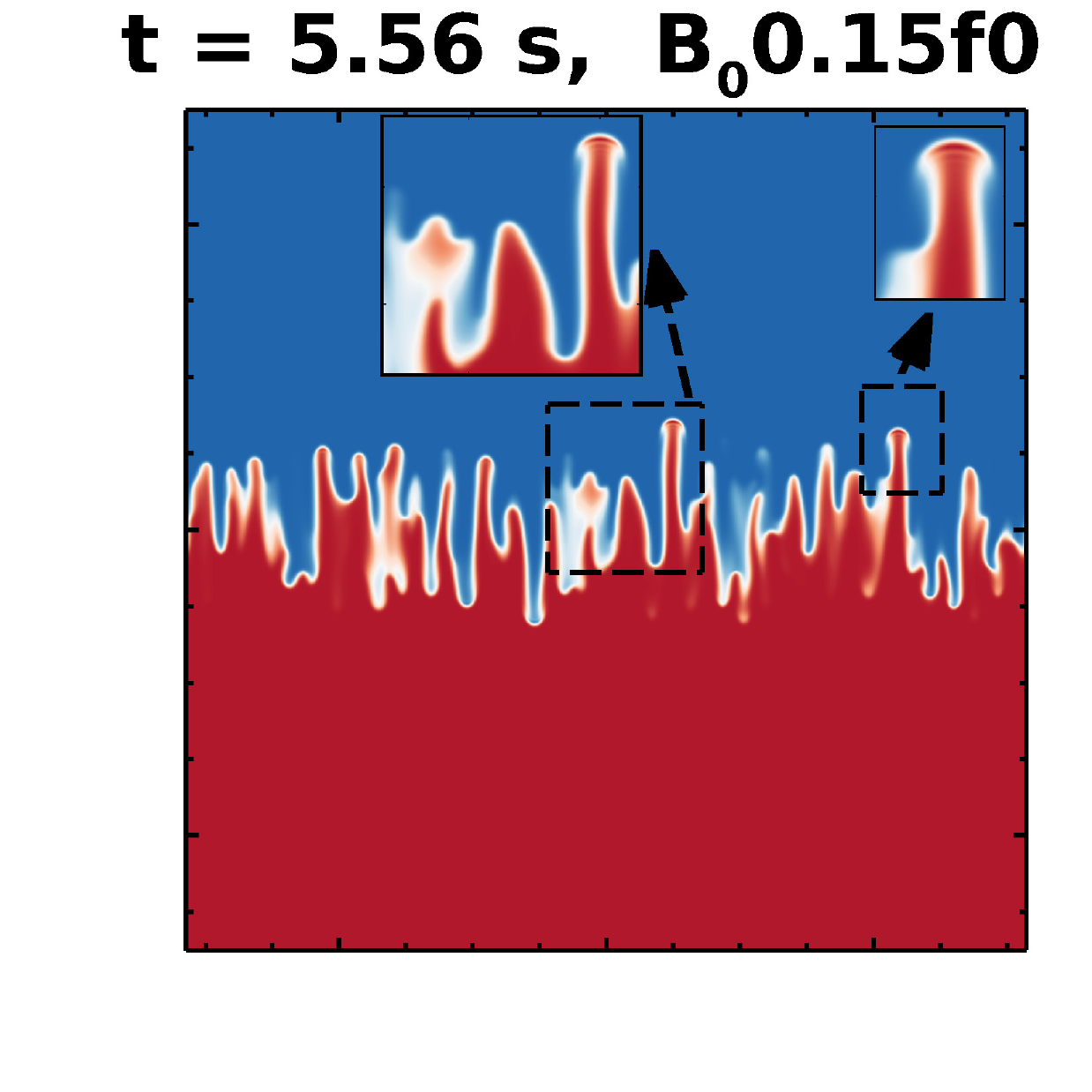}
  	\end{subfigure}
  	\begin{subfigure}{0.245\textwidth}
  		\centering
  		\includegraphics[width=1.0\textwidth,trim={0cm 0cm 1cm 0cm},clip]{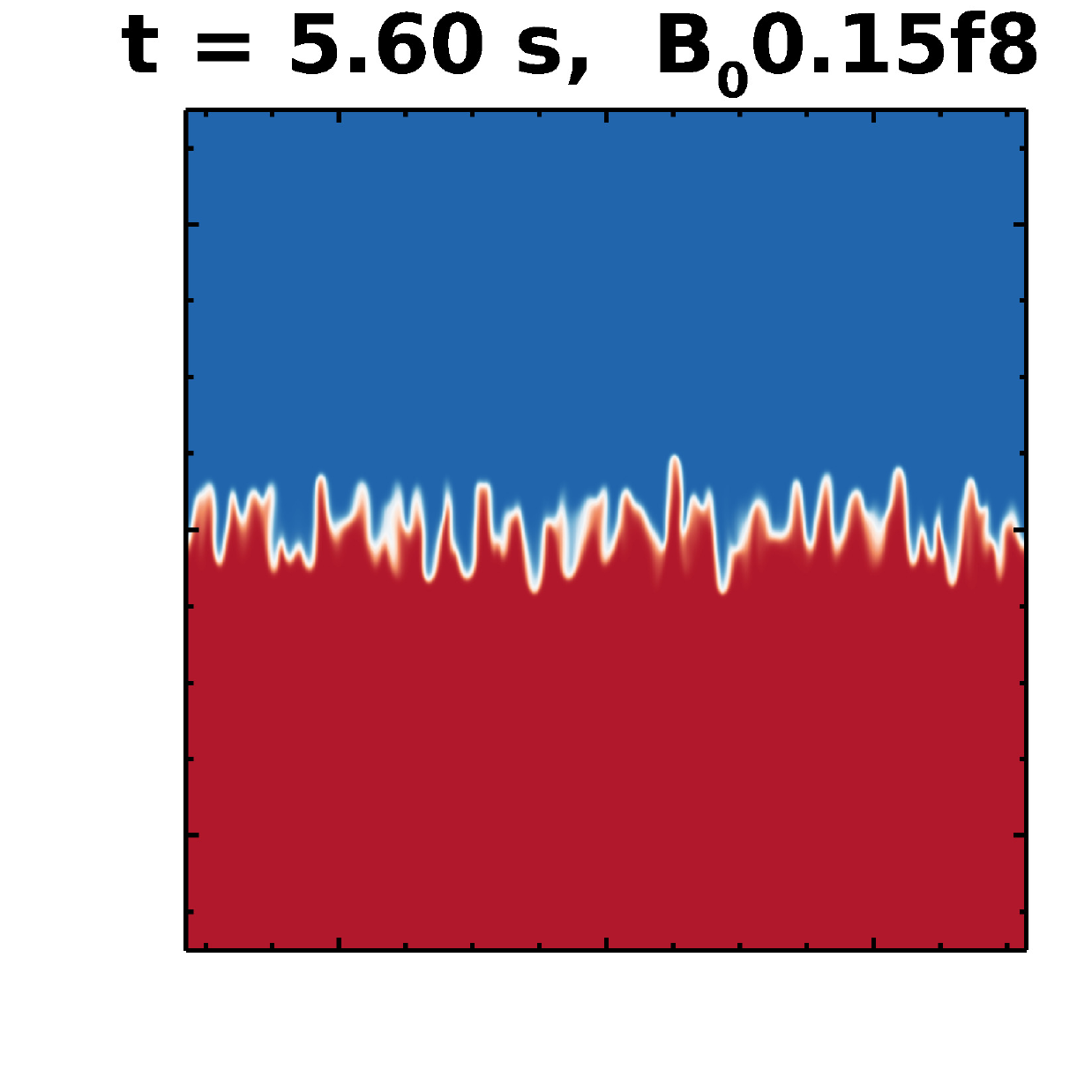}
  	\end{subfigure}   \\
    \begin{subfigure}{0.245\textwidth}
  		\centering
  		\includegraphics[width=1.0\textwidth,trim={0cm 0cm 1cm 0cm},clip]{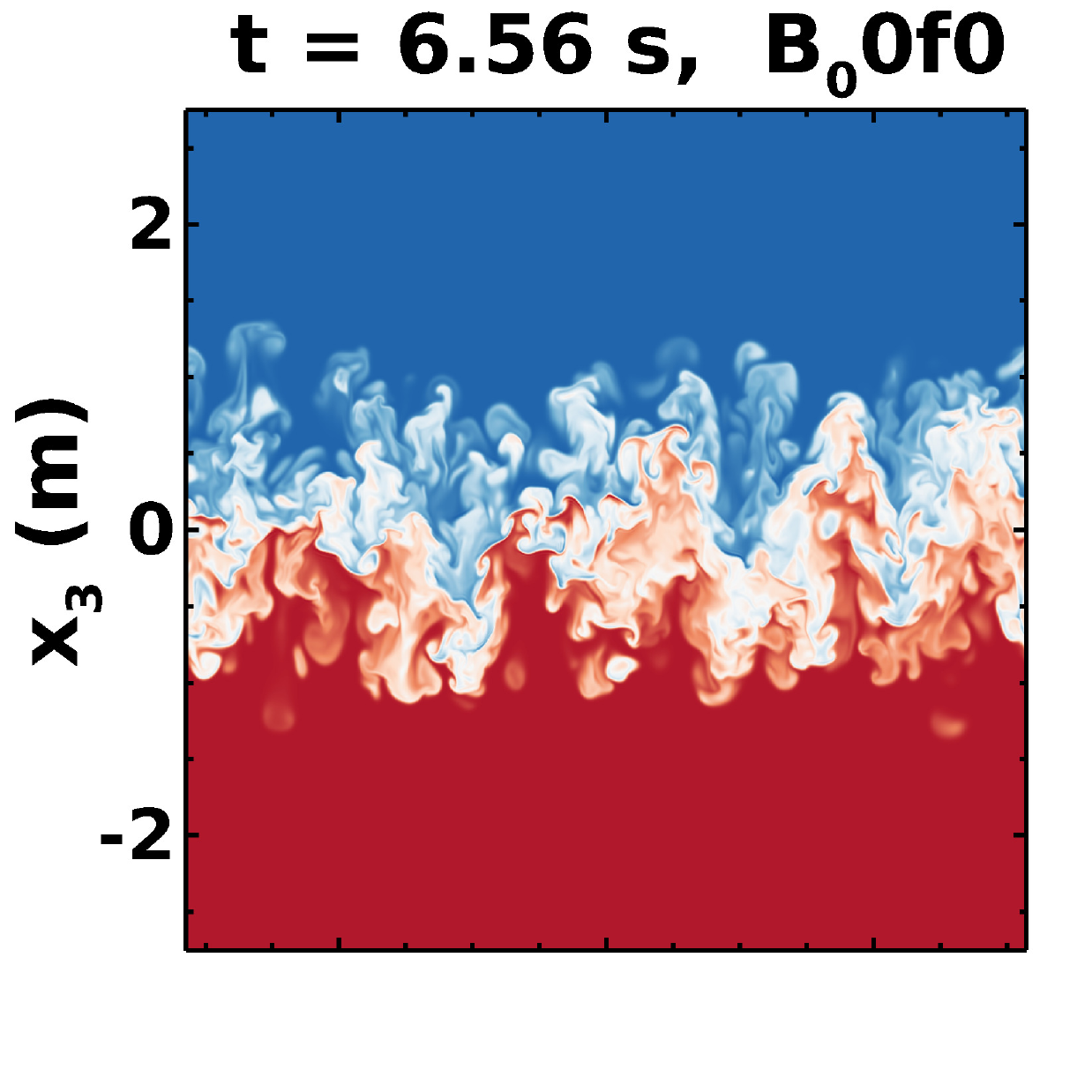}
  	\end{subfigure}
  	\begin{subfigure}{0.245\textwidth}
  		\centering
  		\includegraphics[width=1.0\textwidth,trim={0cm 0cm 1cm 0cm},clip]{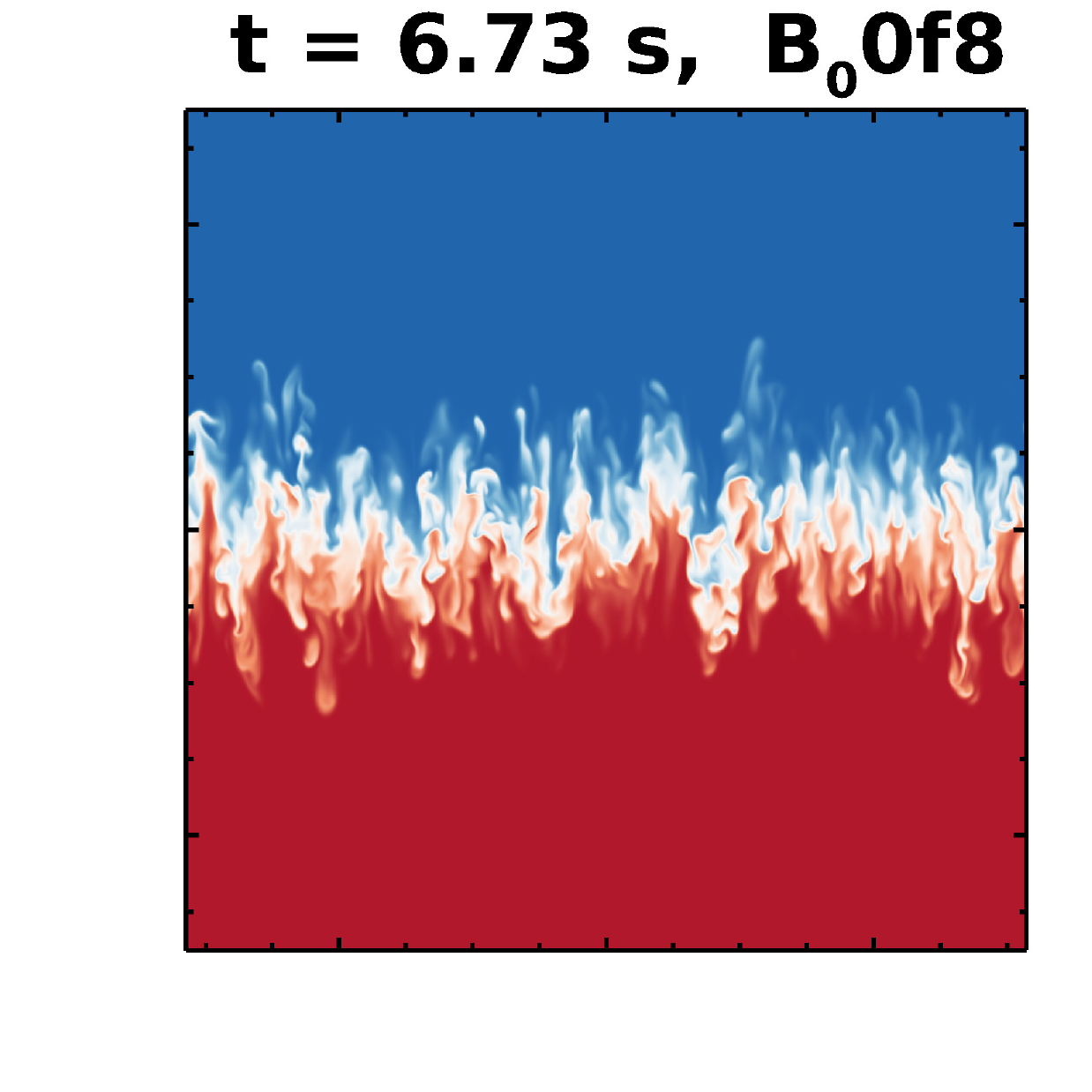}
  	\end{subfigure}
  	\begin{subfigure}{0.245\textwidth}
  		\centering
  		\includegraphics[width=1.0\textwidth,trim={0cm 0cm 1cm 0cm},clip]{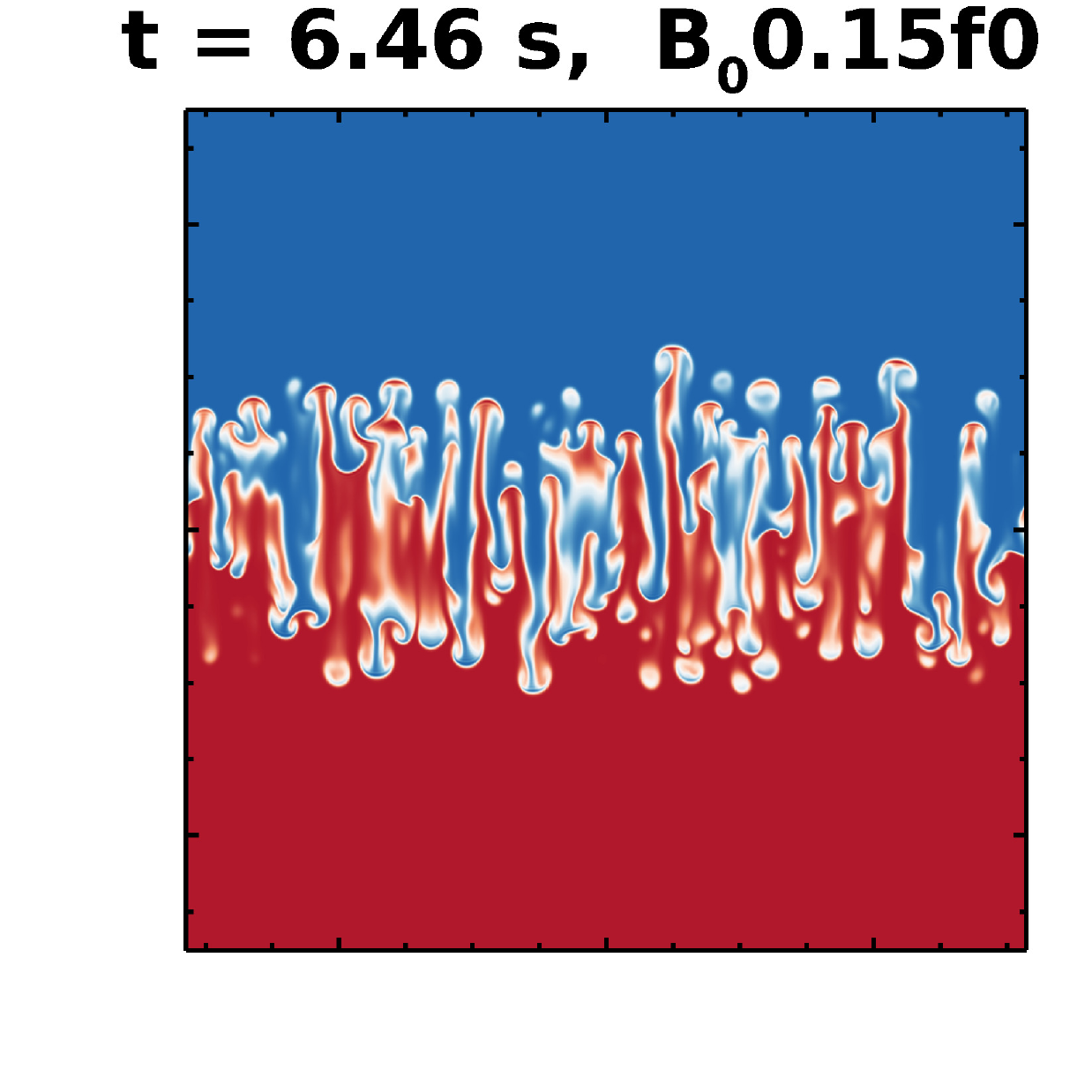}
  	\end{subfigure}
  	\begin{subfigure}{0.245\textwidth}
  		\centering
  		\includegraphics[width=1.0\textwidth,trim={0cm 0cm 1cm 0cm},clip]{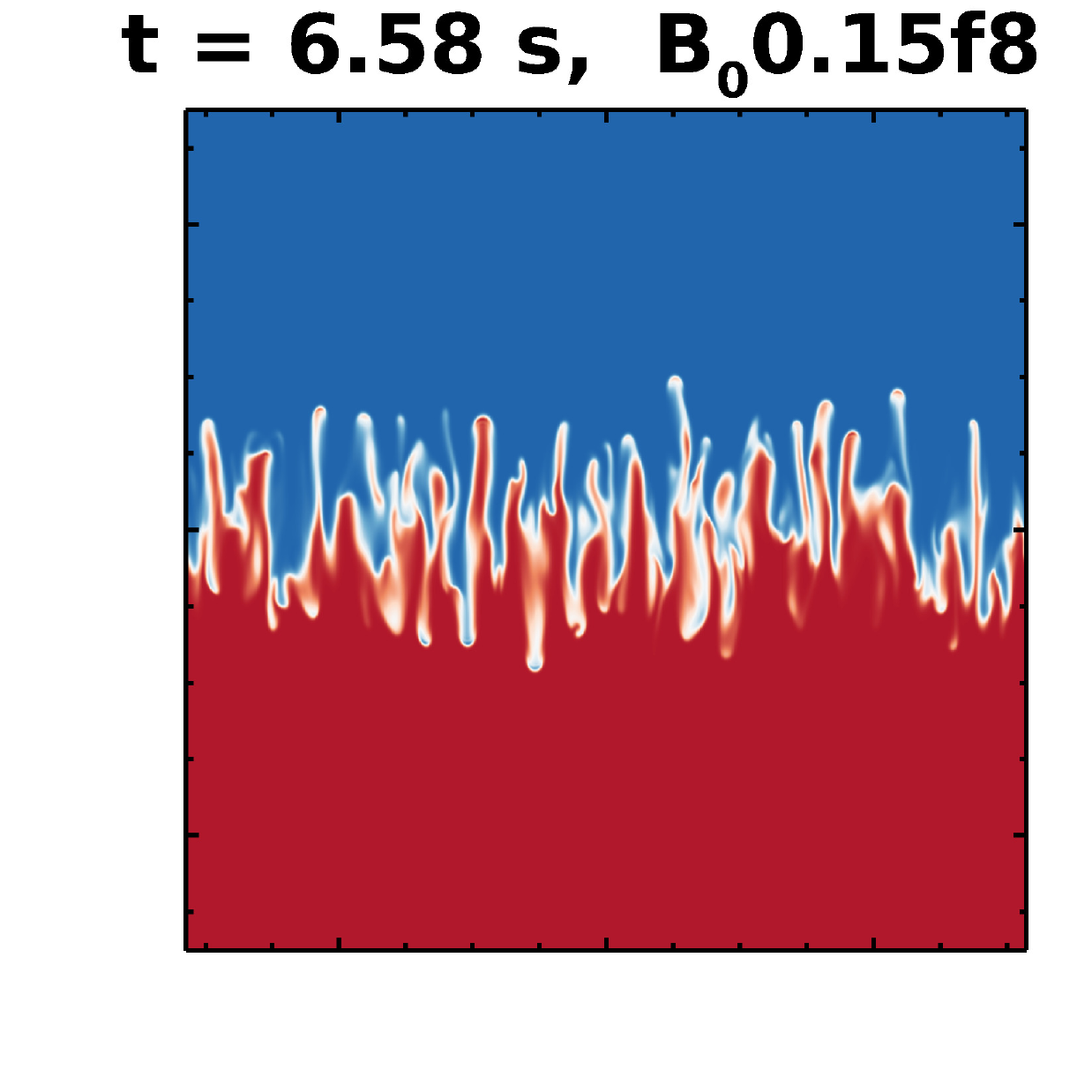}
  	\end{subfigure}  \\
    \begin{subfigure}{0.245\textwidth}
  		\centering
  		\includegraphics[width=1.0\textwidth,trim={0cm 0cm 1cm 0cm},clip]{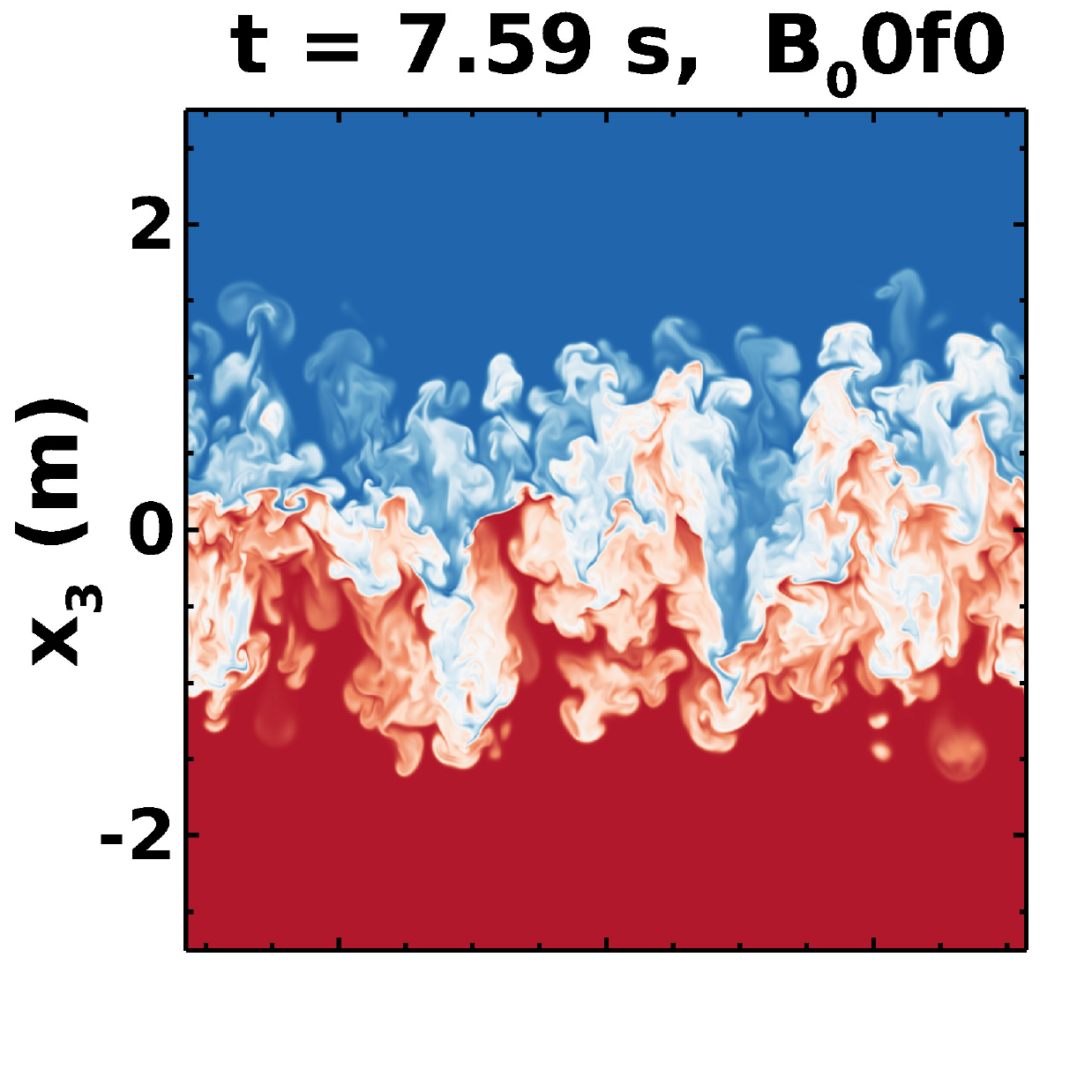}
  	\end{subfigure}
  	\begin{subfigure}{0.245\textwidth}
  		\centering
  		\includegraphics[width=1.0\textwidth,trim={0cm 0cm 1cm 0cm},clip]{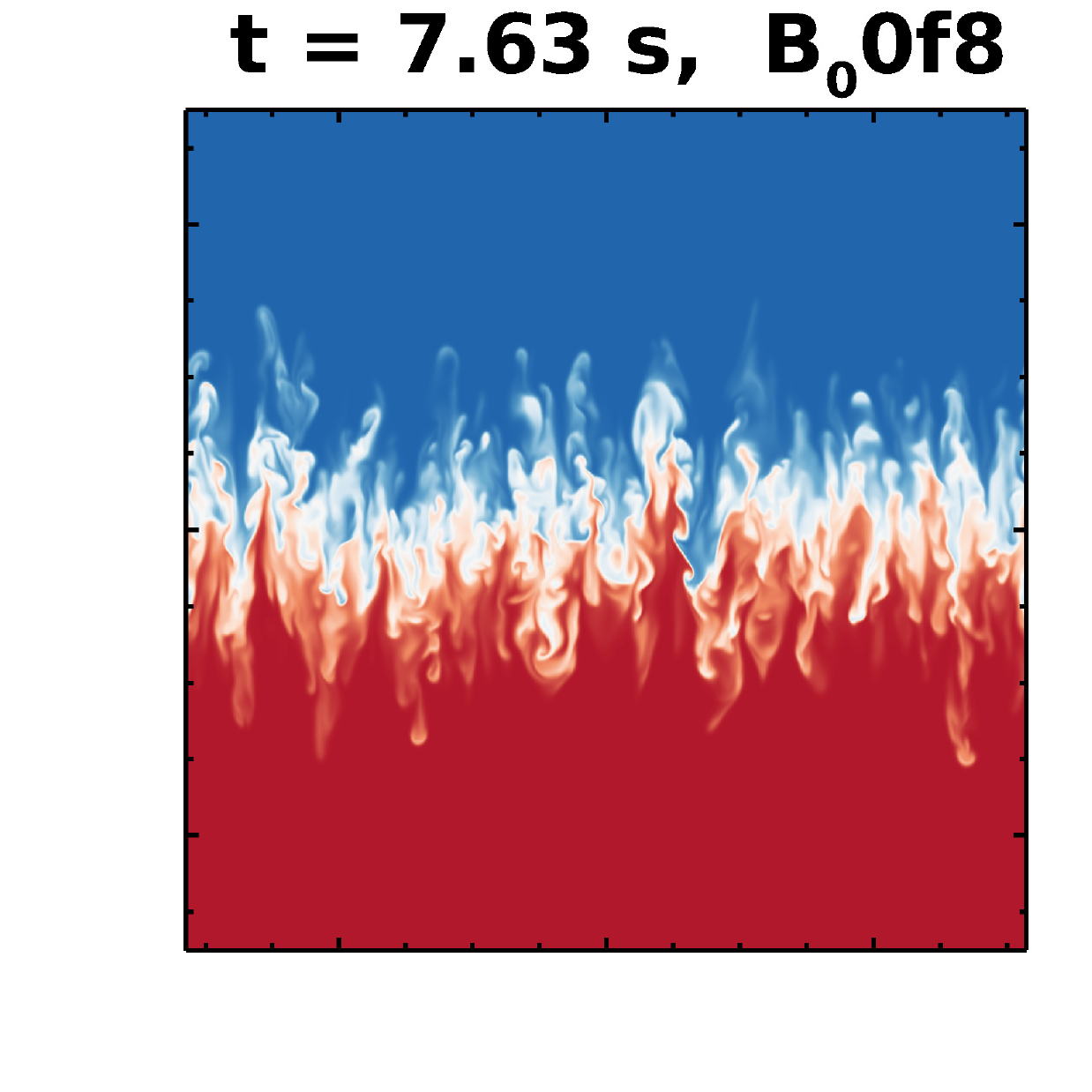}
  	\end{subfigure}
  	\begin{subfigure}{0.245\textwidth}
  		\centering
  		\includegraphics[width=1.0\textwidth,trim={0cm 0cm 1cm 0cm},clip]{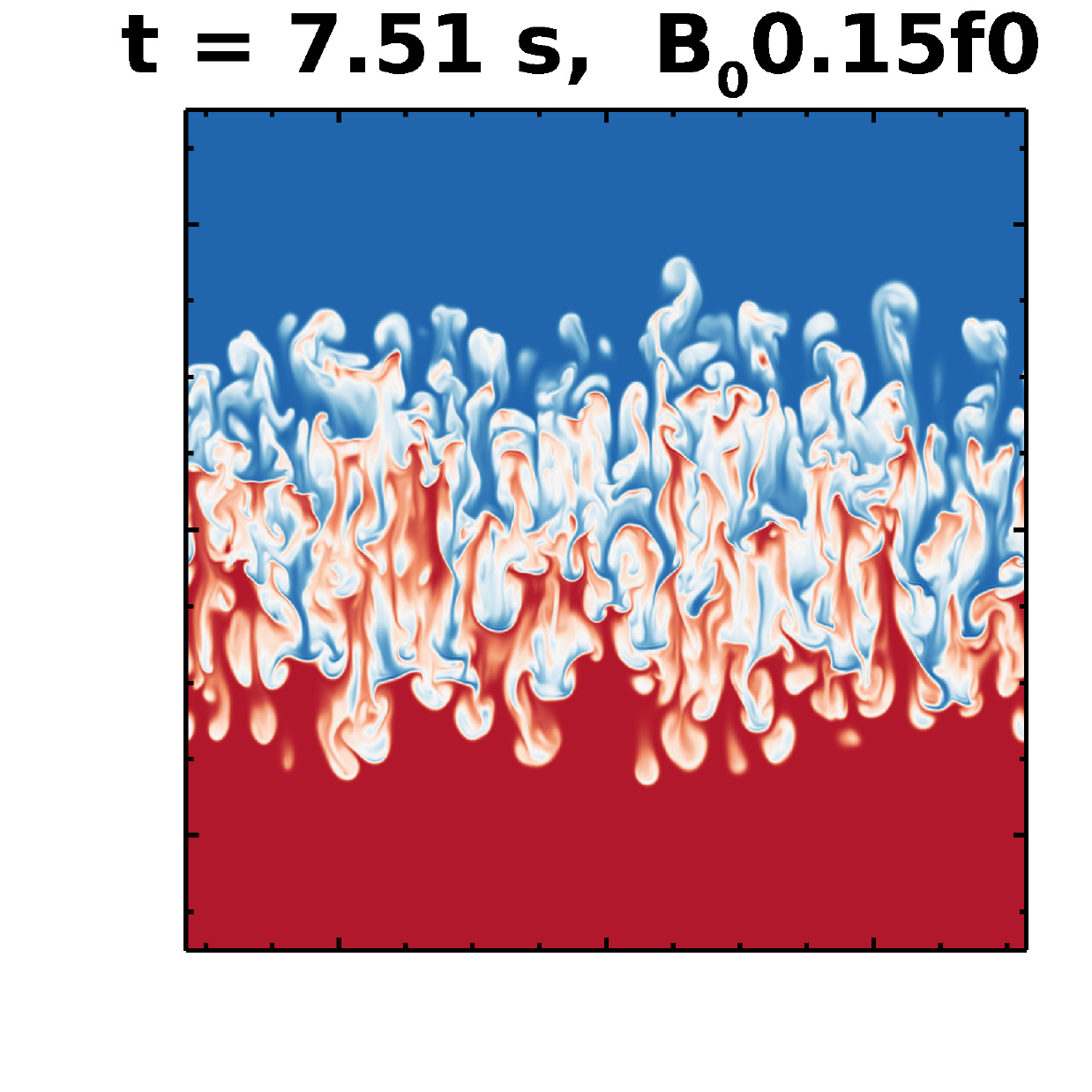}
  	\end{subfigure}
  	\begin{subfigure}{0.245\textwidth}
  		\centering
  		\includegraphics[width=1.0\textwidth,trim={0cm 0cm 1cm 0cm},clip]{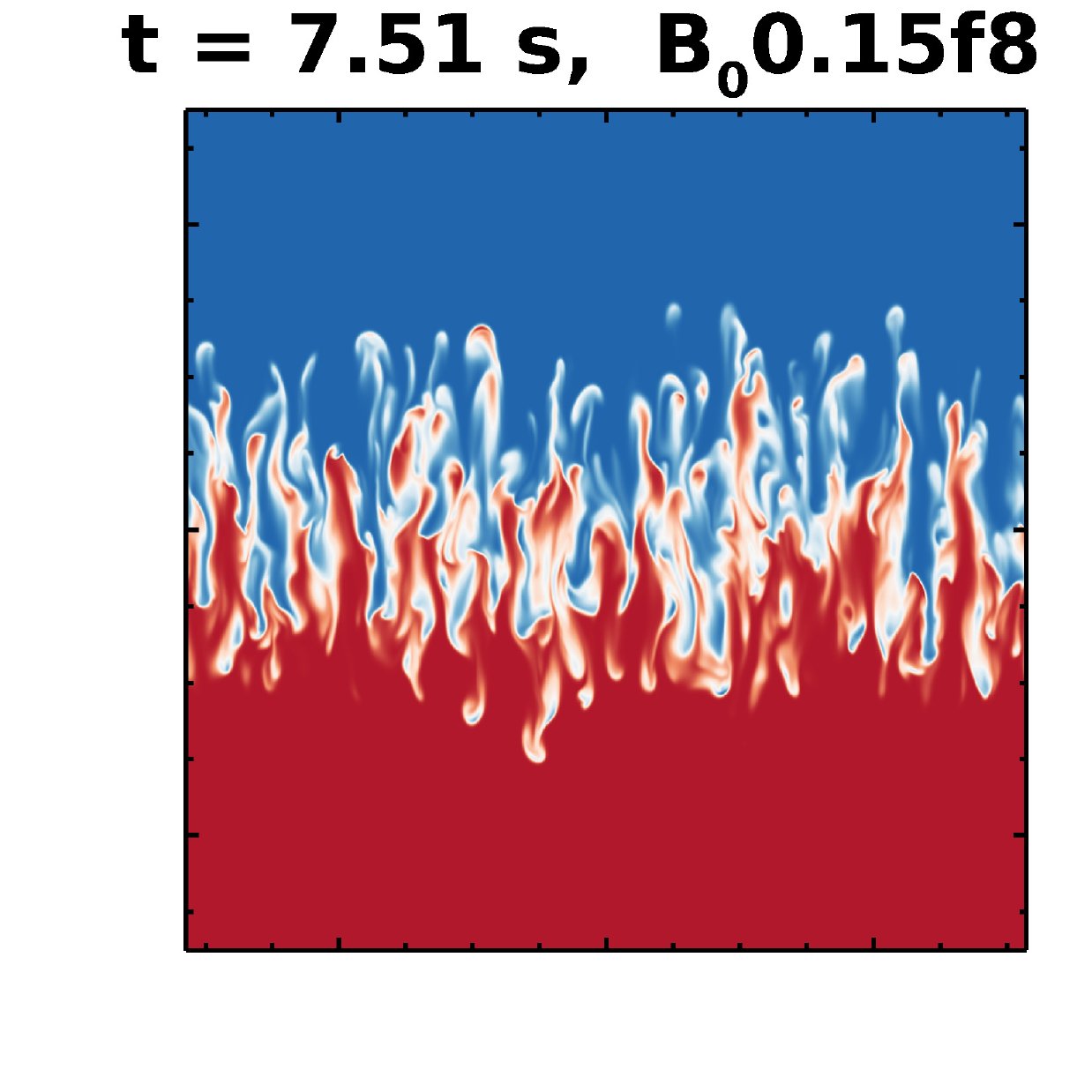}
  	\end{subfigure}  \\
    \begin{subfigure}{0.245\textwidth}
  		\centering
  		\includegraphics[width=1.0\textwidth,trim={0cm 0cm 1cm 0cm},clip]{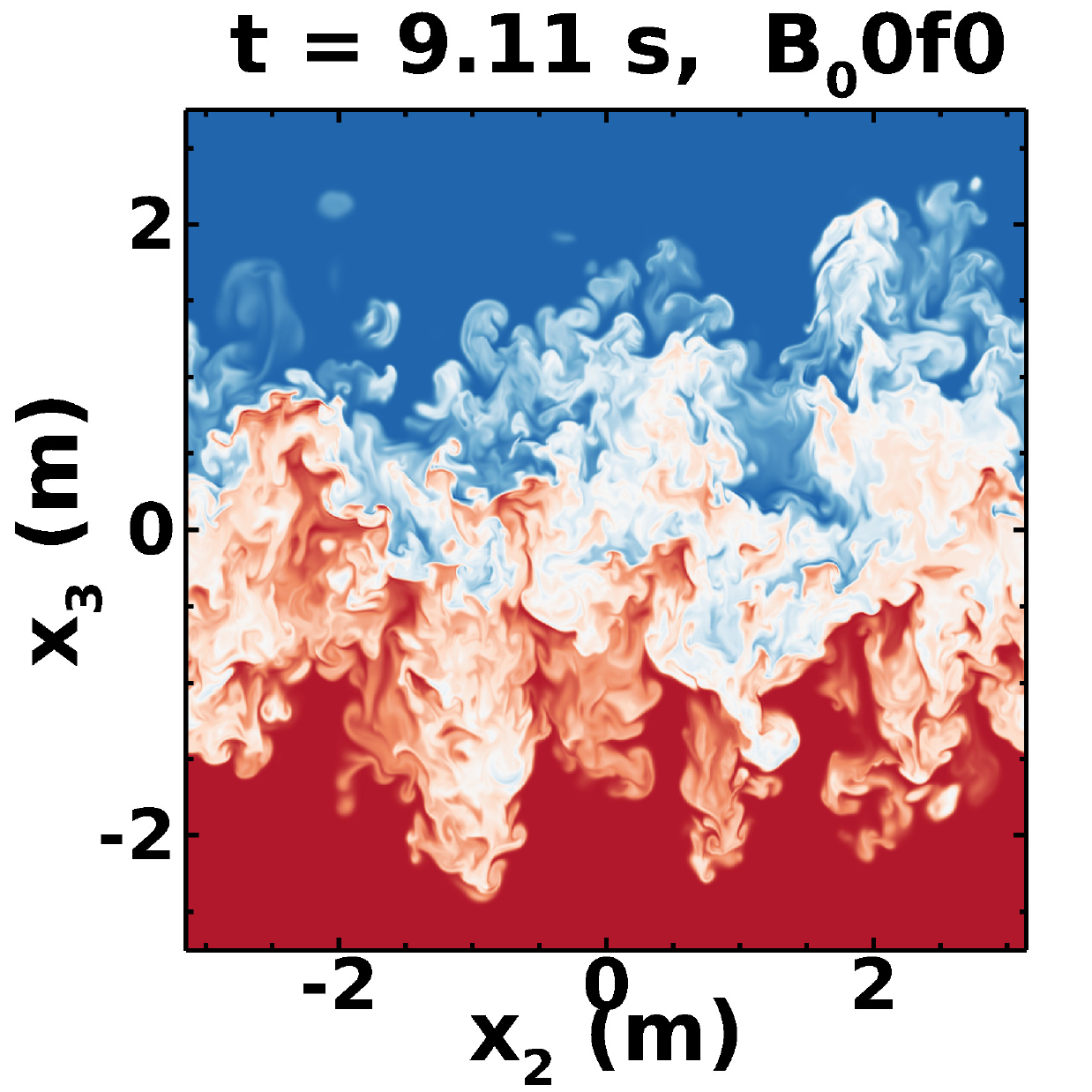}
  	\end{subfigure}
  	\begin{subfigure}{0.245\textwidth}
  		\centering
  		\includegraphics[width=1.0\textwidth,trim={0cm 0cm 1cm 0cm},clip]{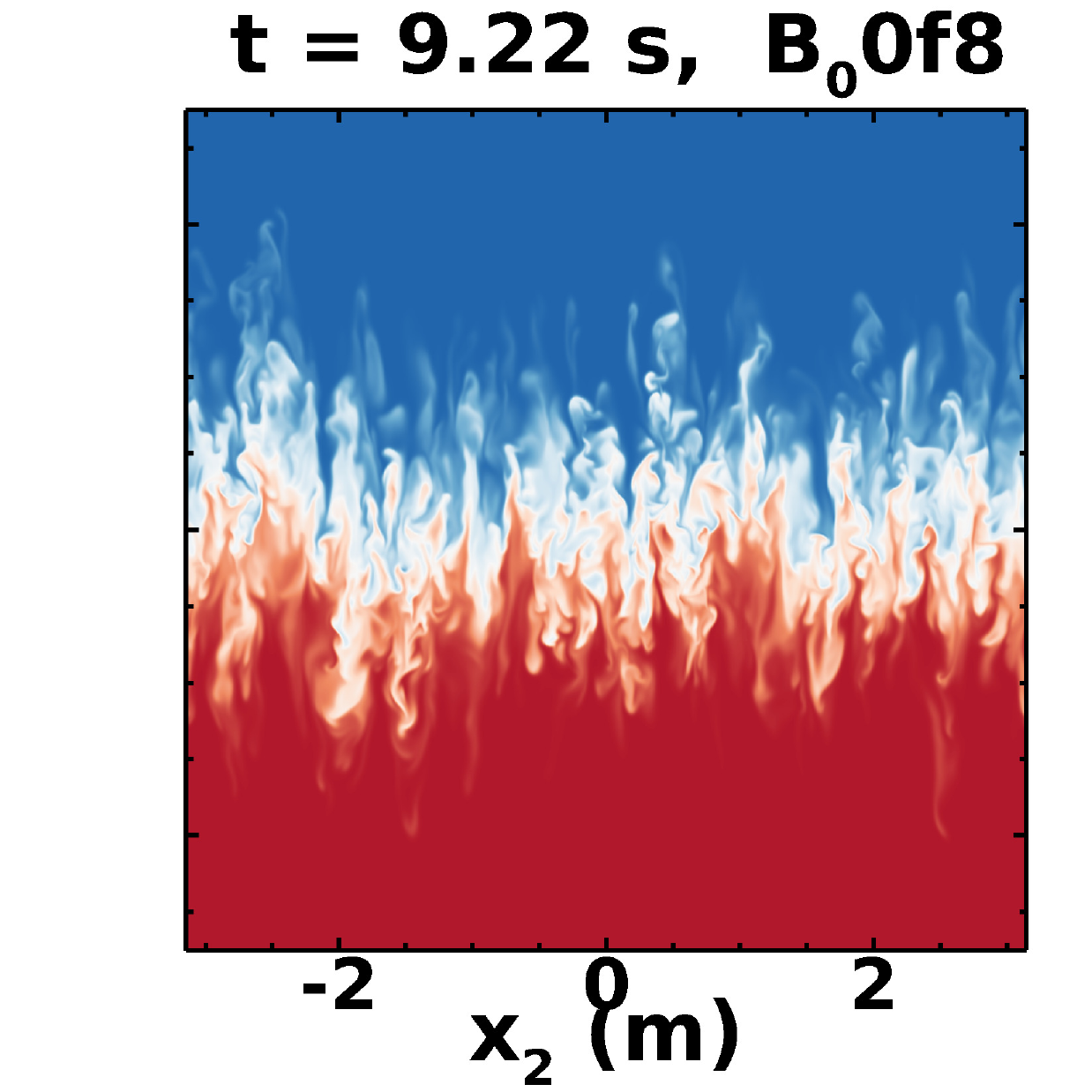}
  	\end{subfigure}
  	\begin{subfigure}{0.245\textwidth}
  		\centering
  		\includegraphics[width=1.0\textwidth,trim={0cm 0cm 1cm 0cm},clip]{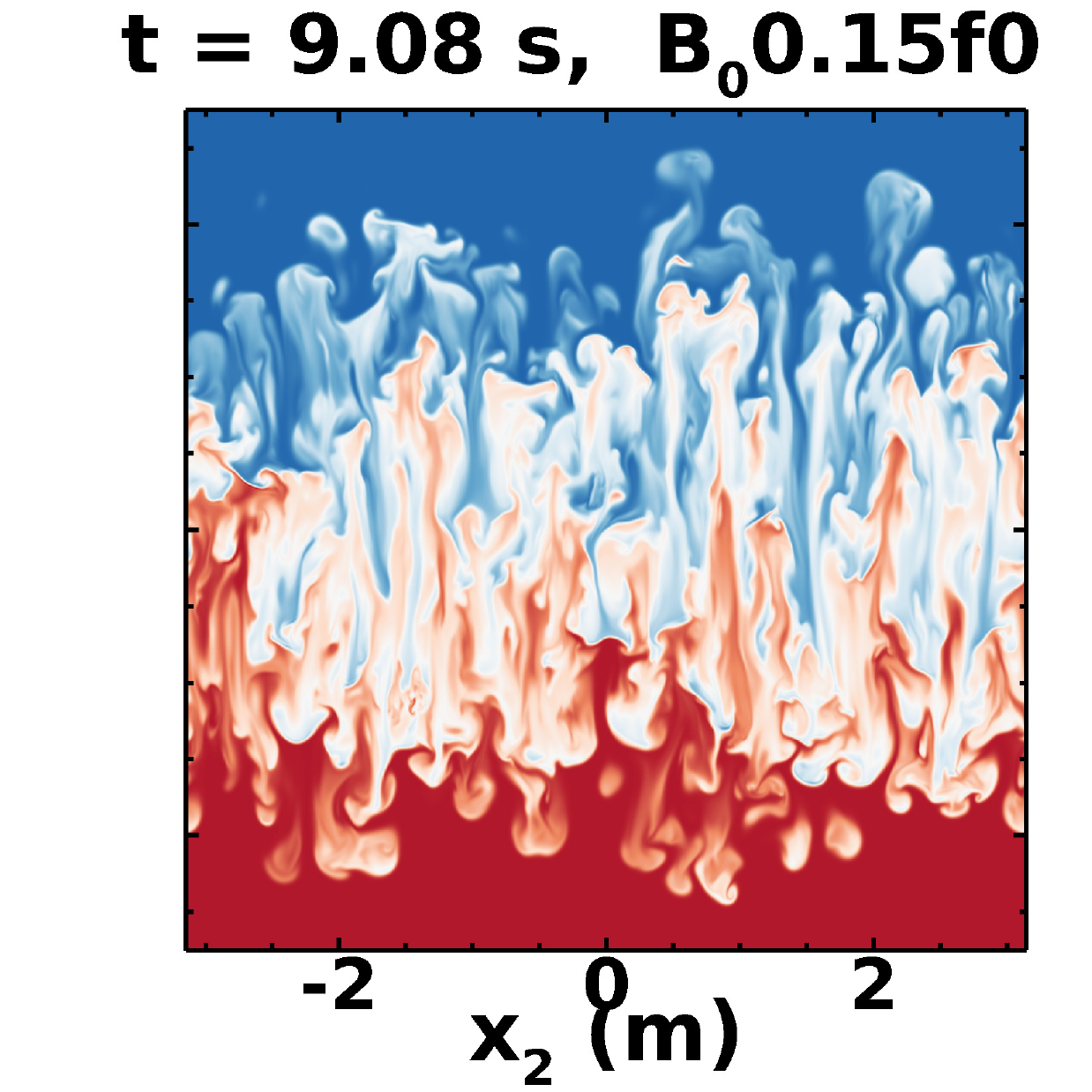}
  	\end{subfigure}
  	\begin{subfigure}{0.245\textwidth}
  		\centering
  		\includegraphics[width=1.0\textwidth,trim={0cm 0cm 1cm 0cm},clip]{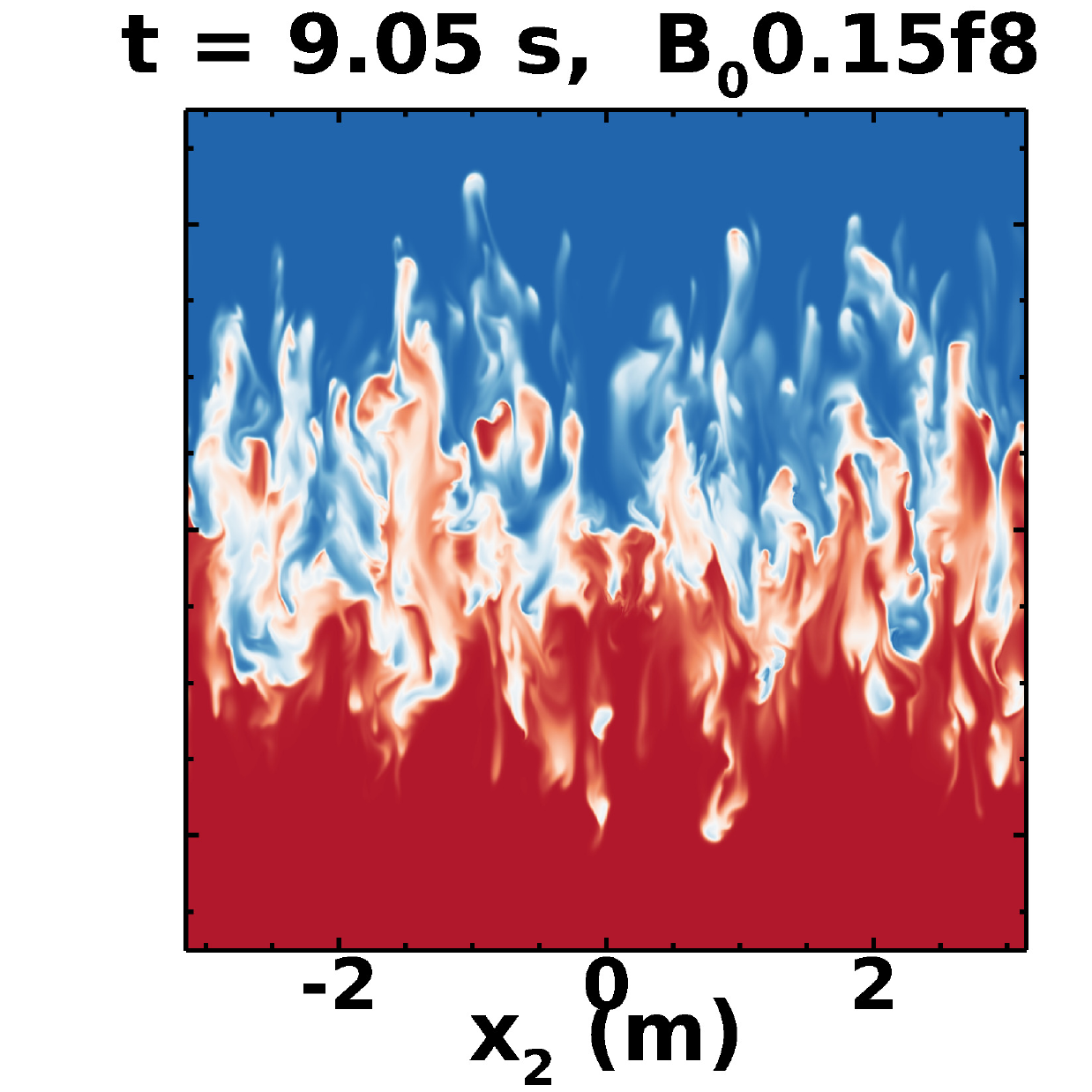}
  	\end{subfigure}\\   
   \begin{subfigure}{0.5\textwidth}
  		\centering
  		\includegraphics[width=1.0\textwidth,trim={0.5cm 1cm 0.5cm 1cm},clip]{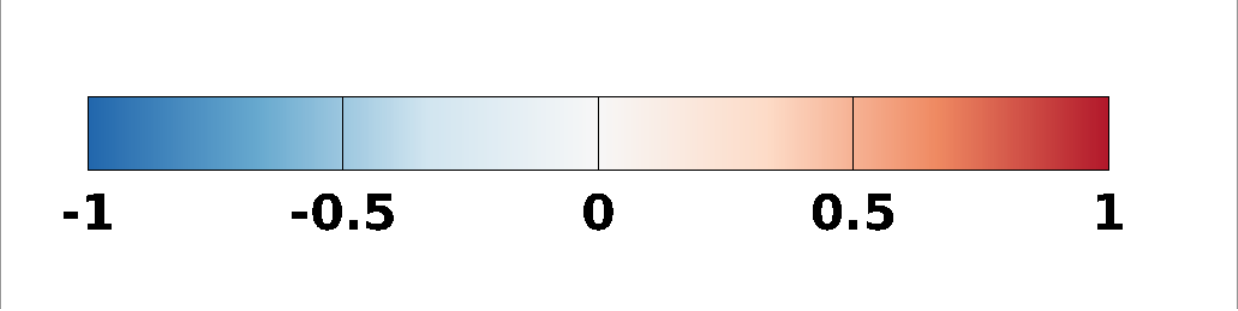}
   		\label{subfig: tempcont}
  	\end{subfigure}   
  	\caption{Instantaneous temperature contours in the vertical $x_2-x_3$ center plane ($x_1=0$) for the (\textit{a}) non-rotating HD case B$_0$0f0 (first left-most column), (\textit{b}) rotating HD case B$_0$0f8 (second column), (\textit{c}) non-rotating MHD case B$_0$0.15f0 (third column), and (\textit{d}) rotating MHD case B$_0$0.15f8 (fourth column) at different time instants. The blue color represents cold fluid ($T=-1$), and the red color represents hot fluid ($T=1$).}
 \label{fig: temp}
\end{figure}

To illustrate the effect of the imposed vertical mean magnetic field and rotation on the formation and evolution of flow structures, we show the snapshots of the temperature field ($T$) at different time instants for the non-rotating HD B$_0$0f0, rotating HD B$_0$0f8, non-rotating MHD B$_0$0.15f0, and rotating MHD B$_0$0.15f8 cases in figures \ref{Temp B00f0}, \ref{Temp B00f8}, \ref{Temp B015f0}, and \ref{Temp B015f8}, respectively. For non-rotating HD case B$_0$0f0, the initial perturbed interface becomes unstable to the downward acting gravitational accelerations ($g$) (or RT instability), resulting in the growth of perturbations. This leads to the emergence of small-scale rising (hot fluid) and sinking (cold fluid) thermal plumes with mushroom-shaped caps at their tips, as depicted in the enlarged view in figure \ref{Temp B00f0} at $t=3.12$s These caps are attributed to the secondary Kelvin–Helmholtz (KH) instability generated by the relative motion between the plumes \citep{sharp1984overview,jun1995numerical,abarzhi2010review,ZHOU20171a,ZHOU20171b}. The plumes start interacting non-linearly with each other, penetrating the opposite region (see figure \ref{Temp B00f0} at $t=5.66$s) and hence, enhancing the heat transfer between the fluids. Eventually, turbulent mixing occurs, and the mixing layer grows continuously due to the conversion of potential energy into turbulent kinetic energy \citep{cabot2006reynolds,boffetta2017incompressible,boffetta2022dimensional}. The evolution of the turbulent mixing layer is illustrated at $t=6.56$s, $7.59$s, and $9.11$s in figure \ref{Temp B00f0}. 
 \captionsetup[subfigure] {textfont=normalfont,singlelinecheck=off,justification=raggedright, labelfont=bf, textfont=bf,font=large}
At a high rotation rate $f=8$ (HD case B$_0$0f8), the formation of mushroom-shaped caps at the plume tips is inhibited, reducing the deformation of the plumes and their horizontal interactions with each other, as depicted by the enlarged view in the figure \ref{Temp B00f8} at $t=4$s. This signifies that the Coriolis force stabilizes the flow, and forms coherent and vertically elongated thermal plumes as apparent at all time instants in figure \ref{Temp B00f8}. These elongated thermal plumes indicate the presence of the Taylor–Proudman constraint imposed by the Coriolis force (as explored later through force analysis), which prevents changes along the rotational axis, resulting in a two-dimensional flow. Additionally, the Coriolis force reduces mixing layer thickness owing to the suppression of fluctuations in the vertical velocity compared to the non-rotating case B$_0$0f0, suggesting a reduction in heat transfer. This can be observed by comparing figure \ref{Temp B00f0} for B$_0$0f0 case with figure \ref{Temp B00f8} for B$_0$0f8 case at the same time instants. The supplementary Movie $1$ and Movie $2$ demonstrate the flow evolution for B$_0$0f0 and B$_0$0f8, respectively. \cite{boffetta2016rotating} and \cite{wei2022small} also report similar effects of rotation.  \\

The inclusion of a mean vertical magnetic field $B_0=0.15$ significantly modifies the formation and growth of the thermal plumes in both the non-rotating ($f=0$) and rotating ($f=8$) cases as demonstrated in figures \ref{Temp B015f0} and \ref{Temp B015f8}, respectively. For the non-rotating MHD case B$_0$0.15f0, in the initial (linear) phase, perturbations are vertically stretched by the mean magnetic field, as depicted in figure \ref{Temp B015f0} at $t=4$s, followed by the emergence of plumes. In contrast to B$_0$0.15f0, the plumes in cases B$_0$0f0 and B$_0$0f8 have already been formed and achieved a significant growth till $t=4$s, signifying the imposed magnetic field delays the onset of the instability. We define the onset of RT instability as the time instant when the mixing layer begins to grow. Since the imposed magnetic field acts tangentially to the sheared interface of the vertically stretched perturbations, the secondary shear instabilities (KH instabilities) at small scales are suppressed (as discussed later), inhibiting the horizontal mixing. Consequently, the vertically stretched perturbations transform to smooth vertically elongated plumes, followed by their rapid growth, as reported by \cite{jun1995numerical} using linear theory and 2-D simulations. Later, the secondary shear instabilities are triggered, which destabilizes the smooth elongated plumes, leading to the emergence of mushroom-shaped caps at their tips, as indicated in the enlarged views at $t\simeq5.56$s in figure \ref{Temp B015f0}. The elongated plumes signify collimation of flow along the magnetic field lines in the vertical direction and indicate that the imposed magnetic field reduces the interactions between the thermal plumes. With time evolution, the rapid vertical stretching of plumes continues, and the relative velocity becomes significant enough to enhance the strength of secondary shear instabilities, resulting in the clear formation of mushroom-shaped caps, as shown at $t=6.46$s in figure \ref{Temp B015f0}. Eventually, these caps begin to detach from their respective plumes due to vertical stretching followed by the full breakup of the sheared thin plumes into small structures as observed in figure \ref{Temp B015f0} at $t=7.51$s (see supplementary Movie $3$). Consequently, turbulent mixing begins, which remains weaker than in the HD case B$_0$0f0 because the fluid structures are still collimated along the magnetic field lines in the vertical direction, as apparent at $t\simeq9.08$s in figure \ref{Temp B015f0}. However, this flow collimation in the B$_0$0.15f0 case results in the increased height of the mixing layer, enhancing the heat transfer as compared to the HD cases.\\

When rotation $f=8$ is added along with $B_0=0.15$ (MHD case B$_0$0.15f8), the Coriolis force also plays a significant role by adding its stabilization effect which suppresses the growth of the elongated plumes, as can be observed by comparing figures \ref{Temp B015f8} and \ref{Temp B015f0} for B$_0$0.15f8 and B$_0$0.15f0 cases, respectively, at $t=4.0$s and $5.6$s. Additionally, the presence of rotation inhibits the formation of mushroom-shaped caps that were observed in the non-rotating case B$_0$0.15f0 (see a comparison among figures \ref{Temp B015f8} and \ref{Temp B015f0} at $t\simeq6.5$s). The continued vertical stretching owing to the imposed magnetic field results in the thinning of plumes, eventually causing them to break down as illustrated in figure \ref{Temp B015f8} at $t=7.51$s, $9.05$s (see supplementary Movie $4$). The comparison between figures \ref{Temp B015f0} at $t=9.08$s and \ref{Temp B015f8} at $t=9.05$s reveals that the mixing is weaker in the B$_0$0.15f8 than that in B$_0$0.15f0, signifying the suppression of velocity fluctuations and heat transfer by the Coriolis force. \\ 

\captionsetup[subfigure]{textfont=normalfont,singlelinecheck=off,justification=raggedright}  
\begin{figure}
    \centering
    \begin{subfigure}{0.495\textwidth}
        \centering
        \hspace{-1.65cm}\includegraphics[width=1.1\textwidth,trim={0cm 0.0cm 0.0cm 0cm},clip]{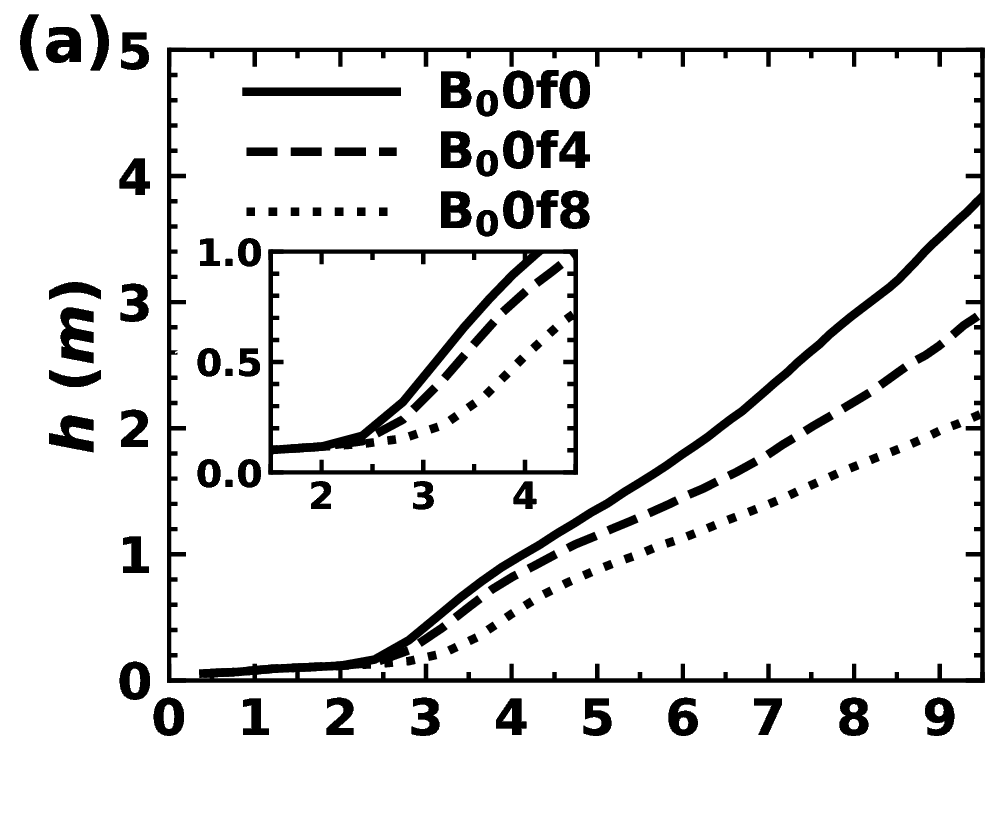}
        \label{ht_a}
    \end{subfigure}
    \begin{subfigure}{0.495\textwidth}
        \centering
        \hspace{-0.735cm}\includegraphics[width=1.1\textwidth,trim={0cm 0.0cm 0.0cm 0cm},clip]{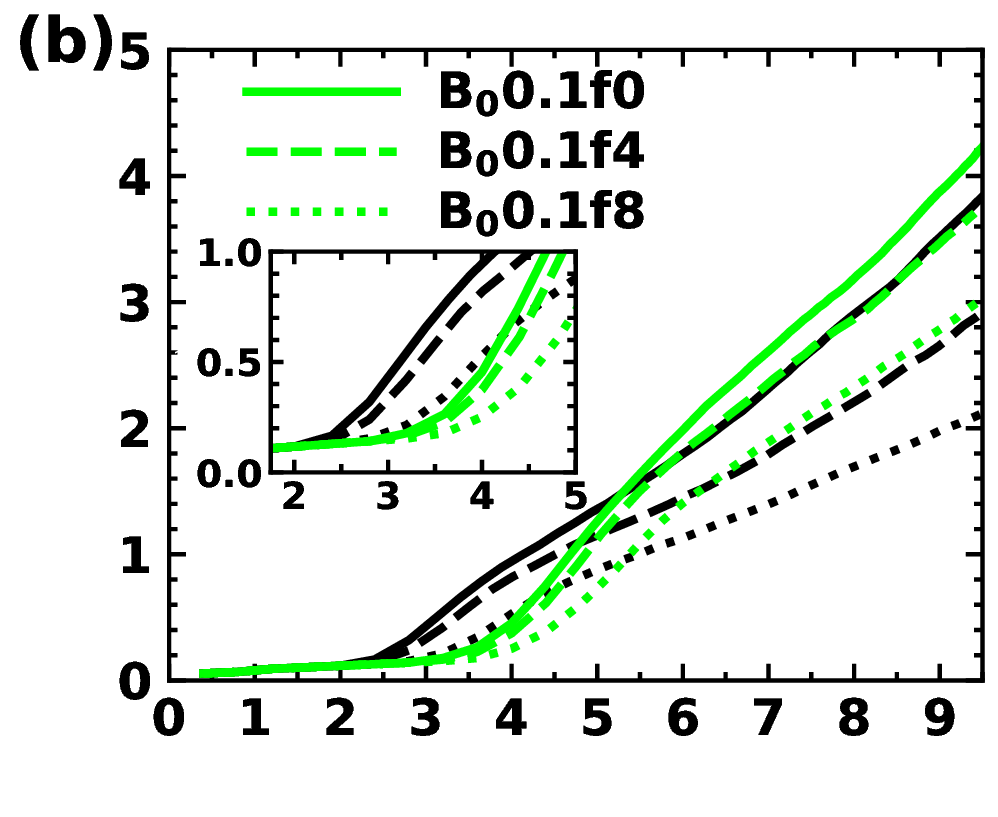}\hspace{-0.735cm}
        \label{ht_b}
    \end{subfigure}
    \\
    \begin{subfigure}{0.495\textwidth}
        \centering
        \hspace{-1.65cm}\includegraphics[width=1.1\textwidth,trim={0cm 0.0cm 0.0cm 0cm},clip]{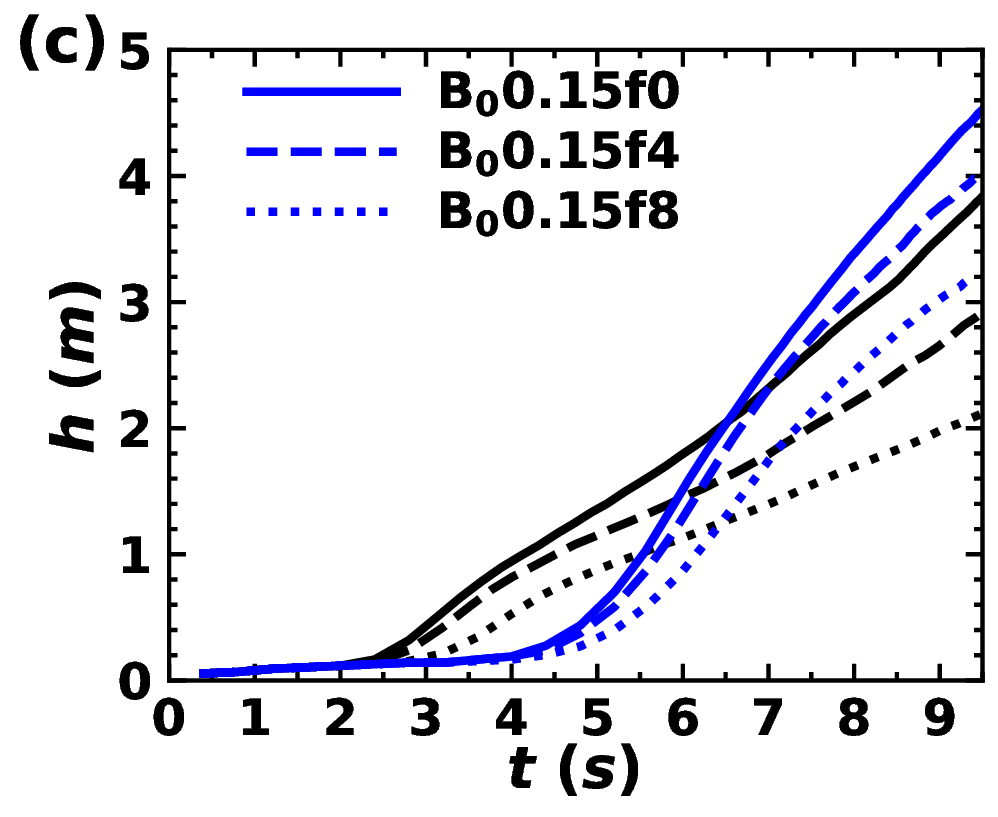}
        \label{ht_c}
    \end{subfigure}
    \begin{subfigure}{0.495\textwidth}
        \centering
        \hspace{-0.735cm}\includegraphics[width=1.1\textwidth,trim={0cm 0.0cm 0.35cm 0cm},clip]{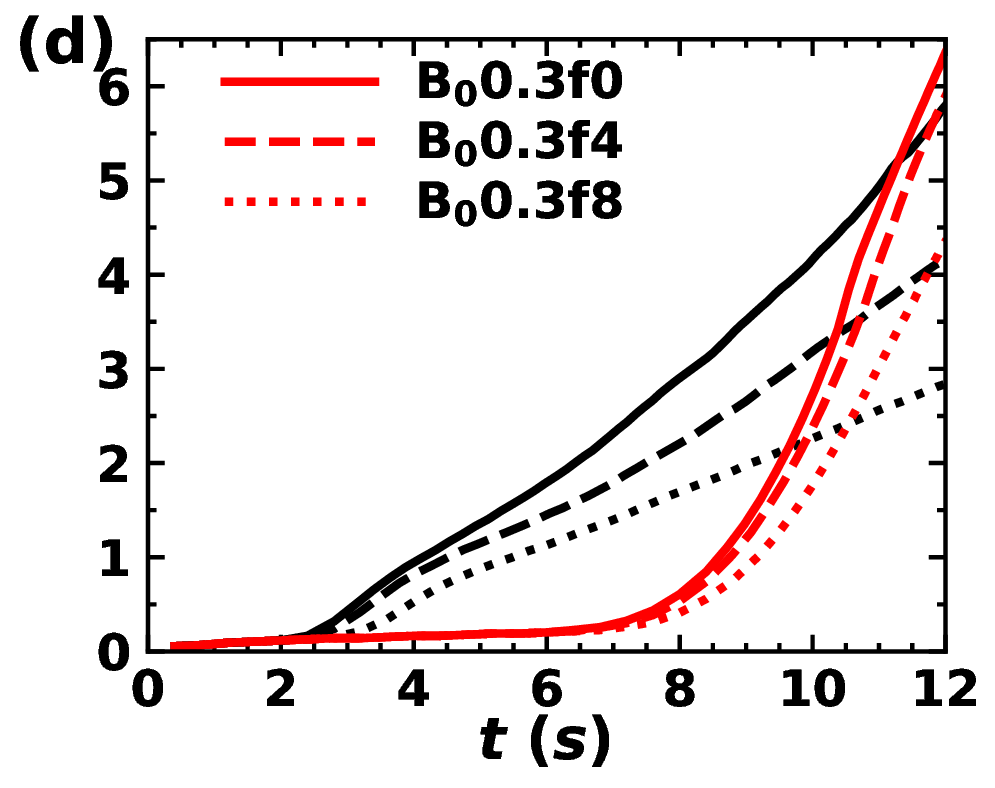} \hspace{-0.735cm}
        \label{ht_d}
    \end{subfigure}    
    \caption{Temporal evolution of the mixing layer height $h$ for non-rotating and rotating (\textit{a}) HD cases, (\textit{b}) HD and MHD at $B_0=0.1$ cases, (\textit{c}) HD and MHD at $B_0=0.15$ cases, and (\textit{d}) HD and MHD at $B_0=0.3$ cases. Insets in (\textit{a}) and (\textit{b}) correspond to the onset of RT instability.}
    \label{fig: h}
 \end{figure}

To quantitatively assess the flow evolution under the influence of magnetic field and rotation, we evaluate the height of the mixing layer ($h$) as a function of time in figure \ref{fig: h}. The mixing layer height is calculated based on the threshold value of $x_3$ at which $\overline{ T}(x_3,t)$ reaches a fraction $r$ of the maximum value, such that $\overline{ T}(\pm x_3,t)=\mp r \Theta$, where $r=0.9$ \citep{boffetta2016rotating}. We show the temporal evolution of $h$ for the non-rotating and rotating HD cases in figure \ref{fig: h}a, and compare this evolution between the HD cases and MHD cases for $B_0=0.1$, 0.15, and 0.3 in figures \ref{fig: h}b, \ref{fig: h}c, and \ref{fig: h}d, respectively. For the non-rotating HD case B$_0$0f0, $h$ starts to grow at $t\simeq2.5$s with the emergence of thermal plumes, followed by the rapid growth in the turbulent mixing region (see figure \ref{fig: h}a). With the addition of a rotation rate of $f=4$ (i.e., HD case B$_0$0f4), the growth rate of $h$ is suppressed compared to the B$_0$0f0 case. This suppression becomes stronger as $f$ increases from $4$ to $8$ (B$_0$0f8 case) due to the increased stabilization effect exerted by the Coriolis force. Additionally, $h$ starts increasing at $t\simeq3-3.5$s (see inset of figure \ref{fig: h}a) for the B$_0$0f8 case, whereas for B$_0$0f0 and B$_0$0f4 the growth starts at $t\simeq2.5$s, signifying that the Coriolis force delays the onset of the instability at high rotation rates. \\

When the vertical mean magnetic field $B_0=0.1$ is imposed in the absence of rotation B$_0$0.1f0, $h$ starts growing at $t\simeq3.5-4$s signifying the delay in the onset of instability compared to the HD B$_0$0f0 case where $h$ started growing at $t\simeq2.5$s (see figure \ref{fig: h}b). We can observe a rapid increase in $h$ for B$_0$0.1f0, which eventually exceeds the $h$ for B$_0$0f0. The formation of vertically elongated plumes by $B_0$, or the flow collimation along the vertical magnetic field lines, which tends to inhibit plumes interactions, is the reason behind the enhanced growth of $h$ for the B$_0$0.1f0 case compared to the B$_0$0f0 case. \\

Inclusion of rotation, $f=4$, with $B_0=0.1$ (B$_0$0.1f4 case) reduces the growth of $h$ compared to the non-rotating case B$_0$0.1f0. This reduction becomes more pronounced when $f=8$ (figure \ref{fig: h}b). This decrease in $h$ is attributed to the stabilizing effect exerted by the Coriolis force, which suppresses the growth of the vertically elongated plumes. Owing to the elongated plumes in MHD cases, $h$ for B$_0$0.1f4 and B$_0$0.1f8 remain higher than the corresponding HD B$_0$0f4 and B$_0$0f8 cases, respectively. 
For $f=0$, with $B_0=0.15$ (figure \ref{fig: h}c) and $B_0=0.3$ (figure \ref{fig: h}d), $h$ begins to grow at $t\simeq 4-4.5$s and $t\simeq 7-7.5$s, respectively, revealing that the delay in the onset of instability increases with an increase in the strength of the imposed magnetic field which is consistent with the predictions of linear theory in the sense that the growth of instability reduces as $B_0$ increases in the linear phase  \citep{chandrasekhar1968hydrodynamic,jun1995numerical}. However, at later times, after experiencing rapid growth, $h$ becomes larger with an increase in $B_0$ from $0.15$ to $0.3$ (B$_0$0.15f0 and B$_0$0.3f0), compared to the cases with $B_0=0,\  0.1$ for $f=0$ (see comparison among figures \ref{fig: h}b, \ref{fig: h}c and \ref{fig: h}d). This enhancement is attributed to the increased tendency of $B_0$ ($=0.15,\,0.3$) to collimate the flow along the vertical magnetic field lines, resulting in stronger vertical stretching of the plumes. Similar to the rotating MHD cases B$_0$0.1f4 and B$_0$0.1f8, the growth of $h$ decreases for B$_0$0.15f4 and B$_0$0.15f8 (B$_0$0.3f4 and B$_0$0.3f8) cases compared to the corresponding non-rotating MHD B$_0$0.15f0 (B$_0$0.3f0) cases, due to the stabilization effect of the Coriolis force. Since the vertical elongation of the plumes is stronger at $B_0=0.15$ than at $B_0=0.1$, $h$ is higher in the B$_0$0.15f4 and B$_0$0.15f8 cases compared to the B$_0$0.1f4 and B$_0$0.1f8 cases, respectively (see comparison between figures \ref{fig: h}b and \ref{fig: h}c). \\

\begin{figure}
 	\centering
 	\begin{subfigure}{0.495\textwidth}
  		\centering
        \caption{}  \label{fig: u3rms B015f0}
  		\includegraphics[width=1.0\textwidth,trim={0cm 0cm 0cm 0cm},clip]{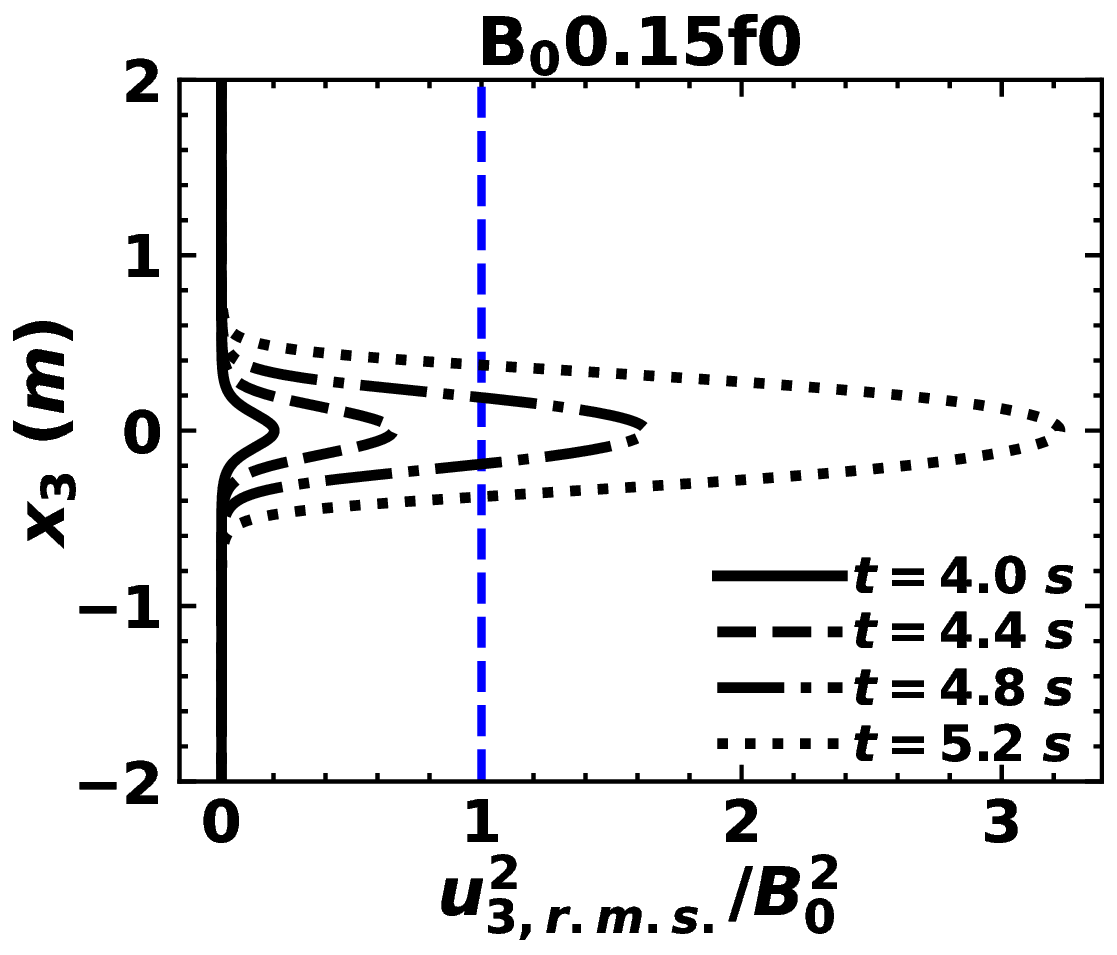}
  	\end{subfigure}
   \hfill
    \begin{subfigure}{0.495\textwidth}
  		\centering
        \caption{}  \label{fig: u3rms B03f0}
  		\includegraphics[width=1.0\textwidth,trim={0cm 0cm 0cm 0cm},clip]{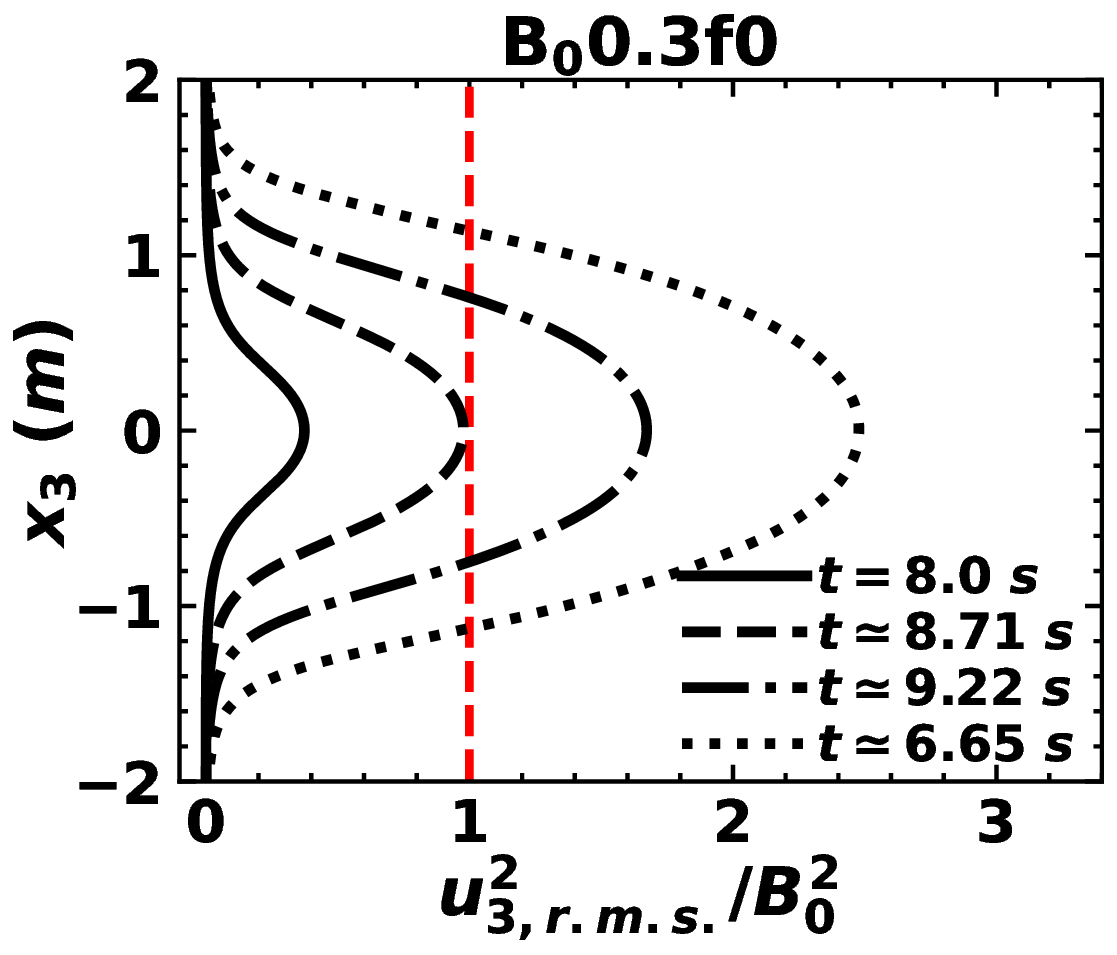}   		
  	\end{subfigure}
    \caption{Vertical variation of the squared $r.m.s.$ vertical velocity $u_{3,r.m.s.}$ normalized by the squared mean vertical magnetic field $B_0$, scaled as the Alfv\`{e}n velocity, for the non-rotating MHD cases (\textit{a}) B$_0$0.15f0 and (\textit{b}) B$_0$0.3f0 at different time instants.}
    \label{fig: u3rms}
\end{figure}

To further understand the suppression of secondary small-scale shear instabilities in the initial phase of RT instability owing to the vertically imposed magnetic field, we plot the vertical profiles of the squared $r.m.s.$ vertical velocity ($u_{3,r.m.s.}^2$) normalized by the $B_0^2$ (scaled as the Alfv\`{e}n velocity), for the non-rotating MHD cases B$_0$0.15f0 and B$_0$0.3f0 in figures \ref{fig: u3rms B015f0} and \ref{fig: u3rms B03f0}, respectively, at different time instants. \cite{jun1995numerical} proposed that for the vertically imposed mean magnetic field case, if the square of the relative velocity of the plumes is smaller than the squared Alfv\`{e}n velocity, then the secondary shear instabilities (KH instabilities) are suppressed, and there is no flow transverse to the gravity vector. For B$_0$0.15f0, figure \ref{fig: u3rms B015f0} illustrates that $u_{3,r.m.s.}^2/B_0^2 < 1$ till $t=4.4$s suggesting that although the perturbations at the interface stretch to take the shape of elongated plumes, secondary shear instabilities are not triggered. At $t = 4.8$s, and beyond the plumes are destabilized by the onset of secondary shear instabilities as also indicated in the enlarged views in figure \ref{Temp B015f0} at $t=5.56$s (see Supplementary Movie $3$). For B$_0$0.3f0, $u_{3,r.m.s.}^2/B_0^2 < 1$ until $t=8.71$s and is $> 1$ beyond it. In this case, the instability is already triggered by $t=8.71$s, and the height of the mixing layer is $h\simeq1.2$m at $t=8.71$s (figure \ref{fig: h}d). In contrast to the B$_0$0.15f0 case, $h$ is approximately $3$ times larger for the B$_0$0.3f0 until $u_{3,r.m.s.}^2/B_0^2 < 1$. This signifies that for B$_0$0.3f0, the plumes have already reached greater heights than B$_0$0.15f0 before triggering secondary shear instabilities. Consequently, large, smooth, vertically elongated plumes form in B$_0$0.3f0 (see supplementary Movie $5$).  \\

\captionsetup[subfigure]{textfont=normalfont,singlelinecheck=off,justification=raggedright, labelfont=bf, textfont=bf,font=large}

 \captionsetup[subfigure]{textfont=normalfont,singlelinecheck=off,justification=raggedright}
 \begin{figure}
    \centering
    \begin{subfigure}{0.495\textwidth}
        \centering
         \hspace{-1.65cm}\includegraphics[width=1.1\textwidth,trim={0cm 0.0cm 0.0cm 0cm},clip]{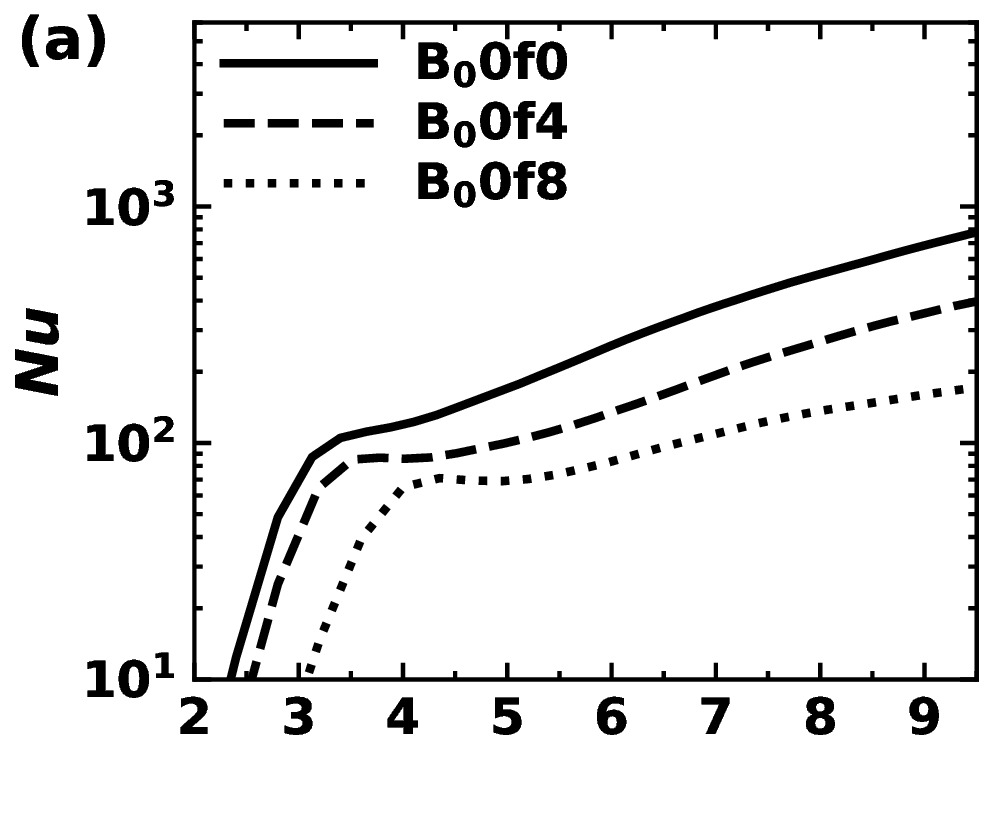}
         \label{Nu_t_a}
    \end{subfigure}
    \begin{subfigure}{0.495\textwidth}
        \centering
        \hspace{-0.735cm} \includegraphics[width=1.1\textwidth,trim={0cm 0.0cm 0.0cm 0cm},clip]{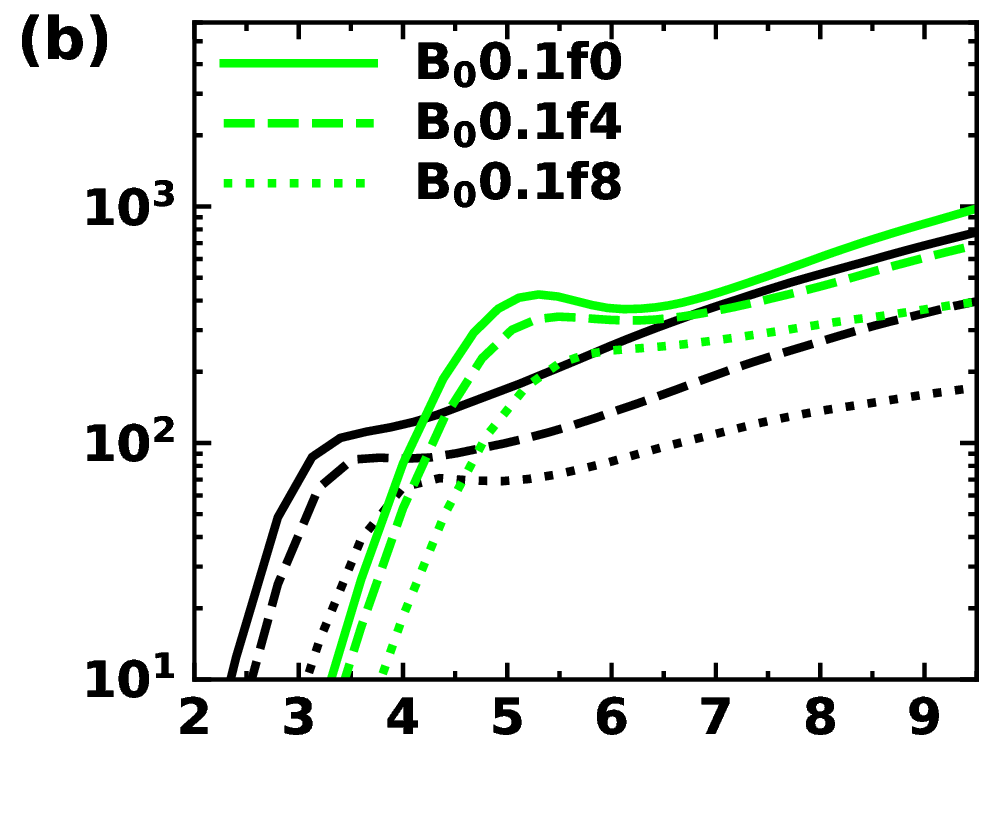} \hspace{-0.735cm}
         \label{Nu_t_a}
    \end{subfigure}
    \\
    \begin{subfigure}{0.495\textwidth}
        \centering
        \hspace{-1.65cm} \includegraphics[width=1.1\textwidth,trim={0cm 0.0cm 0.0cm 0cm},clip]{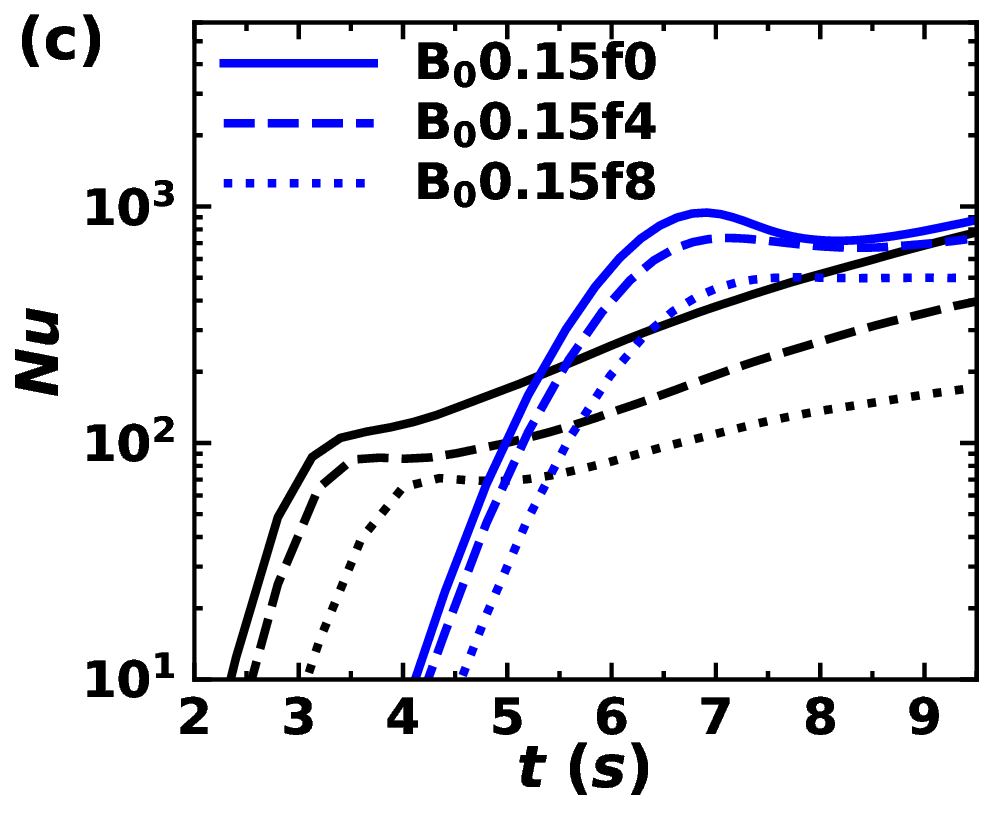}
         \label{Nu_t_a}
    \end{subfigure}
    \hfill
    \begin{subfigure}{0.495\textwidth}
        \centering
        \hspace{-0.735cm} \includegraphics[width=1.1\textwidth,trim={0cm 0.0cm 0.35cm 0cm},clip]{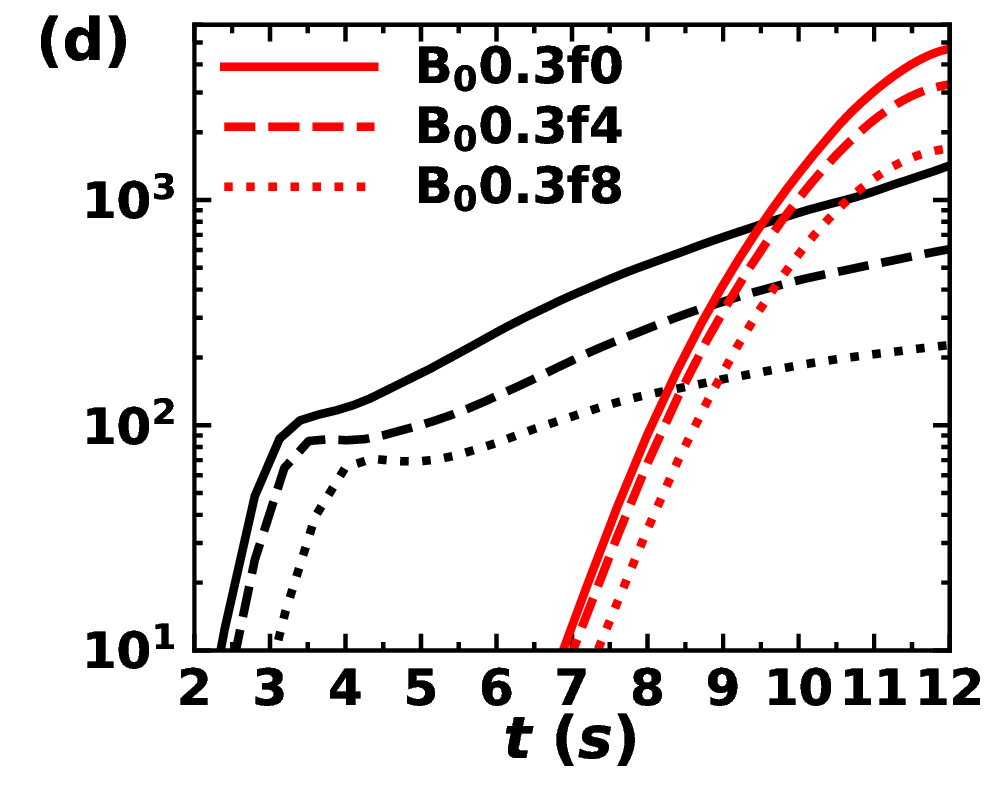} \hspace{-0.735cm} 
         \label{Nu_t_a}
    \end{subfigure}    
    \caption{Temporal evolution of the Nusselt number ($Nu$, defined in equation \ref{Nu eq1}) 
    for non-rotating and rotating (\textit{a}) HD cases, (\textit{b}) HD and MHD for $B_0=0.1$ cases, (\textit{c}) HD and MHD for $B_0=0.15$ cases, and (\textit{d}) HD and MHD for $B_0=0.3$ cases.}
    \label{fig: Nu}
 \end{figure}

The evolving thermal plumes transport heat, and to quantify this heat transfer efficiency, we plot the Nusselt number ($Nu$), defined as the ratio of total heat transfer (convective and conductive) to conductive heat transfer, as a function of time in figure \ref{fig: Nu}. We compute $Nu$ as \citep{boffetta2009kolmogorov,boffetta2010statistics}
 \begin{equation}
     \label{Nu eq1}
      Nu=1+\frac{\langle u_3' T' \rangle h}{\kappa \Theta},
 \end{equation}
 where, $\langle \cdot \rangle$ denotes the spatial average inside the mixing layer of height $h$ (defined in equation \ref{average p31}). Figure \ref{fig: Nu}a depicts the temporal evolution of $Nu$ for the non-rotating and rotating HD cases. Figures \ref{fig: Nu}b, \ref{fig: Nu}c, and \ref{fig: Nu}d, respectively, compare the $Nu$ evolution between the HD cases and MHD cases for $B_0=0.1$, 0.15, and 0.3. The $Nu$ increases with the growth and subsequent mixing of the thermal plumes for all the cases. The effect of the Coriolis force on the suppression of the growth of $h$ and fluctuations of vertical velocity ($u_3'$) and temperature ($T'$), 
 results in a decrease in heat transfer. Therefore, in the HD cases, the $Nu$ decreases with an increase in rotation rates from $f=4$ to $8$ compared to the non-rotating B$_0$0f0 case (see figure \ref{fig: Nu}a). Similar results were also reported by \cite{boffetta2016rotating}.\\
 
In the non-rotating MHD cases B$_0$0.1f0 and B$_0$0.15f0, the $Nu$ is enhanced compared to the corresponding HD case B$_0$0f0 (figures \ref{fig: Nu}b and \ref{fig: Nu}c). The imposed $B_0$ delays the onset of instability, and therefore, we see a delay in the rapid increase in $Nu$ for the MHD cases. After this rapid increase, the growth of $Nu$ for the MHD cases decreases significantly. Beyond this slowdown period, the $Nu$ again increases at a rate similar to the HD cases. The vertically elongated plumes owing to the effect of the vertically imposed $B_0$ are efficient in transferring heat between the bottom hot fluid and the upper cold fluid with limited horizontal mixing, resulting in the initial rapid growth and enhancement of $Nu$ for the B$_0$0.1f0 and B$_0$0.15f0 cases. After the rapid growth, the mushroom-shaped caps formed at the tips of elongated plumes undergo breakdown and detach from their respective plumes, as observed in figure \ref{Temp B015f0} at $t\simeq6.46$s and $t\simeq7.51$s for B$_0$0.15f0 case. This results in a decrease in the vertical velocity and temperature fluctuations, which further slows down the evolution of the $Nu$ between $t\simeq5.25-6.25$s in B$_0$0.1f0 and $t\simeq6.8-8.4$s in B$_0$0.15f0. Eventually, the complete breakup of the plumes occurs, leading to fluid mixing. Consequently, $Nu$ begins to increase again, which remains higher than the corresponding HD case B$_0$0f0 due to the larger mixing layer height $h$ in the MHD cases B$_0$0.1f0 and B$_0$0.15f0 (see figures \ref{fig: h}b and \ref{fig: h}c, respectively). The decrease in vertical velocity and temperature fluctuations during the breakdown and detachment of the mushroom-shaped caps represents the transition from the initial regime of unbroken elongated plumes to the mixing regime of broken small-scale structures. The heat transfer enhancement (increase in $Nu$) becomes more efficient as $B_0$ increases from 0.15 to 0.3 (compare figures \ref{fig: Nu}c and \ref{fig: Nu}d) due to the stronger vertical stretching of elongated plumes resulting in a significant increase in vertical velocity fluctuations and mixing layer height, and limited horizontal mixing.\\

In the MHD cases with $B_0=0.1,\,0.15$ and $0.3$, the increase in rotation rates from $f=4$ to $f=8$ acts to inhibit the growth of $h$ and breakdown of vertically elongated plumes thereby suppressing fluctuations in both vertical velocity and temperature. Consequently, $Nu$ decreases with increasing $f(=4,\,8)$ in MHD cases $B_0=0.1,\,0.15$ and $0.3$ compared to the corresponding non-rotation MHD ($f=0$) cases (see figures \ref{fig: Nu}b, \ref{fig: Nu}c and \ref{fig: Nu}d). For B$_0$0.1f4 and B$_0$0.15f4, $Nu$ follows a trend similar to that of B$_0$0.1f0 and B$_0$0.15f0, respectively, in the sense that during the transition from initial regime to mixing regime, $Nu$ first experiences a sudden drop, then it slows down and approaches nearly a constant value before increasing again. However, in the B$_0$0.1f8 and B$_0$0.15f8 cases, $Nu$ does not experience a sudden drop during the slowdown. This behavior is attributed to the stronger influence of the Coriolis force at $f=8$ which results in a gradual breaking of the vertically stretched plumes instead of a sudden disintegration. \\

Interestingly, for the non-rotating and rotating MHD cases at $B_0=0.1, 0.15$, and $0.3$, the $Nu$ remains higher than the corresponding HD cases. The enhancement in $Nu$ for the rotating MHD cases over the corresponding HD cases is caused by the vertical stretching and collimation of flow structures along the vertical magnetic lines, resulting in increased $h$ and $u_3'$ in the presence of $B_0$. Note that the $Nu$ at $f=4,8$ for the $B_0=0.3$ cases is even higher than the $B_0=0.1, 0.15$ cases at $f=0$ until the end of the simulations for $B_0=0.3$ cases (compare figures \ref{fig: Nu}b, \ref{fig: Nu}c and \ref{fig: Nu}d). This is attributed to the relatively stronger vertical stretching of the plumes by $B_0=0.3$ than by $B_0=0.1, 0.15$ resulting in larger $h$, $u_3'$, and $T'$. Therefore, even with stronger rotation rates ($f=8)$, the heat transfer increases significantly at higher values of the vertically imposed mean magnetic field. This indicates that imposed magnetic fields can mitigate the instability-suppressing effects of rotation for efficient heat transfer. \\
 
\captionsetup[subfigure]{textfont=normalfont,singlelinecheck=off,justification=raggedright}
   \begin{figure}
    \centering
    \begin{subfigure}{0.48\textwidth}
    \centering
        \includegraphics[width=1.0\textwidth,trim={0cm 0cm 0cm 0cm},clip]{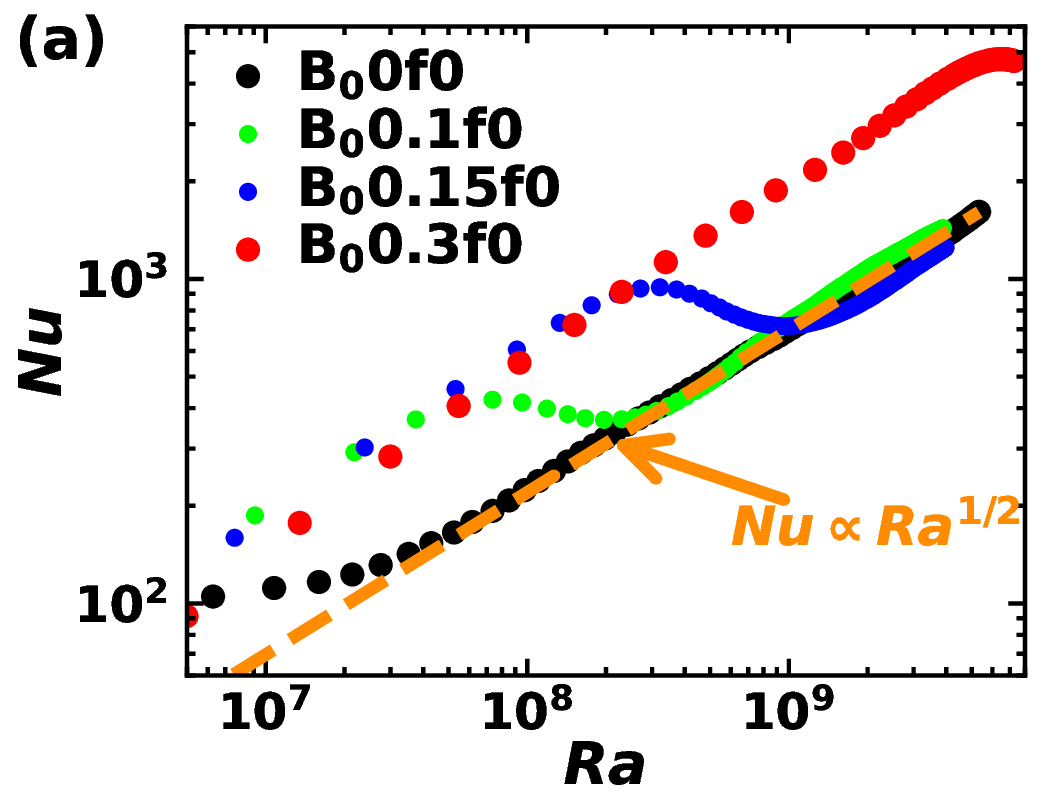}
        \label{fig: Nu Raf0}        
   	\end{subfigure}
    \hfill
    \begin{subfigure}{0.48\textwidth}
   		\centering
   		\includegraphics[width=1.0\textwidth,trim={0cm 0cm 0cm 0cm},clip]{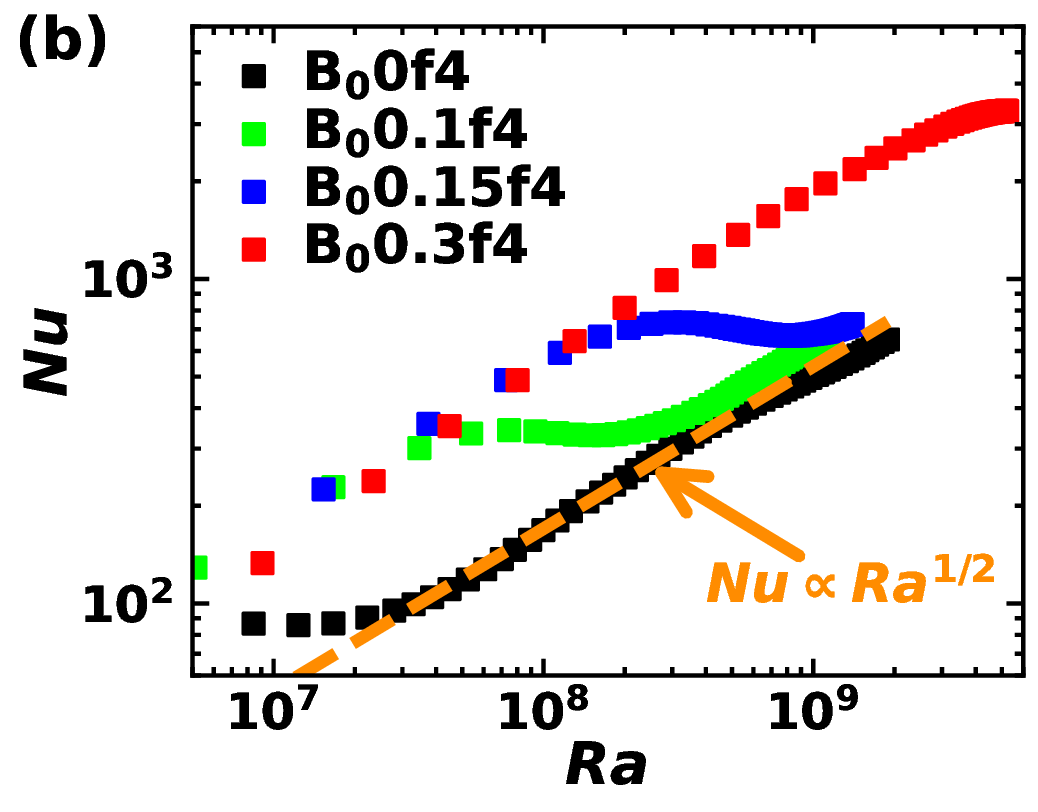}
        \label{fig: Nu Raf4}
   	\end{subfigure}\\
    \begin{subfigure}{0.5\textwidth}
   		\centering
   		\includegraphics[width=1.0\textwidth,trim={0cm 0cm 0.0cm 0cm},clip]{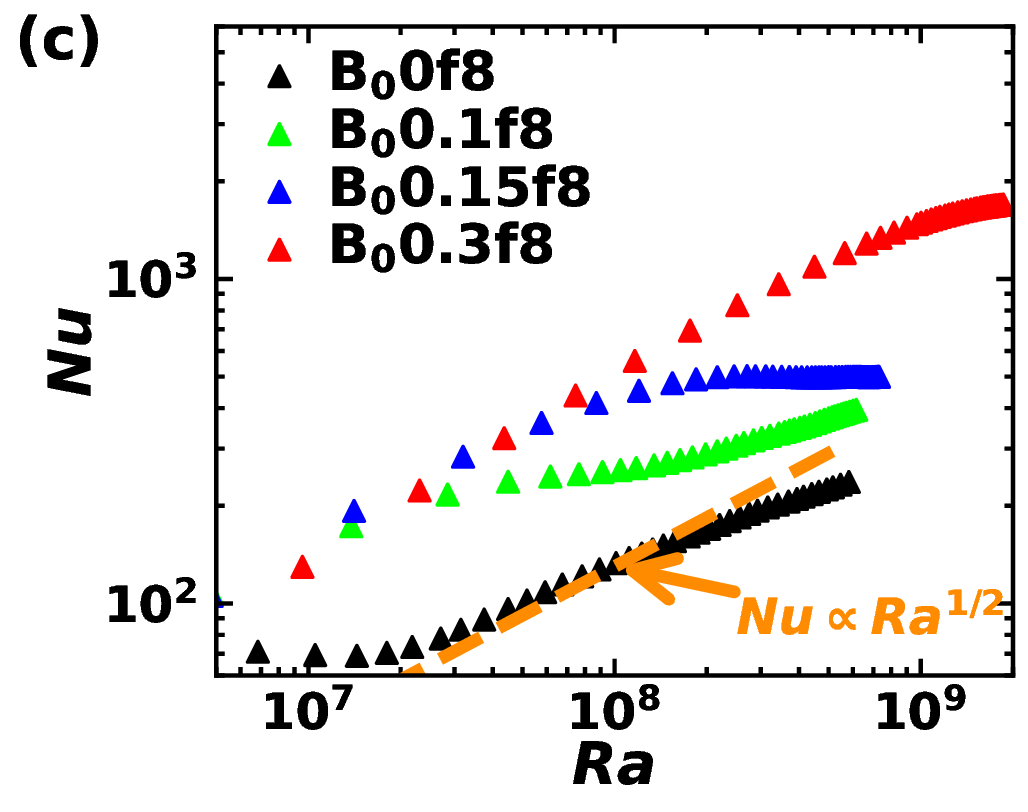}
       \label{fig: Nu Raf8}
   	\end{subfigure}
 \caption{Variation of the Nusselt number, $Nu$ (defined in equation \ref{Nu eq1}), as a function of Rayleigh number, $Ra$ ($\propto h^3$), for the HD and MHD (\textit{a}) non-rotating ($f=0$), (\textit{b}) rotating at $f=4$, and (\textit{c}) rotating at $f=8$ cases. Dashed orange lines represent the ultimate state predictions, i.e., $Nu \propto Ra^{1/2}$ scaling for $Pr=1$ \citep{boffetta2010statistics,boffetta2016rotating}. }
 \label{fig: Nu vs Ra}
\end{figure}

We further investigate the Nusselt number ($Nu$) as a function of the Rayleigh number ($Ra$) to assess the correlation between $u_3^{\prime}$ and $T^{\prime}$ and the presence of the ultimate state scaling of thermal convection ($Nu \simeq Ra^{1/2} Pr^{1/2}$) \citep{boffetta2010statistics,boffetta2016rotating}. The $Ra$ is a dimensionless measure of the temperature difference that forces the system and is defined in terms of the mixing layer height $h$ as $Ra=\beta g \Theta h^3/ (\nu \kappa)$ \citep{boffetta2009kolmogorov,boffetta2010statistics}. Since the calculation of $Nu$ involves $\langle u_3' T' \rangle$ and $h$ (equation \ref{Nu eq1}), plotting it as a function of $Ra$ ($\propto h^3$), i.e., at a fixed $h$, allows us to establish the correlation between $u_3^{\prime}$ and $T^{\prime}$ under the effect of an imposed magnetic field and rotation. Therefore, the $Nu$ as a function of $Ra$ ($\propto h^3$) is plotted in figure \ref{fig: Nu vs Ra}a, \ref{fig: Nu vs Ra}b, and \ref{fig: Nu vs Ra}c for the HD and MHD cases at $f=0$ (B$_0$0f0, B$_0$0.1f0, B$_0$0.15f0 and B$_0$0.3f0), at $f=4$ (B$_0$0f4, B$_0$0.1f4, B$_0$0.15f4 and B$_0$0.3f4), and at $f=8$ (B$_0$0f8, B$_0$0.1f8, B$_0$0.15f8 and B$_0$0.3f8), respectively. We can observe a significant enhancement in $Nu$ before the breakdown of plumes for the MHD cases compared to HD cases at $f=0$ in figure \ref{fig: Nu vs Ra}a. This enhancement is attributed to the vertical stretching of plumes owing to the imposed $B_0$, increasing the correlation between the vertical velocity ($u_3'$) and temperature ($T^{\prime}$) fluctuations. After the plumes breakdown, $Nu$ decreases for B$_0$0.1f0 and B$_0$0.15f0 and becomes comparable to that of B$_0$0f0. This decrease in $Nu$ signifies that the correlation between $u_3'$ and $T'$ reduces when the flow structures break down to produce mixing. This further indicates the suppression of the velocity fluctuations by the imposed $B_0$ in the mixing phase (as shown later). For the B$_0$0.3f0 case, $Nu$ remains higher than the B$_0$0f0, B$_0$0.1f0 and B$_0$0.15f0 cases due to strong vertical stretching of the elongated plumes and shows a small drop at $Ra\simeq7-8\times 10^9$ due to the beginning of plumes breakdown. In figure \ref{fig: Nu vs Ra}a, for HD case B$_0$0f0 and MHD cases B$_0$0.1f0 and B$_0$0.15f0, our DNS confirms the presence of the ultimate state regime, $Nu \simeq Ra^{1/2} Pr^{1/2}$. The $Pr=1$ for our simulations, and therefore $Nu \propto Ra^{1/2}$ similar to \cite{boffetta2012ultimate,boffetta2017incompressible}. In the presence of rotation at $f=4$ and $f=8$, figures \ref{fig: Nu vs Ra}b and \ref{fig: Nu vs Ra}c, respectively, demonstrate an enhancement of $Nu$ for the MHD cases over the HD case. This shows that the correlation between $u_3'$ and $T'$ is higher in the rotating MHD cases than in the corresponding rotating HD cases because the vertical stretching caused by the imposed $B_0$ mitigates the Coriolis force's effect of reducing the velocity fluctuations. The deviation from the ultimate state scaling $Nu \propto Ra^{1/2}$ can be observed with rotation $f=4$ and $f=8$ HD and MHD cases with $B_0=0.1,0.15$. \cite{boffetta2016rotating} reported a similar deviation from the ultimate state scaling in the presence of rotation (HD) without a magnetic field. The ultimate state scaling is invalid for both the non-rotating and rotating MHD cases at $B_0=0.3$ because turbulent mixing of flow structures is not triggered until the end time of simulations for these cases.\\ 
 
 \captionsetup[subfigure]{textfont=normalfont,singlelinecheck=off,justification=raggedright}
\begin{figure}
    \centering
  	\begin{subfigure}{0.495\textwidth}
    \centering
    \caption{}  \label{fig: tke2a}
   		\includegraphics[width=1.0\textwidth,trim={0cm 0cm 0cm 0cm},clip]{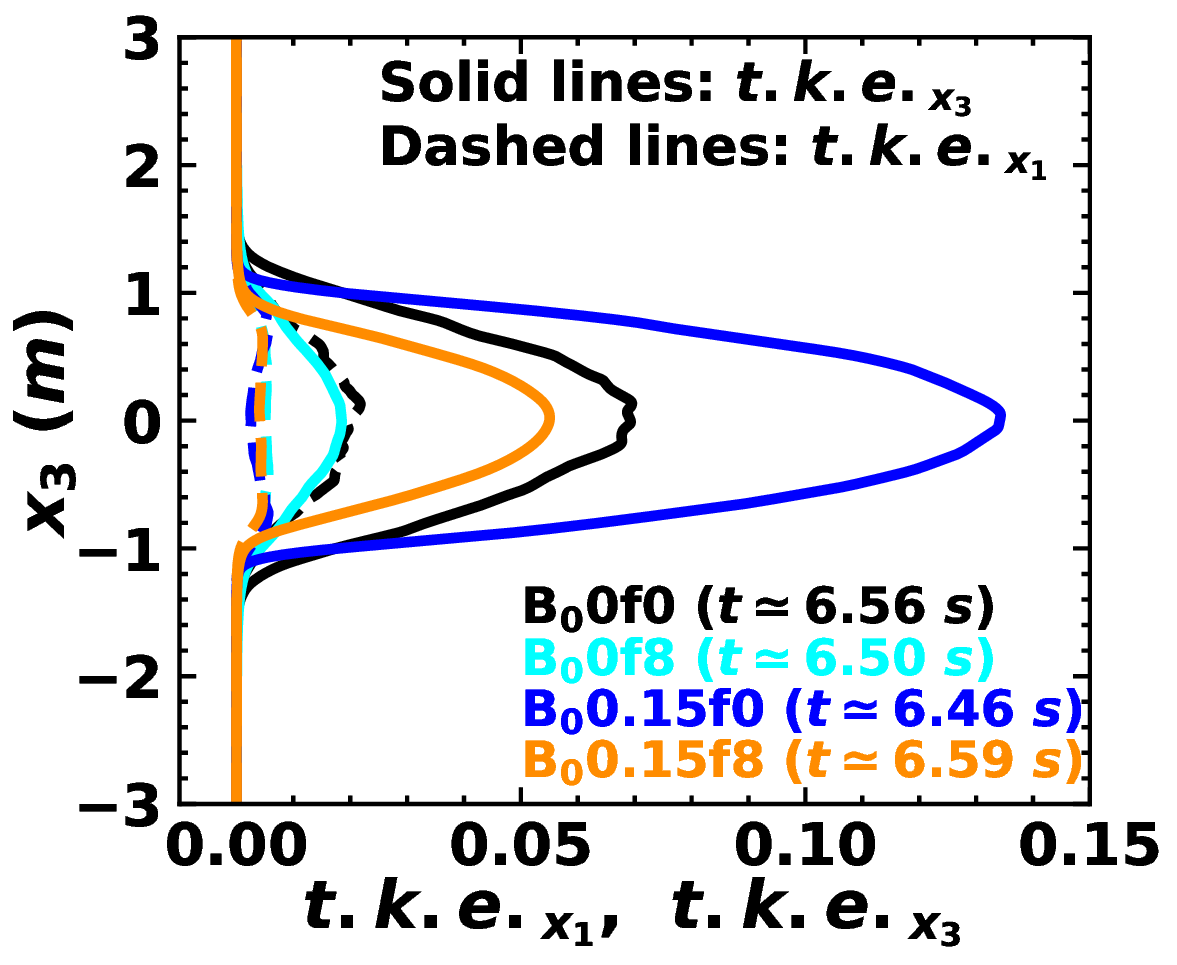}   		
   	\end{subfigure}
    \begin{subfigure}{0.495\textwidth}
   		\centering
        \caption{}  \label{fig: tke2b}
   		\includegraphics[width=1.0\textwidth,trim={0cm 0cm 0cm 0cm},clip]{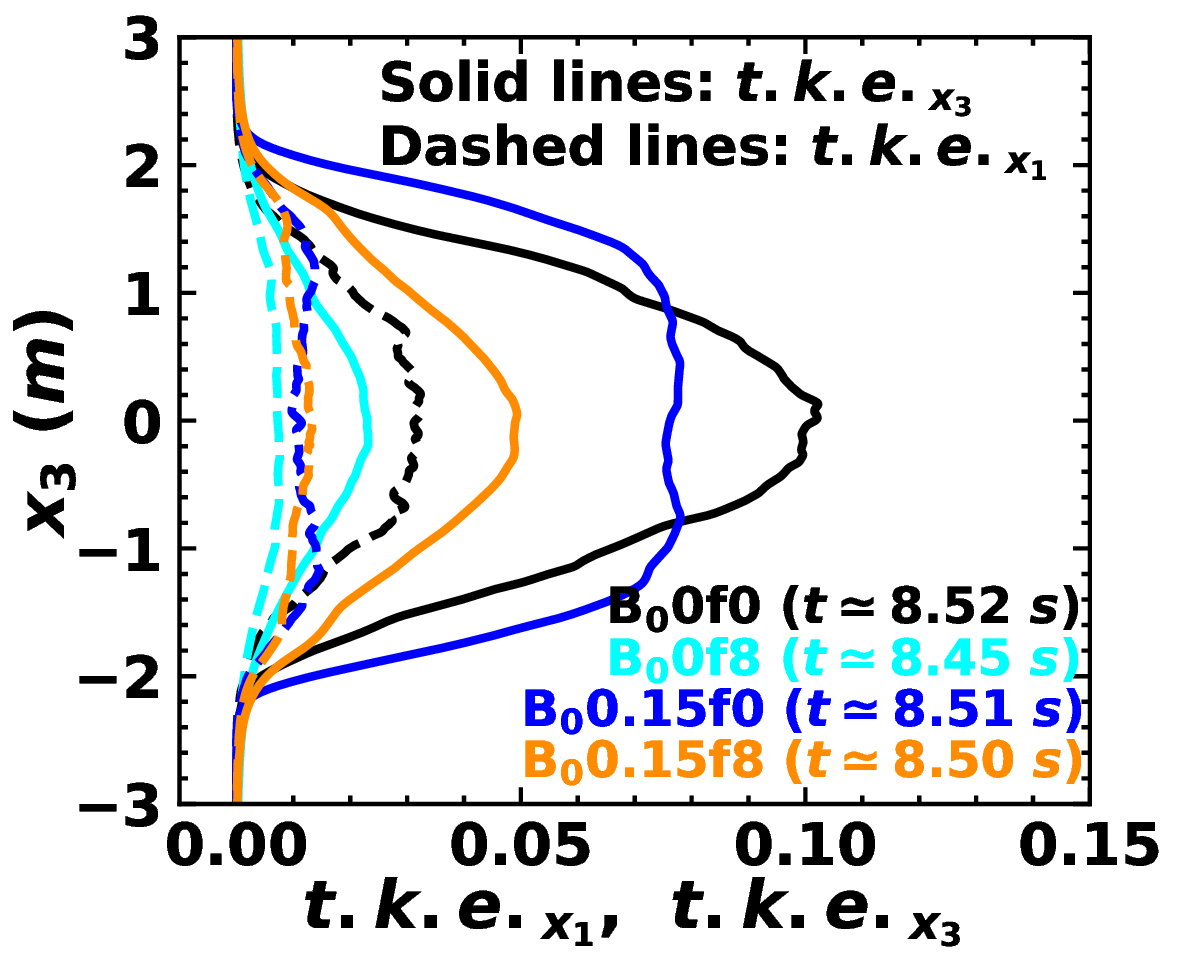}
   	\end{subfigure}
 \caption{Vertical profiles of the horizontal ($x_1$) component of horizontally averaged turbulent kinetic energy ($t.k.e._{x_1}$) and vertical ($x_3$) component of horizontally averaged turbulent kinetic energy ($t.k.e._{x_3}$) for the HD cases B$_0$0f0 and B$_0$0f8, and MHD cases B$_0$0.15f0 and B$_0$0.15f8 at time (\textit{a}) $t\simeq6.5$s and (\textit{b}) $t\simeq8.5$s. Solid lines denote $t.k.e._{x_3}$, while dashed lines denote $t.k.e._{x_1}$. }
 \label{fig: tke2}
\end{figure}

To further support the reasons for the enhancement of $Nu$ in the non-rotating and rotating MHD cases compared to the corresponding HD cases, we plot the vertical profile of the horizontal ($x_1$) and vertical ($x_3$) components of the horizontally averaged turbulent kinetic energy ($t.k.e.$) for the HD cases B$_0$0f0 and B$_0$0f8 and MHD cases B$_0$0.15f0 and B$_0$0.15f8 at $t\simeq6.5$s in figure \ref{fig: tke2a} and at $t=8.5$s in figure \ref{fig: tke2b}. The horizontal component of $t.k.e.$ (denoted by $t.k.e._{x_1}$) and vertical component of $t.k.e.$ (denoted by $t.k.e._{x_3}$) are defined as 
 \begin{equation}
  \label{tke}
    t.k.e._{x_1} = \frac{ \overline{u_1'^2}}{2}; \quad t.k.e._{x_3} = \frac{ \overline{u_3'^2}}{2}, 
 \end{equation}
where, $t.k.e._{x_1}$ and $t.k.e._{x_3}$ are functions of $(x_3,t)$. For all cases, figures \ref{fig: tke2a} and \ref{fig: tke2b} demonstrate the dominance of $t.k.e._{x_3}$ over $t.k.e._{x_1}$, indicating that the flow is primarily driven in the vertical direction, as expected due to vertically downward acting gravity normal to the interface. Figure \ref{fig: tke2a} at $t\simeq6.5$s shows that the $t.k.e._{x_3}$ is significantly smaller for B$_0$0f8 than that for the B$_0$0f0 case due to the presence of the Coriolis force, resulting in smaller $Nu$ in B$_0$0f8 compared to B$_0$0f0. 
For B$_0$0.15f0 and B$_0$0.15f8, $t.k.e._{x_3}$ is significantly larger than the corresponding $t.k.e._{x_1}$. This signifies that in the presence of an imposed magnetic field, the elongated plumes exhibit higher fluctuations in vertical velocity with limited horizontal mixing. Therefore, the elongated plumes are efficient in transferring heat between two fluids. In the MHD cases B$_0$0.15f0 and B$_0$0.15f8, $t.k.e._{x_3}$ is higher than in the corresponding HD case B$_0$0f0 and B$_0$0f8. This implies an enhancement in $Nu$ for the MHD cases compared to the corresponding HD cases. For the HD case B$_0$0f0, $t.k.e._{x_3}$ is higher than $t.k.e._{x_1}$, but $t.k.e._{x_1}$ exhibits significantly larger values compared to the MHD B$_0$0.15f0 and B$_0$0.15f8 cases. This indicates relatively stronger horizontal mixing in B$_0$0f0 than in B$_0$0.15f0 and B$_0$0.15f8. Smaller $t.k.e._{x_3}$ for B$_0$0.15f8 compared to the B$_0$0.15f0 case also corroborates the smaller $Nu$ in B$_0$0.15f8 compared to B$_0$0.15f0. \\
 
Figure \ref{fig: tke2b} depicts that at $t=8.5$s, for B$_0$0.15f0, the $t.k.e._{x_3}$ is smaller near the mid-plane ($x_3=0$m) compared to the B$_0$0f0 case. However, towards the plume tips ($x_3\simeq \pm 1.5$m) the $t.k.e._{x_3}$ for B$_0$0.15f0 is higher compared to  B$_0$0f0 indicating a wider mixing layer height $h$ for B$_0$0.15f0, as also evident from figure \ref{fig: h}c. Since $Nu$ also depends on $h$ (equation \ref{Nu eq1}), the $Nu$ is higher for B$_0$0.15f0 than the B$_0$0f0 case in the mixing phase (after the plume breakdown) primarily as a consequence of the increased $h$. 
The comparison between figures \ref{fig: tke2a} and \ref{fig: tke2b} reveals that $t.k.e._{x_3}$ is reduced at $t\simeq8.5$s (mixing phase) compared to that at $t\simeq6.5$s (during plume breakdown) for the MHD case B$_0$0.15f0, and this reduction is small for B$_0$0.15f8. Therefore, the imposed $B_0$ suppresses the velocity fluctuations in the mixing phase after the plume breakdown. Consequently, a decrease in $Nu$ was observed for the MHD cases in figures \ref{fig: Nu vs Ra}a, \ref{fig: Nu vs Ra}b and \ref{fig: Nu vs Ra}c, where $Nu$ is plotted as a function of $Ra(\propto h^3)$, during the mixing phase. Although, for B$_0$0.15f8 case, the reduction in $t.k.e._{x_3}$ is small, it remains higher and spans over the wider mixing layer ($h$) than the corresponding HD case. Owing to this, the $Nu$ is enhanced both as a function of time and $Ra(\propto h^3)$ in the mixing phase of the rotating MHD cases as compared to the corresponding HD cases, as observed in figures \ref{fig: Nu}b, \ref{fig: Nu}c, \ref{fig: Nu}d and figures \ref{fig: Nu vs Ra}b, \ref{fig: Nu vs Ra}c.\\
 
From the above discussion, it is apparent that the thermal plumes' formation, evolution, and disintegration under the influence of rotation and imposed magnetic field drive the heat transfer phenomenon. Therefore, we analyze the dynamic balance between the different forces to understand the behavior of these thermal plumes. We calculate the horizontally averaged $r.m.s.$ (using equation \ref{rms}) of each force i.e., the inertia ($\boldsymbol{F}_I$), the advection ($\boldsymbol{F}_A$), the pressure gradient ($\boldsymbol{F}_P$), the Coriolis ($\boldsymbol{F}_C$), the Lorentz ($\boldsymbol{F}_L$), the buoyancy ($\boldsymbol{F}_B$), and the viscous ($\boldsymbol{F}_V$), in momentum equation \ref{momentum1} \citep{guzman2021force,naskar_pal_2022a,naskar_pal_2022b}. We also define the $r.m.s.$ horizontal forces as $F_{x_1x_2,\,r.m.s.}=\left(F_{x_1,\,r.m.s.}^2+F_{x_2,\,r.m.s.}^2\right)^{1/2}$, where $F_{x_1,\,r.m.s.}$ and $F_{x_2,\,r.m.s.}$ denote the components of any $r.m.s.$ force in the $x_1$ and $x_2$ directions. The Lorentz force exerted by the magnetic field acts in all three directions, the Coriolis force, due to rotation about the vertical axis, acts only in the horizontal directions, and the buoyancy force is purely vertical due to the gravity vector. \\
 
\begin{figure}
  	\begin{subfigure}{0.32\textwidth}
  		\centering
        \label{subfig: B015f0 t65b}
         \hspace{-3.0cm}	\includegraphics[width=1.25\textwidth,trim={0cm 0cm 0cm 0cm},clip]{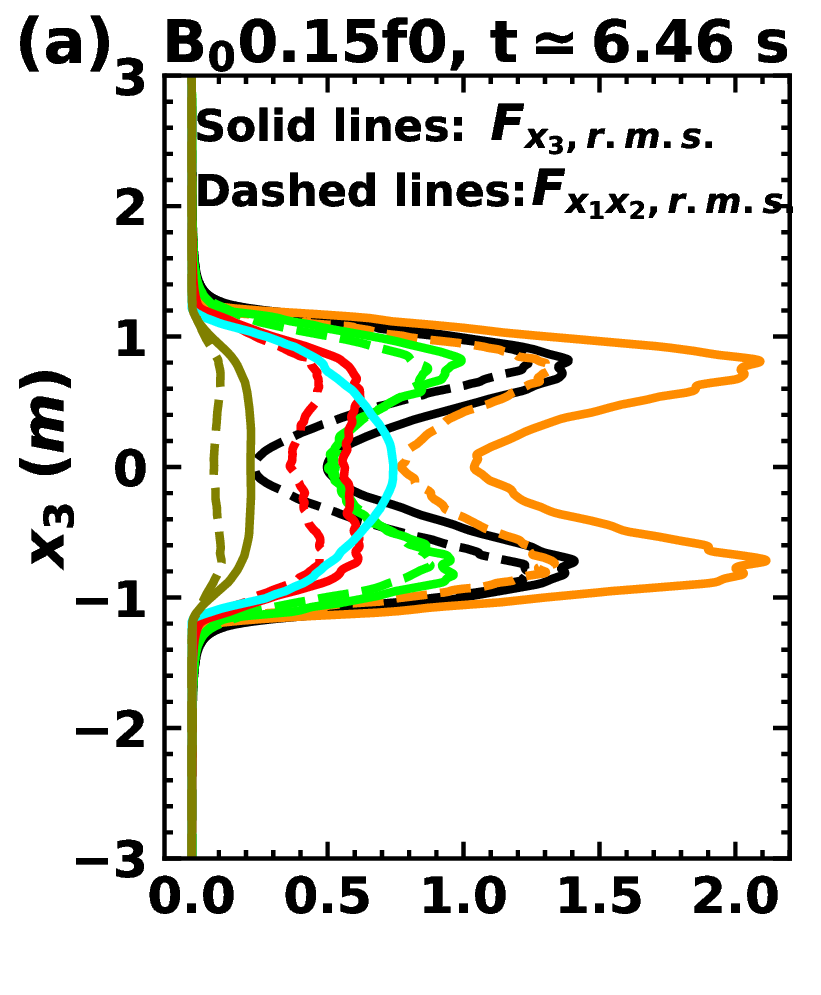}
  	\end{subfigure}
   \begin{subfigure}{0.32\textwidth}
  		\centering
        \label{subfig: B015f4 t65b}
         \hspace{-1.15cm}	\includegraphics[width=1.25\textwidth,trim={0cm 0cm 0cm 0cm},clip]{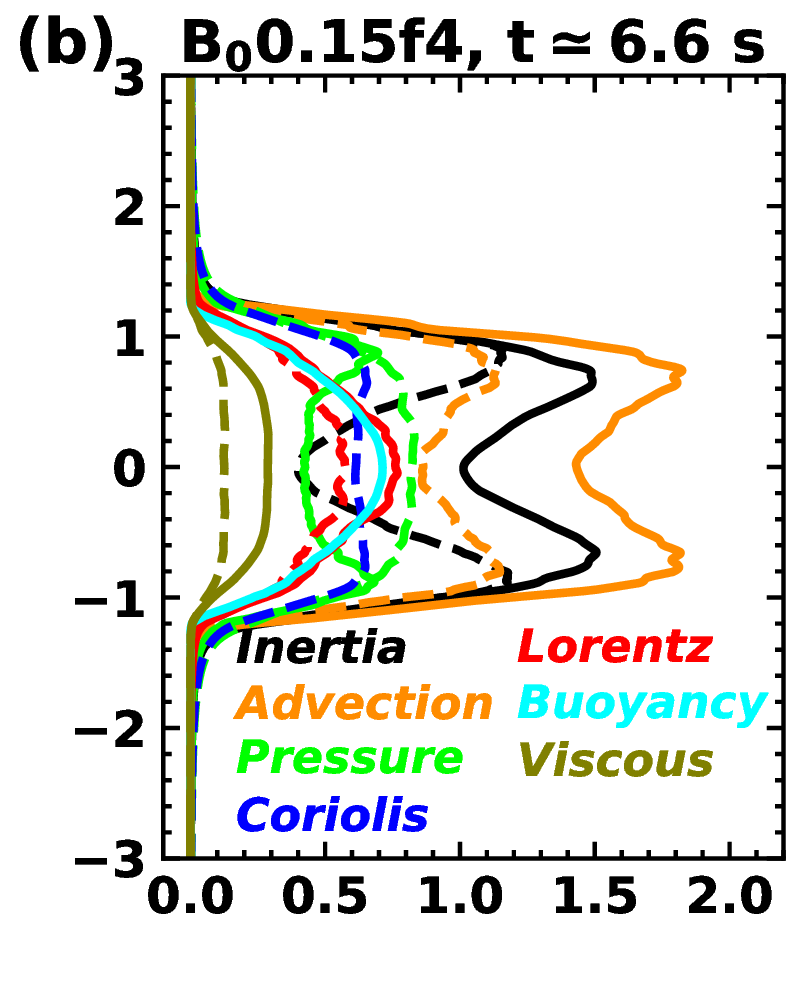}   		
  	\end{subfigure}
  \begin{subfigure}{0.32\textwidth}
    \centering
        \label{subfig: B015f8 t65b}
  	       \hspace{-1.0cm}\includegraphics[width=1.25\textwidth,trim={0cm 0cm 0cm 0cm},clip]{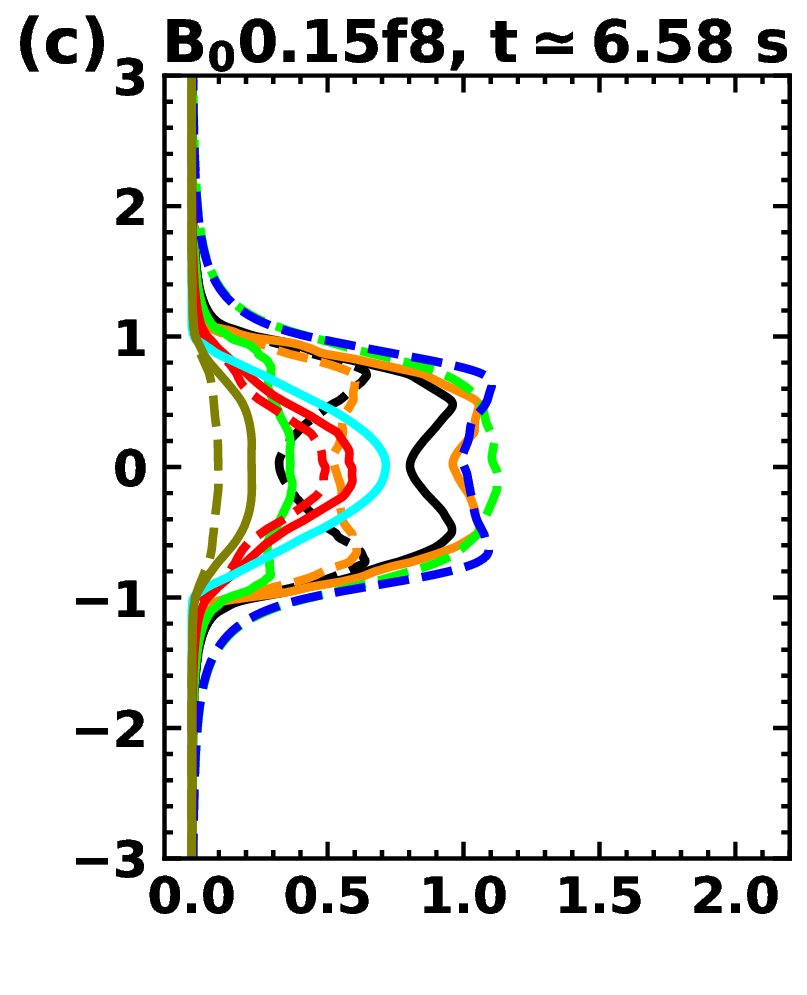}   		
  	\end{subfigure}\\
    \begin{subfigure}{0.32\textwidth}
  		\centering
  \hspace{-3.0cm}	\includegraphics[width=1.25\textwidth,trim={0cm 0cm 0cm 0cm},clip]{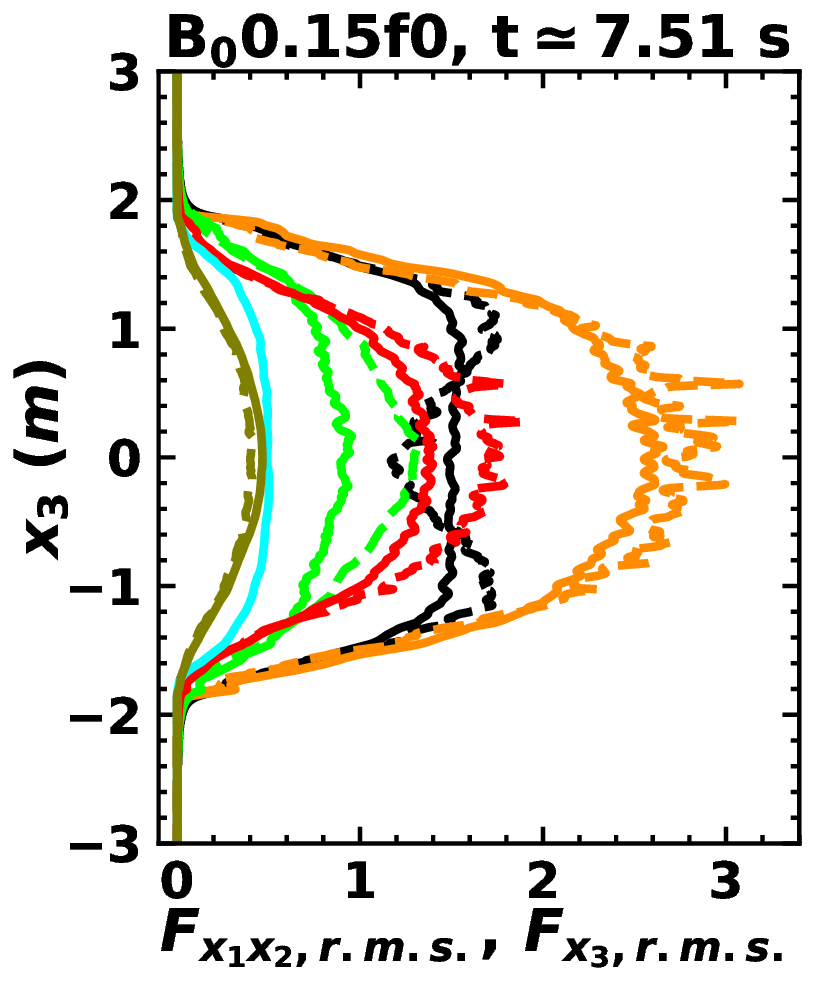}
  	\end{subfigure}
   \begin{subfigure}{0.32\textwidth}
  		\centering
  \hspace{-1.10cm}\includegraphics[width=1.25\textwidth,trim={0cm 0cm 0cm 0cm},clip]{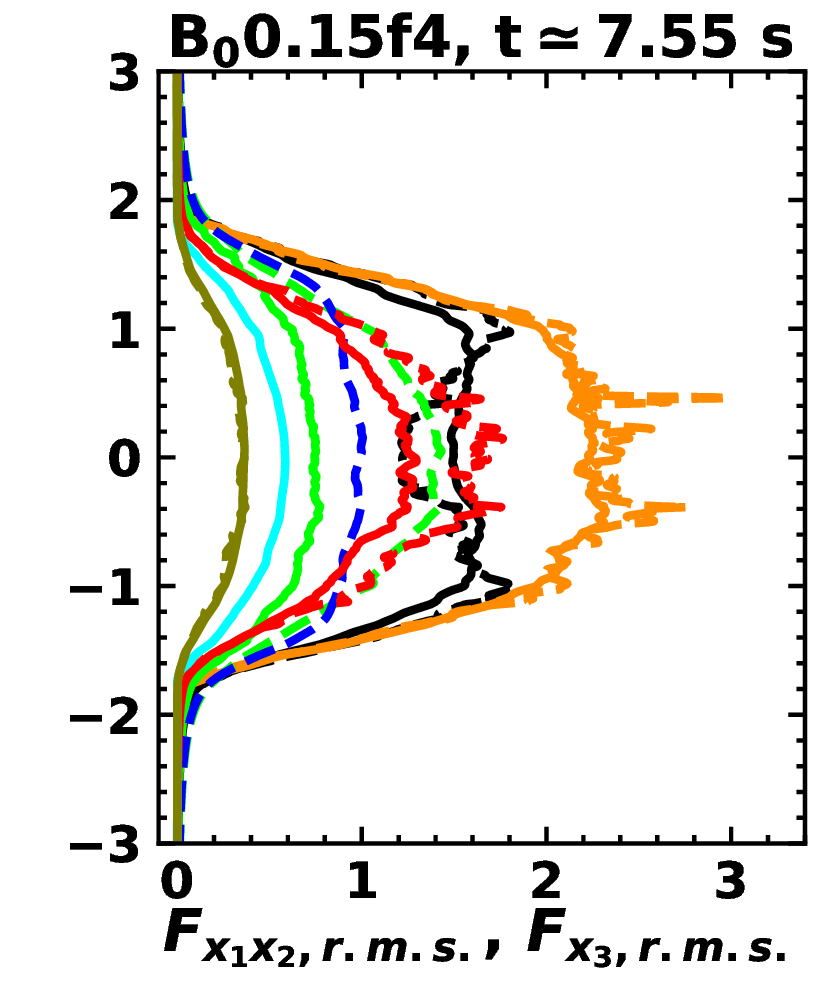}   		
  	\end{subfigure}
  \begin{subfigure}{0.32\textwidth}
        \centering
  \hspace{-1.0cm}\includegraphics[width=1.25\textwidth,trim={0cm 0cm 0cm 0cm},clip]{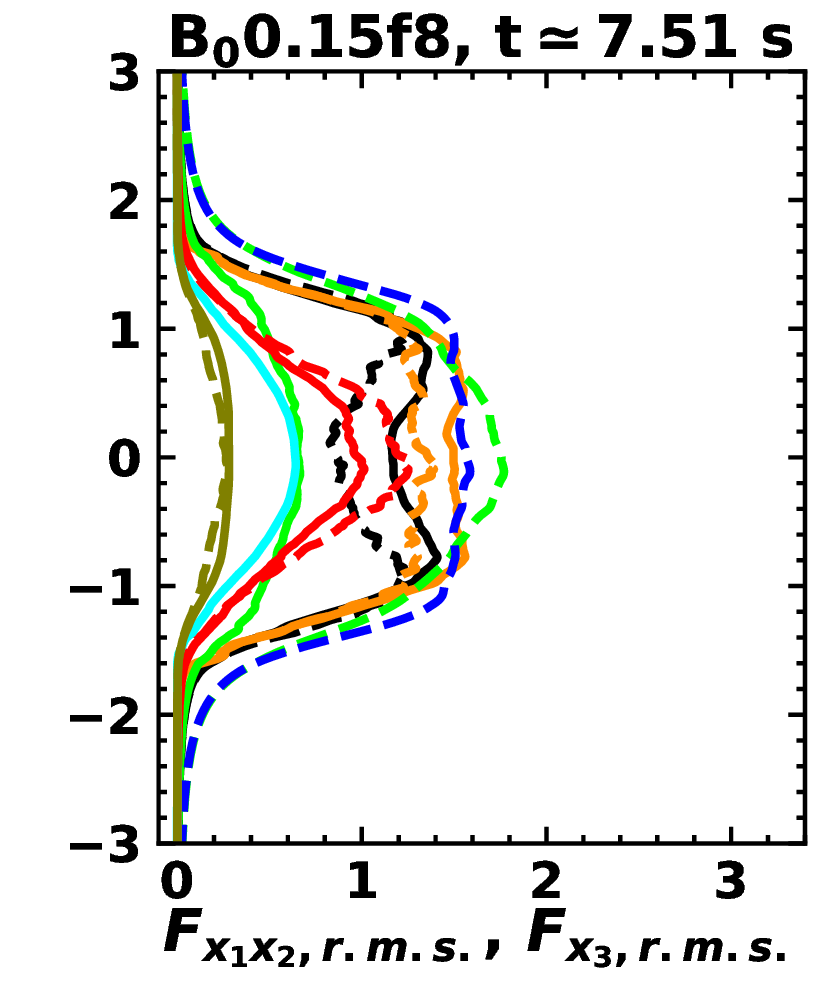}   		
  	\end{subfigure}
    \caption{Instantaneous vertical profiles of the horizontally averaged root mean square ($r.m.s.$) of each term (force) present in the vertical and horizontal component of the momentum equation (\ref{momentum1}). Horizontal forces, $F_{x_1x_2,\,r.m.s.}=\sqrt{F_{x_1,\,r.m.s.}^2+F_{x_2,\,r.m.s.}^2}$, are shown with dashed lines, whereas the vertical forces, $F_{x_3,\,r.m.s.}$, are indicated with solid lines for the (\textit{a}) non-rotating MHD case B$_0$0.15f0 at $t\simeq6.46$s, $7.51$s, (\textit{b}) rotating MHD case B$_0$0.15f4 at $t\simeq6.6$s, $7.55$s, and (\textit{a}) rotating MHD case B$_0$0.15f8 at $t\simeq6.58$s, $7.51$s. }
 \label{fig: force1}
\end{figure}

The instantaneous vertical ($x_3$) variation of $r.m.s.$ horizontal $F_{x_1x_2,\,r.m.s.}$ and $r.m.s.$ vertical $F_{x_3,\,r.m.s.}$ forces for the MHD cases B$_0$0.15f0, B$_0$0.15f4 and B$_0$0.15f8 is illustrated in figures \ref{fig: force1}a, \ref{fig: force1}b and \ref{fig: force1}c, respectively, at time instants $t\simeq6.5$s and $t\simeq7.5$s. These time instants correspond to the breakdown of plumes, which were depicted in figures \ref{Temp B015f0} and \ref{Temp B015f8} for the cases B$_0$0.15f0 and B$_0$0.15f8. We first recall that initially at $t=0$s, the fluids are at rest (i.e., $\boldsymbol{u}(\boldsymbol{x},t)=0$), and the mean magnetic field $B_0$ is imposed vertically. The random perturbations are seeded into the temperature field, which induces velocity fluctuations via the buoyancy term $\beta g T \delta_{i3}$ in the momentum equation \ref{momentum1}. These velocity fluctuations generate magnetic field fluctuations ($\boldsymbol{B}'$) through the stretching term $B_0 \partial u_i/\partial x_j$ in the induction equation \ref{mag eq1}. Subsequently, the Lorentz force due to the total magnetic field ($B_i=B_0\delta_{i3}+B_i'$) present as a source term in momentum equation \ref{momentum1}, alters the flow evolution, which in turn modifies the production of $\boldsymbol{B}'$ in the induction equation. This phenomenon repeats at successive time steps. The $r.m.s.$ of the Lorentz force includes a combined contribution from the mean and fluctuating components of the magnetic field. Since only the mean magnetic field is imposed vertically ($B_0$), the vertical component of the Lorentz force is expected to be higher than its horizontal components until the fluctuating magnetic field becomes significant due to the breakdown of flow structures. Therefore, the horizontal component of the Lorentz force acts to break the plumes. Figure \ref{fig: force1}a for the B$_0$0.15f0 case at $t\simeq6.46$s shows that the vertical component (solid line) of the Lorentz force is higher than the horizontal component (dashed line), signifying that plumes stretching is stronger than plumes breakup (see temperature contours in figure \ref{Temp B015f0} at $t\simeq6.46$s). The strength of the horizontal component of the Lorentz force is slightly higher towards the tips of plumes (i.e., $x_3\simeq \pm 1m$) than towards the center or mid-plane (at $x_3\simeq0m$), indicating that more breakup of elongated plumes occurs at their tips. Consequently, the larger nonlinear interactions near the plume tips result in a significantly higher advection force, which becomes the dominant force among all forces, at the plume tips compared to near the mid-plane. The magnitude of the inertial and the pressure forces is also greater at the plume tips than at the mid-plane. In contrast to the Lorentz, the advection, the inertia, and the viscous forces, the buoyancy force is larger towards the mid-plane than at the plume tips. At $t\simeq7.51$s in figure \ref{fig: force1}a, the horizontal component of the Lorentz force is higher than the vertical component, and both components increase towards the mid-plane, resulting in the breakup of each plume along its entire surface (see temperature contours in figure \ref{Temp B015f0} at $t\simeq7.51$s and supplementary Movie $3$). Therefore, the advection force also increases towards the mid-plane while spanning over the entire height of the mixing layer. The horizontal components of the advection and the pressure forces are also higher than the corresponding vertical components. The magnitudes of the Lorentz and the advection forces are higher at $t\simeq7.51$s than at $t\simeq6.46$s owing to the increased breakup of the plumes.\\ 

 \captionsetup[subfigure]{textfont=normalfont,singlelinecheck=off,justification=raggedright}
 
At $t\simeq6.6$s for the rotating MHD case B$_0$0.15f4, figure \ref{fig: force1}b shows competition between the  Coriolis and the Lorentz forces. The Coriolis force and the horizontal Lorentz force are in close balance near the mid-plane, while the Coriolis force dominates over the horizontal Lorentz force close to the plume tips. Therefore, the breakup of elongated plumes due to the horizontal Lorentz force is inhibited by the Coriolis force. Additionally, the Coriolis force is smaller than the vertical Lorentz force at the mid-plane, while it dominates near the plume tips. Therefore, the growth of the plumes due to the vertical stretching caused by the vertical Lorentz force is suppressed by the Coriolis force, as observed in the plot for the temporal evolution of the mixing layer height $h$ in figure \ref{fig: h}c. The higher values of the vertical components of the Lorentz force than its horizontal components indicate that the vertical stretching of plumes prevails over the plume breakup at $t\simeq6.6$s. As a result, the vertical components of the advection, the inertial, and the viscous forces remain higher than their horizontal counterparts. In contrast, the horizontal component of the pressure force is higher than its vertical component. At $t\simeq7.55$s, both the horizontal and the vertical components of the Lorentz force dominate over the Coriolis force, except near the plume tips. This signifies that the effect of the Coriolis force in inhibiting the growth and breakup of the plumes, except near the plume tips, is mitigated. The magnitude of the horizontal and vertical components of the advection force are comparable. \\

In figure \ref{fig: force1}c for the high rotation rate case B$_0$0.15f8 at $t\simeq6.58$s, the Coriolis force dominates significantly over the horizontal and the vertical components of the Lorentz force, which greatly inhibits the growth and breakdown of plumes (see temperature contours in figure \ref{Temp B015f8} at $t\simeq6.58$s and supplementary Movie $4$). The horizontal component of the pressure force also increases significantly compared to its vertical component. At $t\simeq7.51$s, the horizontal component of the Lorentz force is greater than its vertical component, suggesting that the breakdown is prominent over the vertical stretching of the plumes. However, the breakdown is still retarded by the dominant Coriolis force. Therefore, the vertical plumes with less distortion are visible in figure \ref{Temp B015f8} at $t\simeq7.51$s. We can conclude that with an increase in rotation rates (B$_0$0.15f4 and B$_0$0.15f8), the Coriolis force becomes stronger and dominates significantly over the Lorentz force at high rotation (B$_0$0.15f8) resulting in the suppression of plumes growth and breakup. As a result, the advection force is reduced with increasing rotation rates compared to the non-rotating case B$_0$0.15f0, which can be observed by comparing figures \ref{fig: force1}a, \ref{fig: force1}b and \ref{fig: force1}c for B$_0$0.15f0, B$_0$0.15f4 and B$_0$0.15f8, respectively at $t\simeq6.5$s, $7.5$s. This reduction in advection force causes a decrease in heat transfer, which corroborates the decrease in $Nu$, observed in figure \ref{fig: Nu}c, in the B$_0$0.15f4 and B$_0$0.15f8 cases compared to B$_0$0.15f0. \\

 \captionsetup[subfigure]{textfont=normalfont,singlelinecheck=off,justification=raggedright}
  \begin{figure}
 	\centering
    \begin{subfigure}{0.3\textwidth}
        \centering
        \label{subfig: B0f4 t65}
       \hspace{-3.0cm} \includegraphics[width=1.25\textwidth,trim={0cm 0cm 0cm 0cm},clip]{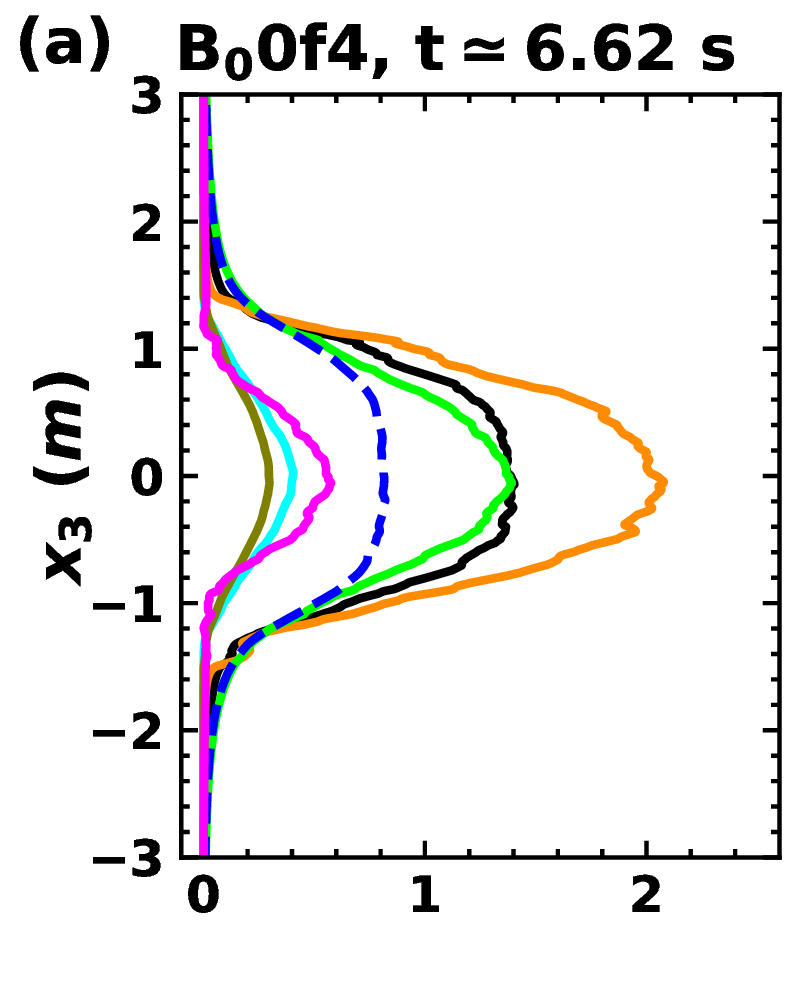}  	
    \end{subfigure}
  \begin{subfigure}{0.3\textwidth}
    \centering
       \label{subfig: B0f8 t65}
  \hspace{-1.25cm}	\includegraphics[width=1.25\textwidth,trim={0cm 0cm 0cm 0cm},clip]{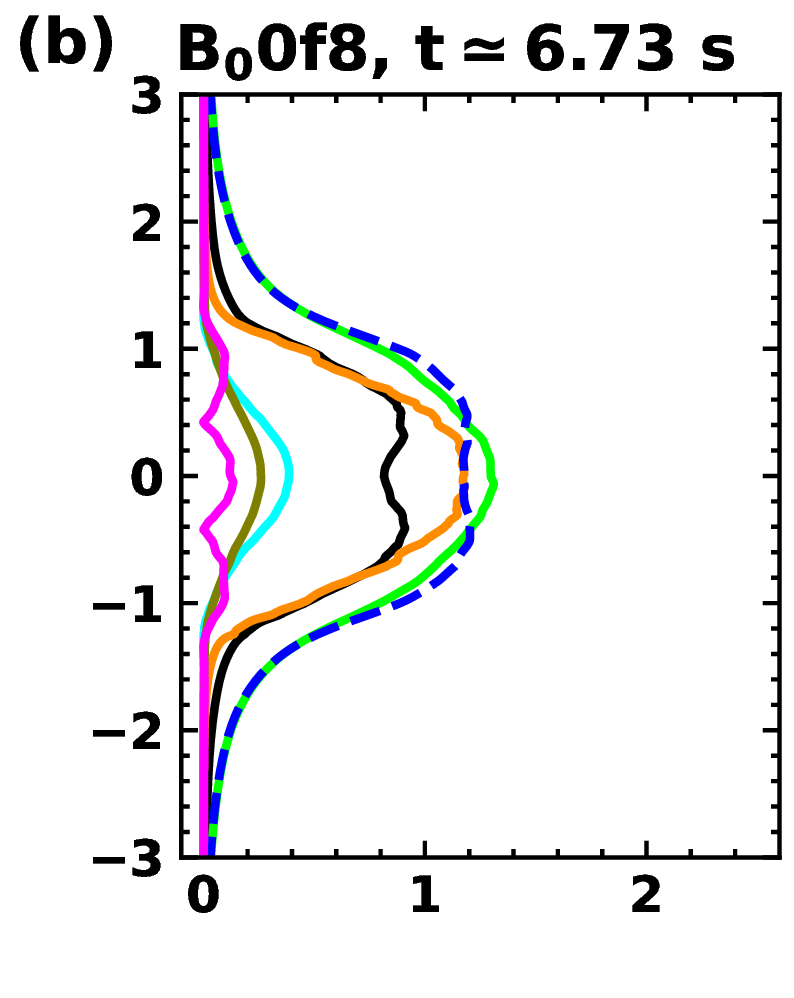}  	
  	\end{subfigure}
  \begin{subfigure}{0.3\textwidth}
    \centering
      \label{subfig: B015f8 t65}
  \hspace{-1.0cm}	\includegraphics[width=1.25\textwidth,trim={0cm 0cm 0cm 0cm},clip]{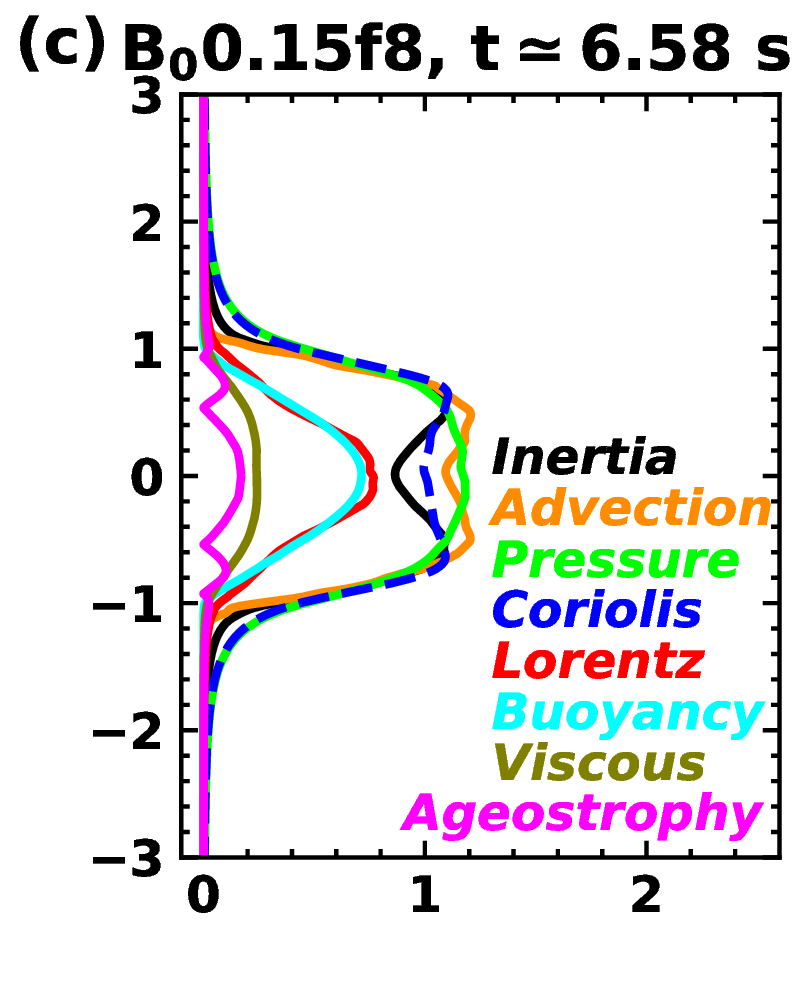}  	
  	\end{subfigure}
  \\
    \begin{subfigure}{0.3\textwidth}
  		\centering
  \hspace{-3.0cm}	\includegraphics[width=1.25\textwidth,trim={0cm 0cm 0cm 0cm},clip]{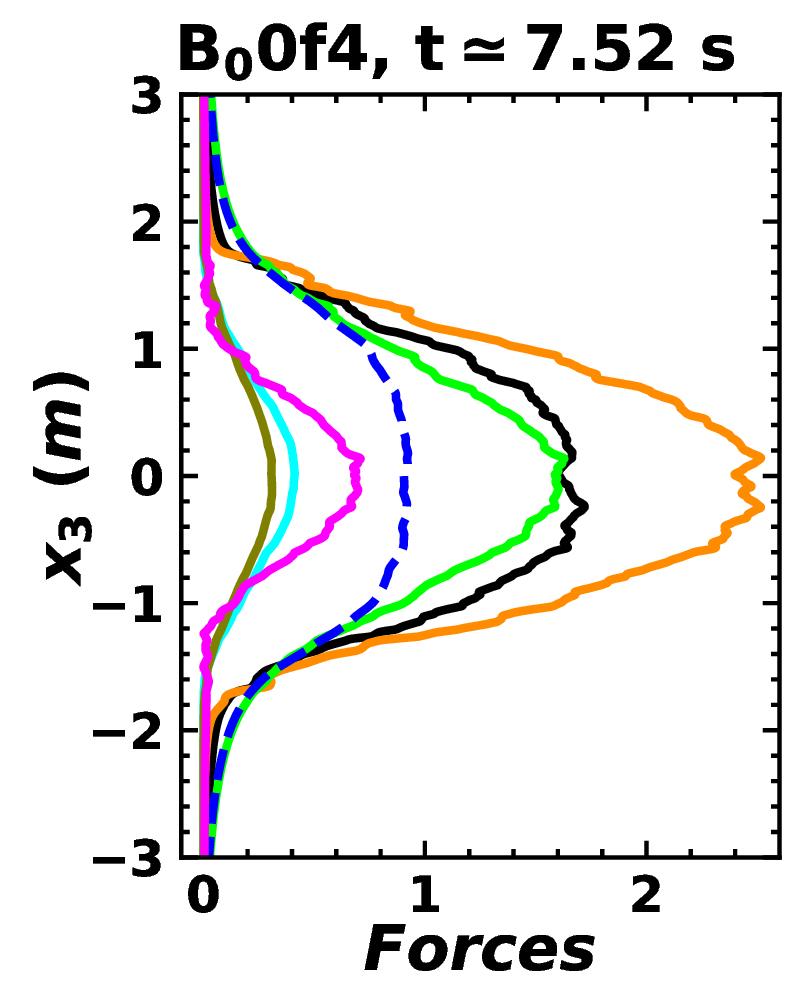}   		
  	\end{subfigure}
  \begin{subfigure}{0.3\textwidth}
    \centering
  \hspace{-1.25cm}\includegraphics[width=1.25\textwidth,trim={0cm 0cm 0cm 0cm},clip]{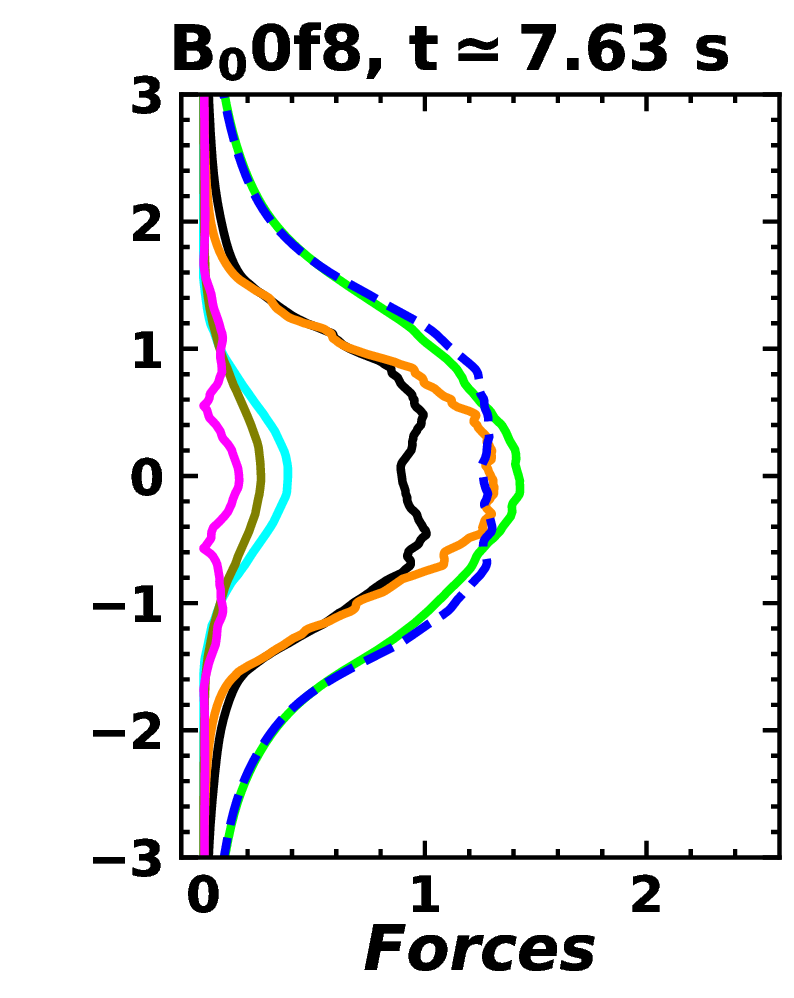}   		
  	\end{subfigure}
  \begin{subfigure}{0.3\textwidth}
    \centering
  \hspace{-1.0cm}	\includegraphics[width=1.25\textwidth,trim={0cm 0cm 0cm 0cm},clip]{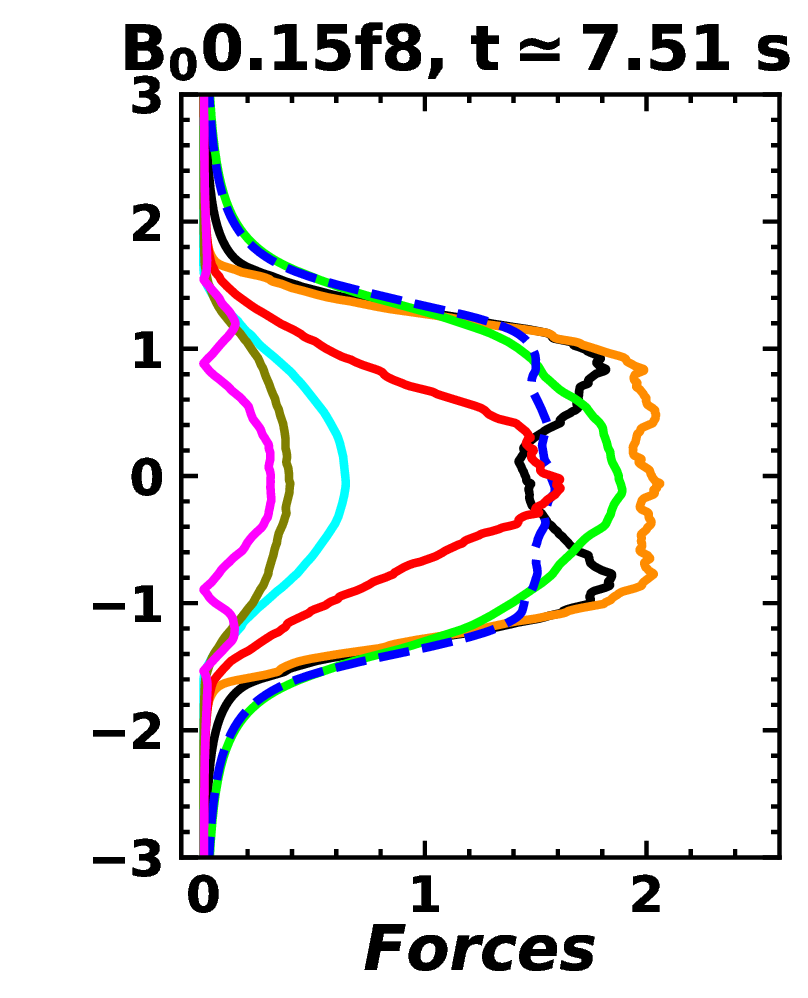}   		
  	\end{subfigure}
    \caption{Instantaneous vertical profiles of the resultant $r.m.s.$ values of each force $\left(F_{r.m.s.}=\sqrt{F_{x_1,\,r.m.s.}^2+F_{x_2,\,r.m.s.}^2+F_{x_3,\,r.m.s.}^2}\right)$ in momentum equation \ref{momentum1} for the rotating (\textit{a}) HD case B$_0$0f4 at $t\simeq6.62$s, $7.52$s, (\textit{b}) HD case B$_0$0f8 at $t\simeq6.73$s, $7.63$s, and (\textit{c}) MHD case B$_0$0.15f8 at $t\simeq6.58$s, $7.51$s. Solid lines of different colors represent the resultant $r.m.s.$ values of the corresponding force. The dashed blue line denotes the horizontal Coriolis force. }
 \label{fig: force2}
\end{figure} 

The vertical variation of the resultant $r.m.s.$ values of each force, computed as $F_{r.m.s.}=\left(F_{x_1,\,r.m.s.}^2+F_{x_2,\,r.m.s.}^2+F_{x_3,\,r.m.s.}^2\right)^{1/2}$, in the momentum equation \ref{momentum1} for the rotating HD cases B$_0$0f4, B$_0$0f8, and MHD cases B$_0$0.15f8 is shown in figure \ref{fig: force2}a, \ref{fig: force2}b, and \ref{fig: force2}c, respectively at two different time instants. The deviations from the geostrophic balance between the Coriolis and the pressure forces are measured by the ageostrophy and is estimated as the difference between these two forces. Figures \ref{fig: force2}a and \ref{fig: force2}b for the HD cases illustrate that at a rotation rate of $f=4$ (B$_0$0f4 case), the ageostrophy is higher, but it reduces significantly when the rotation rate increases to $f=8$ (B$_0$0f8 case). This signifies that the Coriolis and the pressure forces are in close balance in B$_0$0f8. This confirms the presence of the Taylor–Proudman constraint, which results in the vertical alignment of the flow structures as observed in the temperature contours in figure \ref{Temp B00f8}. \cite{boffetta2016rotating} also reported the existence of the Taylor–Proudman constraint in rotating RT instability from flow visualization. In B$_0$0f4, the Coriolis and the pressure forces are the subdominant forces, whereas they become nearly dominant forces among all forces in the B$_0$0f8 case. Therefore, we expect the leading-order geostrophic balance between the Coriolis and the pressure forces to occur at rotation rates of $f>8$, constraining the flow to be two-dimensional satisfying the Taylor–Proudman theorem. \\

For the rotating MHD case B$_0$0.15f8 in figure \ref{fig: force2}c at $t\simeq6.58$s and $t\simeq7.51$s, the ageostrophy is higher than the corresponding HD case B$_0$0f8 (figure \ref{fig: force2}b). This deviation from the geostrophic balance is attributed to the presence of Lorentz force. Additionally, in B$_0$0.15f8 case, the higher ageostrophy at $t\simeq7.51$s compared to that at $t\simeq6.58$s due to the increase in the Lorentz force indicates that the deviation from the geostrophic balance increases as the RT instability grows. Therefore, in contrast to the rotating HD case, the geostrophic balance is not expected at high rotation rates in the MHD cases owing to the mitigation of the effect of the Coriolis force by the Lorentz force. Due to this mitigation of the Coriolis force effect, the advection force at $t\simeq7.51$s in B$_0$0.15f8 case (figure \ref{fig: force2}c) is higher than in the B$_0$0f8 case at $t\simeq7.63$s (figure \ref{fig: force2}b). This results in efficient heat transfer in the rotating MHD B$_0$0.15f8 case compared to the rotating HD B$_0$0f8 case.\\

 \captionsetup[subfigure]{textfont=normalfont,singlelinecheck=off,justification=raggedright}
  \begin{figure}
 	\centering
    \begin{subfigure}{0.49\textwidth}
  		\centering
        \caption{}  \label{fig: budget B0f0}
  		\includegraphics[width=1.0\textwidth,trim={0cm 0cm 0cm 0cm},clip]{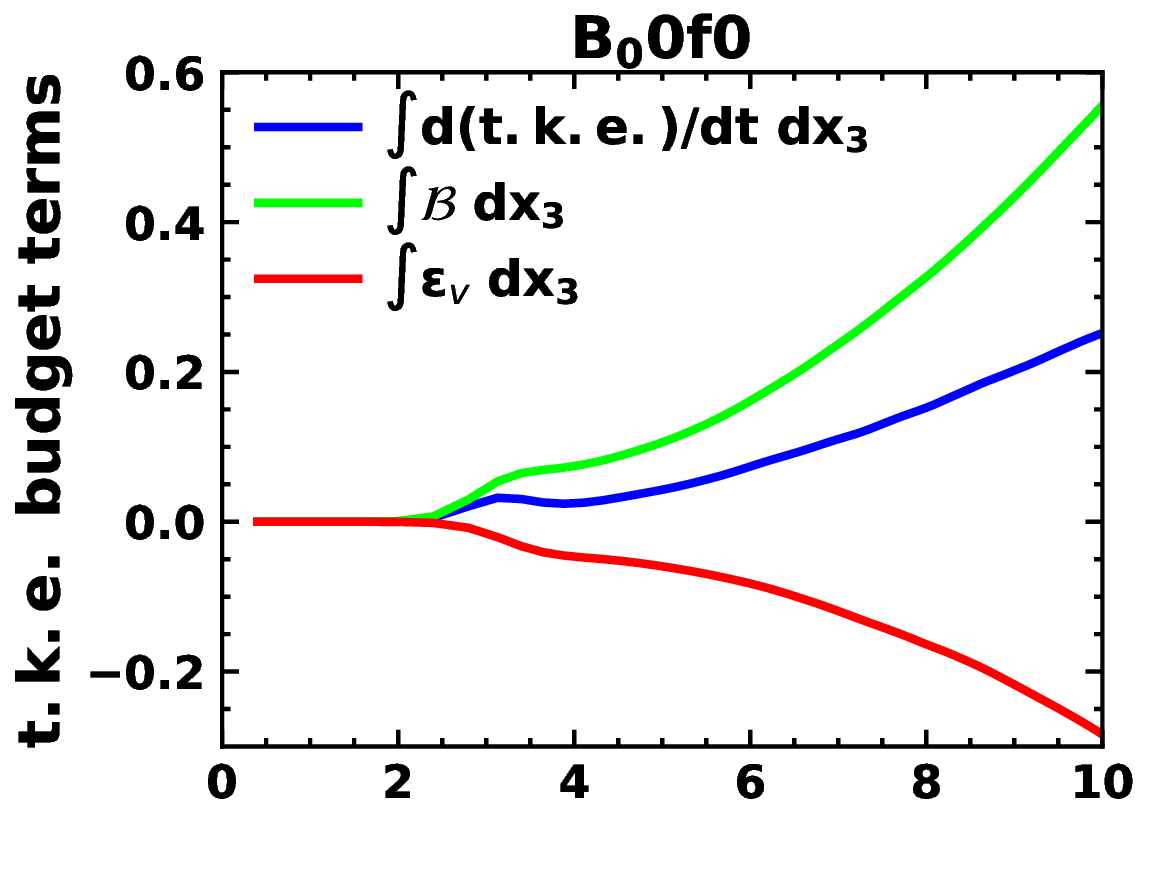}
  	\end{subfigure}
  	\begin{subfigure}{0.49\textwidth}
  		\centering
        \caption{}  \label{fig: budget B01f0}
  		\includegraphics[width=1.0\textwidth,trim={0cm 0cm 0cm 0cm},clip]{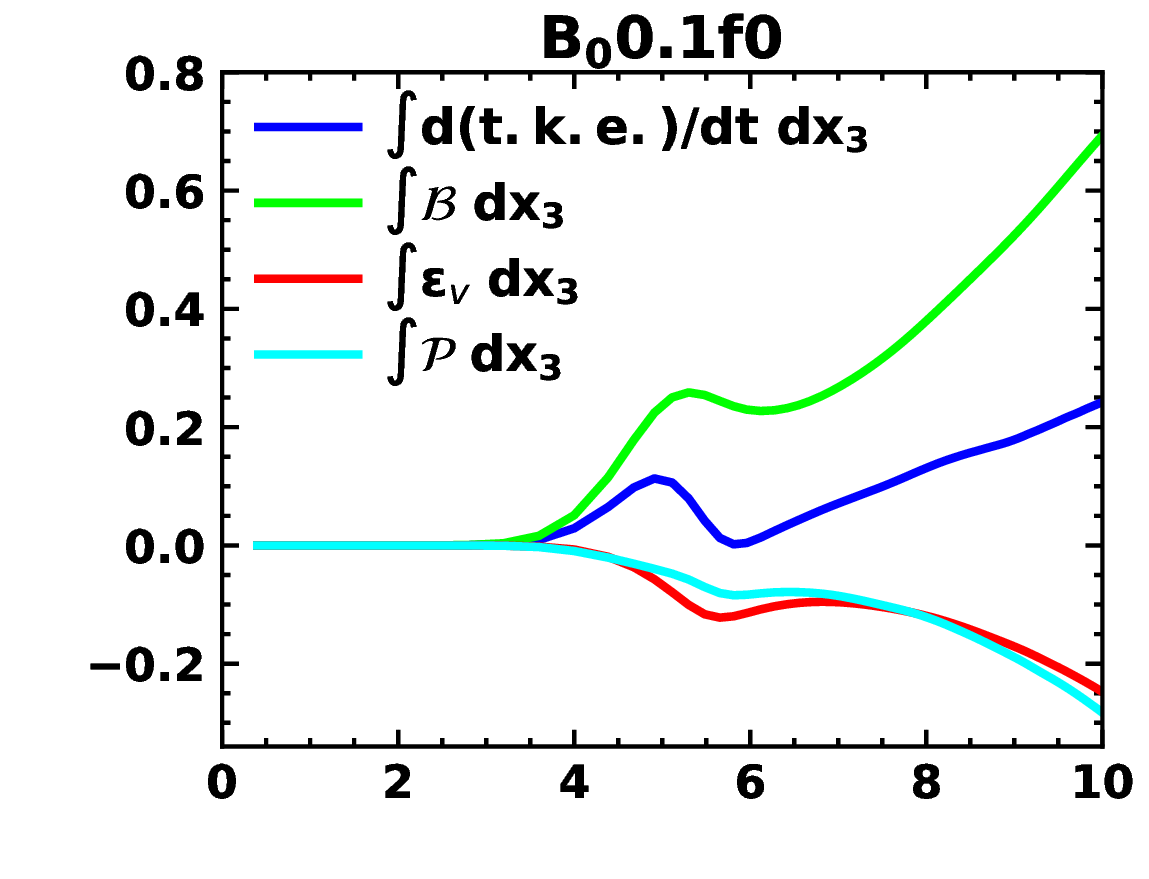}
  	\end{subfigure}
 \\
  \begin{subfigure}{0.49\textwidth}
  		\centering
        \caption{}  \label{fig: budget B01f8}
  		\includegraphics[width=1.0\textwidth,trim={0cm 0cm 0cm 0cm},clip]{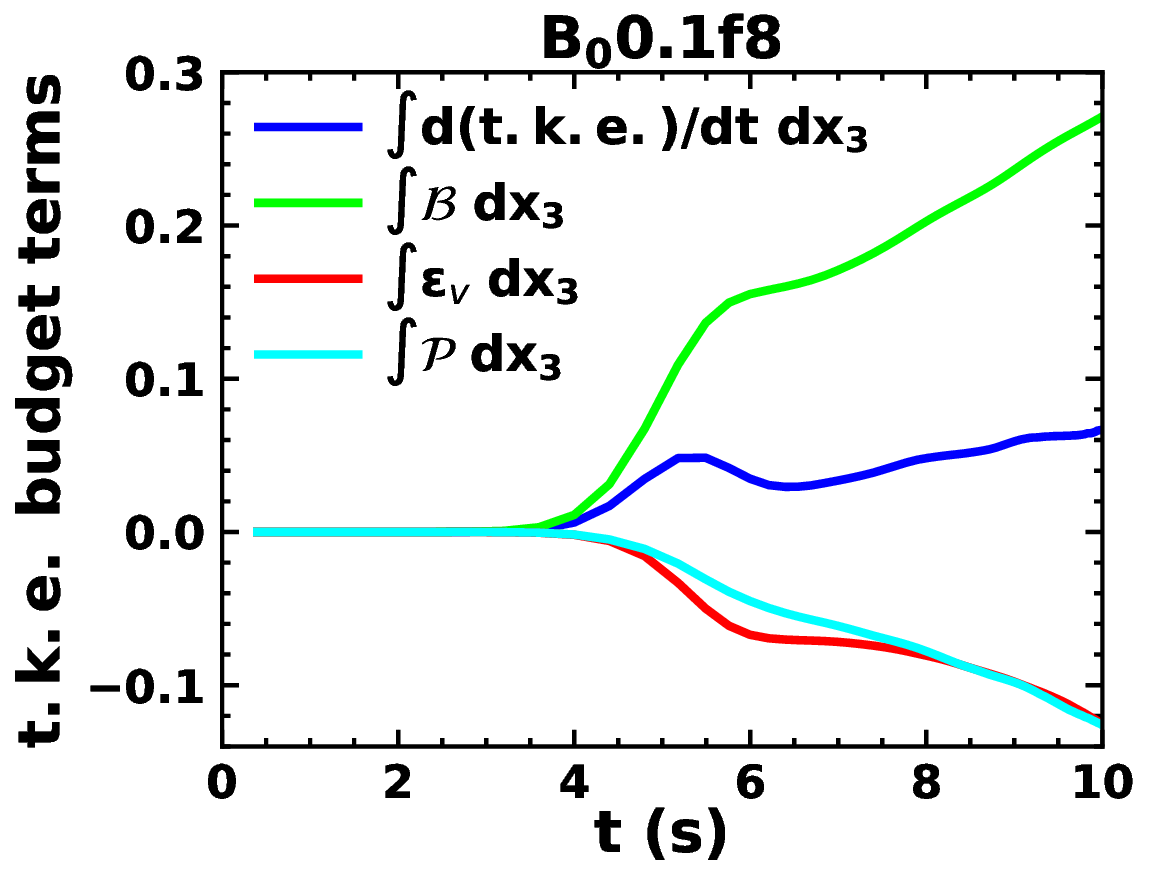}
  	\end{subfigure}
  \begin{subfigure}{0.49\textwidth}
  		\centering
        \caption{}  \label{fig: budget B015f0}
  		\includegraphics[width=1.0\textwidth,trim={0cm 0cm 0cm 0cm},clip]{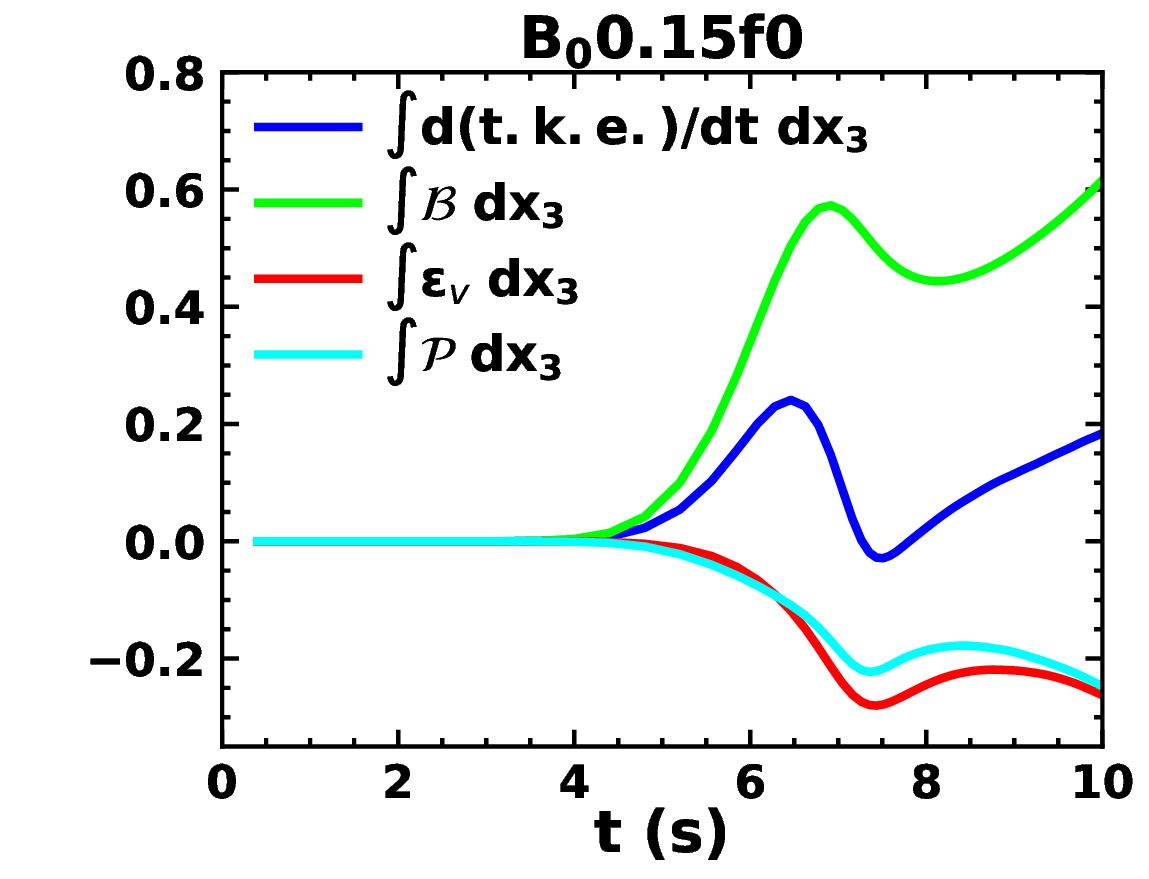}
  \end{subfigure}
 
  	\caption{Temporal evolution of vertically ($x_3$) integrated $t.k.e.$ budget terms of equation \ref{tke budget} for the (\textit{a}) non-rotating HD case B$_0$0f0, (\textit{b}) non-rotating MHD case B$_0$0.1f0, (\textit{c}) rotating MHD case B$_0$0.1f8, and (\textit{d}) non-rotating MHD case B$_0$0.15f0. }
 \label{fig: budget}
\end{figure}
 
The above analysis reveals that the Coriolis and the Lorentz forces significantly impact the growth and breakdown of thermal plumes into small-scale turbulence. Since small-scale turbulence significantly alters the heat transfer phenomenon, we evaluate the turbulent energetics associated with the flow to understand the generation and evolution of turbulent kinetic energy in the presence of the imposed magnetic field and rotation. The temporal evolution equation of the horizontally averaged turbulent kinetic energy, $t.k.e.=\overline{u_i'u_i'}/2$, which is given as follows: 
\begin{equation}
\label{tke budget}
 \frac{dt.k.e.}{dt}=\mathcal{B}-\epsilon_v+\mathcal{S}+\mathcal{P}-\frac{\partial \mathcal{T}_j}{\partial x_j}.
\end{equation}
Here, the buoyancy flux  
\begin{equation}
\label{tke}
\mathcal{B}=\beta g \overline{u_3'T'}
\end{equation}
represents the conversion of available potential energy into $t.k.e.$. The viscous dissipation rate 
\begin{equation}
\label{diss}
    \epsilon_v=2\nu \overline{ s_{ij}' s_{ij}'} \, ;  \quad s_{ij}'=\frac{1}{2}\left( \frac{\partial u_i'}{\partial x_j} + \frac{\partial u_j'}{\partial x_i} \right),
\end{equation}
acts as a sink for $t.k.e.$ that converts the $t.k.e.$ to internal energy. The shear production rate of $t.k.e.$   
\begin{equation}
\mathcal{S}=- \overline{u_i' u_j'} \frac{\partial \overline {u_i}}{ \partial x_j}
\end{equation}
is negligible since no mean velocity field is involved in the present study. The production of $t.k.e.$ due to the work done by the Lorentz force on the velocity field is \citep{naskar_pal_2022b}
\begin{equation}
\mathcal{P}= \mathcal{P}_1 +\mathcal{P}_2+ \mathcal{P}_3,
\end{equation}
where
\begin{equation}
\mathcal{P}_1=-\overline{B_j}\,\overline{B_i'\frac{\partial u_i'}{\partial x_j}}, \quad \mathcal{P}_2=\overline{u_i'B_j'}\,\frac{\partial \overline{B_i}}{\partial x_j} , \quad \mathcal{P}_3=-\overline{B_i'B_j'\frac{\partial u_i'}{\partial x_j}}.
\end{equation}
The energy exchange between the velocity and the magnetic fields occurs through the magnetic production terms $\mathcal{P}_1$, $\mathcal{P}_2$, and $\mathcal{P}_3$.

The transport of $t.k.e.$, which represents the divergence of $t.k.e.$ flux $\mathcal{T}_j$,
\begin{equation}
\frac{\partial\mathcal{T}_j}{\partial x_j} = \frac{\partial}{\partial x_j} \bigg(\overline{ p' u_j' } - 2\nu  \overline{ s_{ij}' u_i' } + \frac{1}{2}\overline{ u_i' u_i' u_j'} - \overline{u_i'B_i'B_j'} - \overline{B_j}\,\overline{u_i'B_i'} \bigg),
\end{equation}
becomes negligible when integrated vertically ($x_3$). The time evolution of the vertically ($x_3$) integrated $t.k.e.$ budget terms of equation \ref{tke budget} are depicted in figures \ref{fig: budget B0f0},  \ref{fig: budget B01f0},  \ref{fig: budget B01f8}, and  \ref{fig: budget B015f0} for the non-rotating HD case B$_0$0f0, non-rotating MHD case B$_0$0.1f0, rotating MHD case B$_0$0.1f8, and non-rotating MHD case B$_0$0.15f0, respectively. In B$_0$0f0 case, the buoyancy flux ($\mathcal{B}$), which produces $t.k.e.$, together with the viscous dissipation rate ($\epsilon_v$) balances ${dt.k.e.}/{dt}$ (see figure \ref{fig: budget B0f0}). In the MHD B$_0$0.1f0, B$_0$0.1f8 and B$_0$0.15f0 cases, a negative value of the term $\mathcal{P}$ signifies that the energy is transferred from the velocity field to the magnetic field (figures \ref{fig: budget B01f0}, \ref{fig: budget B01f8}, and \ref{fig: budget B015f0}). Therefore, the term $\mathcal{P}$ acts as a sink for the $t.k.e.$, which converts the turbulent kinetic energy to turbulent magnetic energy. In the MHD cases, $\mathcal{B}$ together with $\mathcal{P}$ and $\epsilon_v$ balances ${dt.k.e.}/{dt}$. We obtain $t.k.e.$ budget closure in all simulation cases, indicating that the grid spacing considered in the present simulations is adequate to capture the small-scale turbulence. The comparison of figures \ref{fig: budget B01f0} and \ref{fig: budget B01f8} for the B$_0$0.1f0 and B$_0$0.1f8 cases, respectively, shows that the magnitude of all the budget terms (${dt.k.e.}/{dt}$, $\mathcal{B}$, $\epsilon_v$ and $\mathcal{P}$) is reduced in B$_0$0.1f8 compared to B$_0$0.1f0 owing to the instability-suppressing effects of the Coriolis force.\\

\begin{figure*}
    \centering
    \hspace{-2cm}  \includegraphics[width=0.38\textwidth,trim={0.2cm 0.5cm 0.2cm 0.2cm},clip]{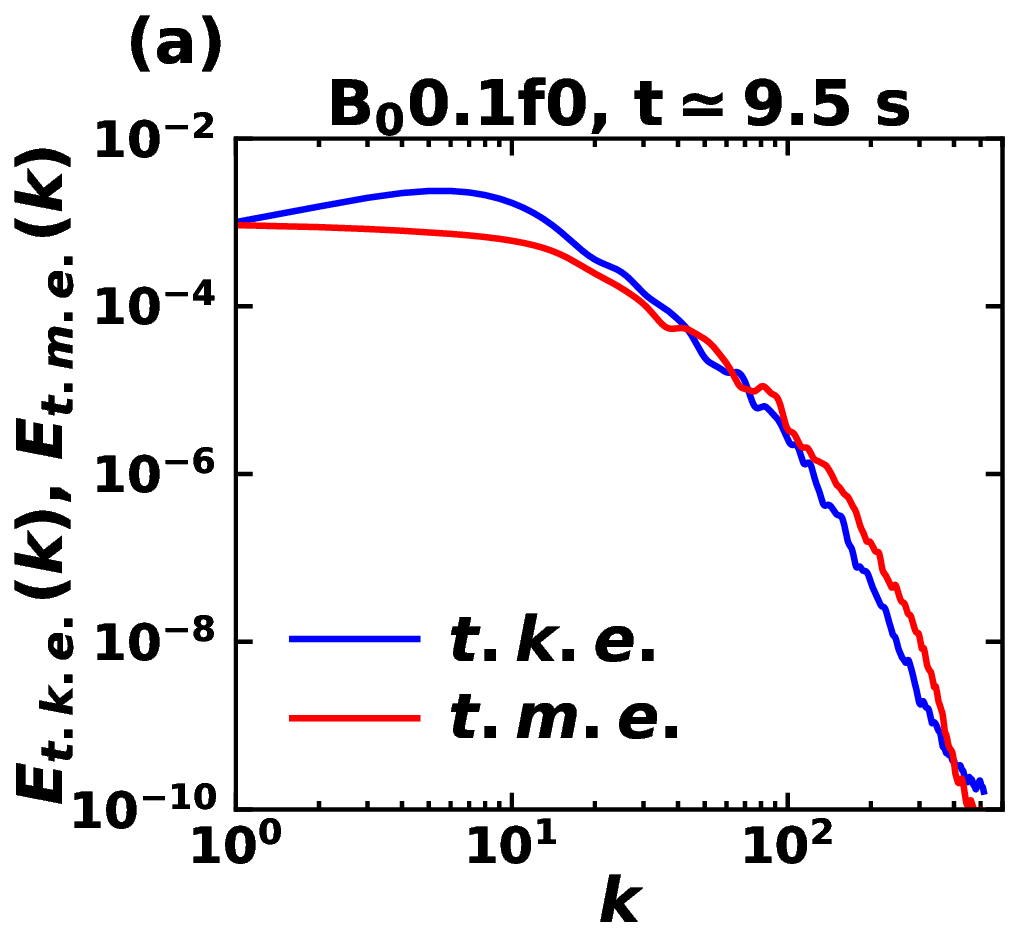} \slabel{fig: spectra Bz01f0}
    \hspace{-0.15cm} \includegraphics[width=0.38\textwidth,trim={0.2cm 0.5cm 0.2cm 0.2cm},clip]{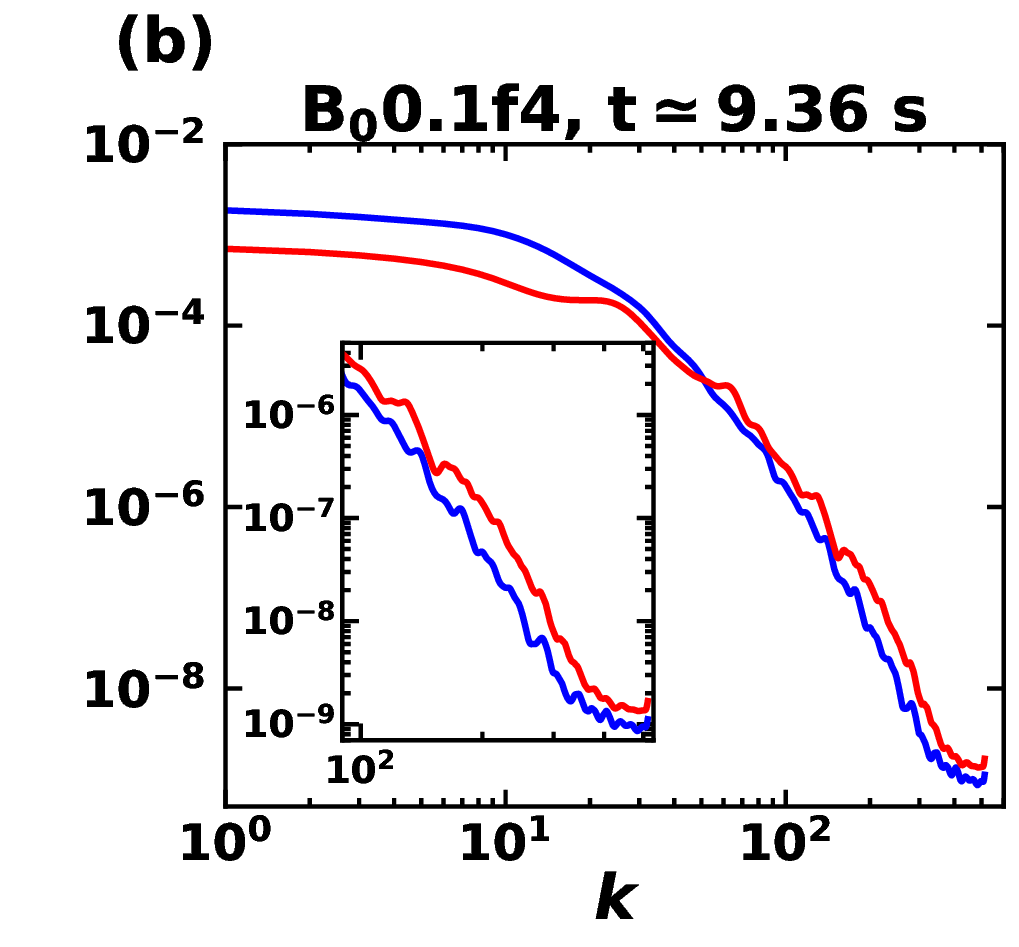} \slabel{fig: spectra Bz01f4}
    \hspace{-0.15cm} \includegraphics[width=0.38\textwidth,trim={0.2cm 0.5cm 0.2cm 0.2cm},clip]{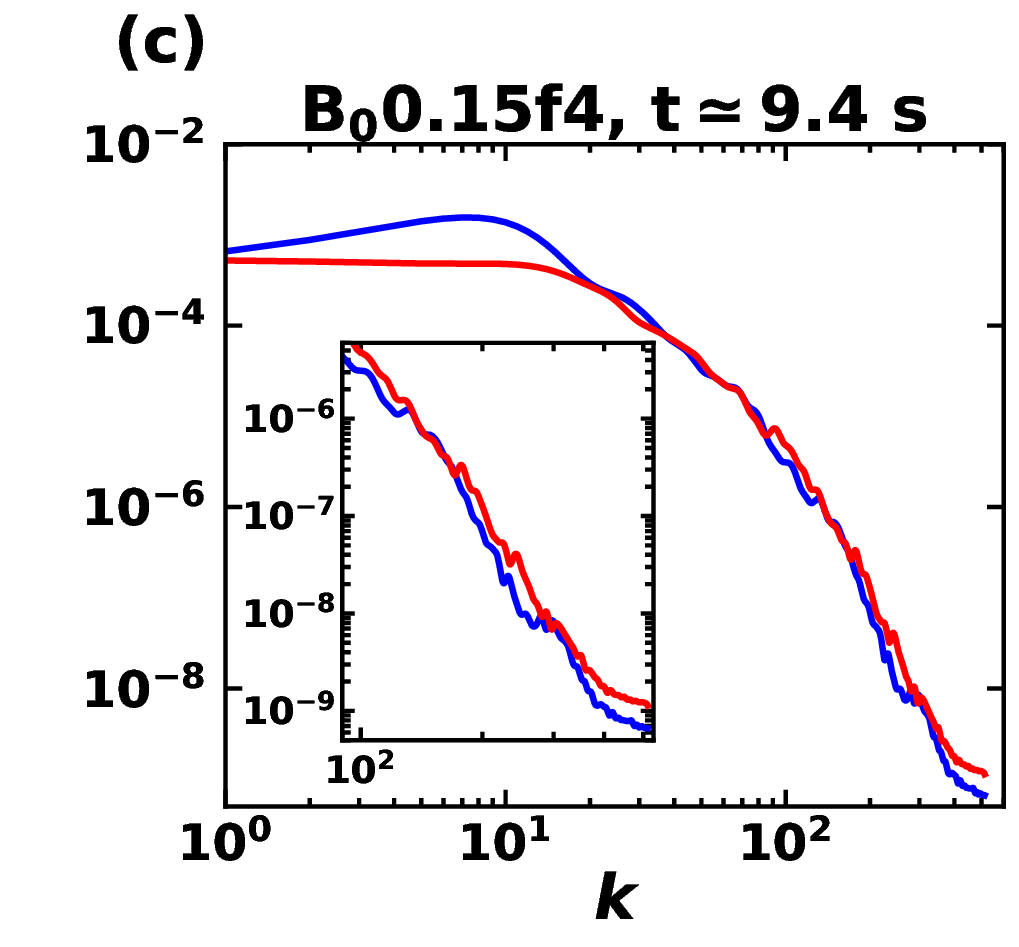}  \slabel{fig: spectra Bz015f4}  
    \hspace{-1.0cm}
    \caption{Horizontal ($x_1$) spectra of turbulent kinetic energy ($E_{t.k.e.}(k)$) and turbulent magnetic energy ($E_{t.m.e.}(k)$) 
    for the MHD cases (\textit{a}) B$_0$0.1f0 at $t\simeq9.5$s, (\textit{b}) B$_0$0.1f4 at $t\simeq9.36$s, and (\textit{c}) B$_0$0.15f4 at $t\simeq9.4$s. $k$ denotes the horizontal ($x_1$) wavenumber. The spectra at large $k$ (small-scale eddies) are magnified in the insets of (\textit{b}) and (\textit{c})}.
 \label{fig: spectra}
\end{figure*}

To corroborate the conversion of the turbulent kinetic energy $t.k.e.$ ($={u_i'u_i'}/2$) to turbulent magnetic energy $t.m.e.$ ($={B_i'B_i'}/2$), we plot the horizontal ($x_1$) spectra of turbulent kinetic energy $t.k.e.$ ($E_{t.k.e.}(k)$) and turbulent magnetic energy $t.m.e.$ ($E_{t.m.e.}(k)$) against horizontal wavenumber $k$ (in $x_1$ direction) in figures \ref{fig: spectra}a, \ref{fig: spectra}b and \ref{fig: spectra}c for the MHD cases B$_0$0.1f0, B$_0$0.1f4, and B$_0$0.15f4, respectively. The spectra are computed at time instant $t\simeq9.36-9.5$s corresponding to the mixing regime. All spectra reveal the presence of $t.m.e.$, which is comparable to $t.k.e.$ at all wavenumbers (or scales). This supports our finding that the energy extracted from the $t.k.e.$ through the magnetic production term $\mathcal{P}$ leads to an increase in the $t.m.e.$ in the flow, as observed from the $t.k.e.$ budget presented for MHD cases in figures \ref{fig: budget B01f0}, \ref{fig: budget B01f8}, and \ref{fig: budget B015f0}. Additionally, all spectra show that the large-scale eddies (small $k$) contain higher $t.k.e.$ than $t.m.e.$. However, the small-scale eddies (large $k$) exhibit higher $t.m.e.$ than $t.k.e.$, indicating a strong energy transfer from $t.k.e.$ to $t.m.e.$ at small scales (see figure \ref{fig: spectra}a and inset in figures \ref{fig: spectra}b and \ref{fig: spectra}c). \\

\section{Conclusions}\label{sec: conclusions}
We have conducted DNS to investigate the thermal convection driven by the Rayleigh-Taylor (RT) instability under the combined influence of the externally imposed vertical mean magnetic field $B_0$ (scaled as the Alfv\`{e}n velocity) and rotation about the vertical axis. The RT instability in the simultaneous presence of magnetic field and rotation plays a crucial role in engineering applications such as the inertial confinement fusion (ICF) implosions \citep{walsh2021biermann,ZHOU20171a}, and astrophysical phenomena such as solar prominences \citep{hillier2017magnetic}, acceleration at the disc-magnetosphere boundary \citep{kulkarni2008accretion}, and many more. Our motivation for this investigation emanates from the absence of any study on the combined influence of the magnetic field and rotation on the evolution of RT instability in the literature. Our DNS provides a comprehensive view of how the interplay between the vertical mean magnetic field and rotation affects the formation and breakup of flow structures and the heat transfer efficiency in the RT configuration. The present simulations can be used as a framework to gain important insights into the engineering and astrophysical phenomena discussed above. \\

We perform the simulations for $B_0=0$, $0.15$, and $0.3$ each at rotation rates of $f=0$, $4$, and $8$. For the non-rotating HD case ($B_0=0$), the non-linear interactions between the thermal plumes with mushroom-shaped caps lead to turbulent mixing and heat transfer between the fluids. In rotating HD cases, with increasing rotation rates from $f=4$ to $8$, the Coriolis force stabilizes the flow by inhibiting the formation of the mushroom-shaped caps at the plume tips and the deformation of plumes, leading to the appearance of coherent vertical plumes. Consequently, the vertical velocity fluctuations are suppressed, resulting in the reduced growth of the mixing layer height $h$ compared to the non-rotating HD case. A slight delay in the onset of RT instability (or the growth of $h$) is observed at $f=8$ compared to the $f=0$ case. In the presence of the imposed $B_0=0.1, 0.15,0.3$ at $f=0$, smooth vertically elongated plumes form owing to the suppression of the secondary shear instabilities (KH instabilities) at small scales, which inhibits horizontal mixing. Later on, the vertically stretched plumes become unstable, leading to the appearance of mushroom-shaped caps at their tips, followed by the detachment of the caps due to continued vertical stretching. Eventually, the full breakdown of thin plumes into small structures occurs, resulting in turbulent mixing. This mixing remains weaker than the corresponding HD case due to the collimation of fluid structures along the vertical magnetic field lines. Therefore, an enhancement in the mixing layer height $h$ is observed compared to the HD cases. We find the delay in the onset of RT instability for $B_0=0.1$ and $0.15$, which becomes significant for $B_0=0.3$, consistent with the previously developed linear theory.
Additionally, the flow collimation along the vertical magnetic field lines becomes stronger with an increase in $B_0$ from 0.1 to 0.15 and 0.3, enhancing $h$. With the inclusion of a high rotation rate of $f=8$ along with $B_0=0.1,0.15,0.3$, the Coriolis force reduces the growth of $h$ compared to the corresponding non-rotating MHD case by suppressing the growth of vertically elongated plumes and inhibits the formation of mushroom-shaped caps at the plume tips. The breakdown of plumes into small-scale turbulence due to the continued vertical stretching by $B_0$ is also suppressed by the Coriolis force. We find that the tendency of the Coriolis force to inhibit the growth of $h$ increases with an increase in $f(=4,8)$ for each $B_0=0.1$, 0.15, and 0.3. Although, for the MHD cases with $B_0=0.1$, 0.15, and 0.3, $h$ is decreased at $f=4,8$ compared to $f=0$, it remains higher than in the corresponding non-rotating and rotating HD cases due to the flow collimation along the vertical magnetic field lines. Even $h$ for $f=8$ MHD is higher than for $f=4$ HD cases, signifying that the vertical stretching caused by the imposed $B_0$ mitigates the Coriolis force’s effect of suppressing the growth of $h$. \\

The Nusselt number $Nu$, quantifying the heat transfer efficiency, is calculated based on the $h$ and correlation between fluctuating vertical velocity $u_3^{\prime}$ and temperature $T^{\prime}$, as a function of time. In the HD cases, $Nu$ decreases with an increase in $f$ from 4 to 8 compared to $f=0$ owing to the suppression of $h$ and $u_3^{\prime}$ by the Coriolis force. In the non-rotating MHD cases for $B_0=0.1, 0.15$, $Nu$ is enhanced compared to the non-rotating HD case. This heat transfer enhancement during the initial regime of unbroken elongated plumes is primarily caused by the vertical stretching of the plumes, which increases $u_3^{\prime}$, signifying the elongated plumes are efficient in transporting heat between hot and cold fluids with limited horizontal mixing. However, in the mixing regime, the heat transfer enhancement is attributed to the collimation of the flow structures along the vertical magnetic field lines, resulting in increased $h$ compared to the non-rotating HD case. With the addition of rotation into $B_0=0.1,0.15$, $Nu$ reduces with increasing $f(=4,8)$ as compared to the corresponding non-rotation MHD cases. The reason for this reduction in heat transfer is the instability-suppressing effect of the Coriolis force, which inhibits the growth and breakdown of the vertically stretched plumes, decreasing $h$ and $u_3'$. We find a significant enhancement in $Nu$ for stronger $B_0(=0.3)$ in non-rotating and rotating cases compared to the corresponding cases for $B_0=0.1$ and 0.15. Moreover, $Nu$ for $B_0=0.3$ at $f=4,8$ until the end of the simulations remains higher than at $f=0,4,8$ for $B_0=0,0.1,0.15$. Interestingly, for the rotating MHD cases, $Nu$ remains higher than the corresponding rotating HD cases because of the vertical stretching and collimation of flow structures along the vertical magnetic lines leading to the increased $h$ and $u_3^{\prime}$ in MHD cases. This indicates that the imposed vertical mean magnetic field mitigates the instability-suppressing effect of the Coriolis force, resulting in efficient heat transfer.\\

For the non-rotating MHD cases, the enhancement in $Nu$ as a function of Rayleigh number ($Ra$) or at a fixed $h$ ($Ra\propto h^3$) compared to the non-rotating HD case is observed during the initial regime of unbroken elongated plumes signifying the increase in the correlation between $u_3^{\prime}$ and $T^{\prime}$. However, in the mixing regime, $Nu$ decreases and becomes comparable to the non-rotating HD case, suggesting a decrease in the correlation between $u_3^{\prime}$ and $T^{\prime}$. In the rotating MHD cases, $Nu$ (as a function of $Ra$) remains higher than the corresponding HD cases due to the higher correlation between $u_3^{\prime}$ and $T^{\prime}$. We observe the presence of the ultimate state regime $Nu\simeq Ra^{1/2}Pr^{1/2}$, where $Pr=1$, in the non-rotating HD and MHD cases for $B_0=0.1,0.15$. However, HD and MHD cases at $f=4,8$ show the deviation from ultimate state scaling. \\

The vertical profiles of the horizontal ($x_1$) and vertical ($x_3$) components of the turbulent kinetic energy ($t.k.e.$) (denoted by $t.k.e._{x_1}$ and $t.k.e._{x_3}$, respectively) reveals the suppression of velocity fluctuations by the Coriolis force in HD and MHD cases compared to the corresponding non-rotating cases. For the non-rotating MHD cases, we find the higher $t.k.e._{x_3}$ than for the corresponding HD case in the initial regime of unbroken elongated plumes, while $t.k.e._{x_1}$ remains negligible. This signifies that the vertically elongated plumes exhibit higher $u_3'$ with limited horizontal mixing, leading to enhancement in heat transfer. In the mixing regime, $t.k.e._{x_3}$ decreases compared to its initial regime and the non-rotating HD case while spanning over larger $h$. Consequently, the imposed $B_0$ acts to suppress the turbulent mixing due to the collimation of flow along the vertical magnetic field lines. Therefore, the enhancement in $Nu$ in the mixing regime is caused by the increased $h$ and not by $u_3'$. In the mixing regime of the rotating MHD cases at $f=8$, the decrease in $t.k.e._{x_3}$ compared to its initial regime is small but remains larger than the corresponding HD case. As a result, the increase in $Nu$ in the rotating MHD cases compared to the corresponding HD cases is attributed to the increase in both $u_3^{\prime}$ and $h$. \\

In the non-rotating MHD cases, the vertical component of the Lorentz force continues to dominate over the horizontal component until the complete breakdown of plumes occurs. This dominance leads to stronger vertical stretching of the plumes. The horizontal component exhibits a larger magnitude towards the plume tips than towards the mid-plane, resulting in the breakup of plume tips and an increase in the advection force near the plume tips. Eventually, the dominance of the horizontal component over the vertical component of the Lorentz force causes the complete breakup of the plumes. A similar interplay between the horizontal and vertical components of the Lorentz force is observed in rotating MHD cases, except for a change in their variation along the vertical direction due to the presence of the Coriolis force. In MHD cases at $f=4$, the Coriolis force dominates the horizontal and vertical components of the Lorentz force towards the plume tips, inhibiting the breakup of plume tips. At $f=8$, the Coriolis force becomes significantly stronger than the horizontal and vertical components of the Lorentz force along the entire surface of the plumes, which inhibits the growth of elongated plumes and their breakdown, resulting in a decrease in the advection force and the heat transfer. In HD cases at $f=8$, a close geostrophic balance between the Coriolis and the pressure forces is achieved, confirming the presence of the Taylor-Proudman constraint. However, the MHD case at $f=8$ does not exhibit a geostrophic balance due to the presence of the Lorentz force, which mitigates the effect of the Coriolis force. This mitigation of the Coriolis force effect by the Lorentz force causes an increase in the advection force and the heat transfer in rotating MHD cases compared to the corresponding HD cases.\\

For the non-rotating and rotating MHD cases, the $t.k.e.$ budget reveals that some portion of the $t.k.e.$ produced by the buoyancy flux converts to turbulent magnetic energy ($t.m.e.$) while the remaining portion converts to the internal energy of the system through viscous dissipation. The horizontal ($x_1$) spectra demonstrate the higher $t.k.e$ than $t.m.e.$ is contained in the large scales, while $t.m.e.$ exceeds $t.k.e.$ at small scales indicating this energy conversion. \\

\backsection[Supplementary data]{\label{SupMat} Supplementary movies are available at https://doi.org/**.****/jfm.***...\\}

\backsection[Acknowledgements]{We want to thank the support and the resources provided by PARAM Sanganak under the National Supercomputing Mission, Government of India at the Indian Institute of Technology, Kanpur are gratefully acknowledged.\\}

\backsection[Declaration of interests]{The authors report no conflict of interest.\\}

\backsection[Author contributions]{The authors contributed equally to analyzing data and reaching conclusions and in writing the paper.}


\appendix
\section{Solver validation}\label{appA} 
In this appendix, we perform simulations to replicate the results of non-magnetic rotating convection (RC) and rotating dynamo convection (DC) of \cite{stellmach2004cartesian} in a rotating plane layer of an electrically conducting fluid. The horizontal fluid layer of depth $d$ is contained between two parallel plates, heated from the bottom and cooled from the top (see \cite{naskar_pal_2022a} for details). In the self-sustained dynamo process, the magnetic field is generated by the convective motion of electrically conducting fluids, which amplifies a small magnetic perturbation by electromagnetic induction. We perform RC simulations for Rayleigh number $Ra=1.4\times10^7$, Ekman number $E=5 \times 10^{-5}$ and Prandtl number $Pr=1$, along with the DC simulations for magnetic Prandtl number $Pr_m=2.5$ with free-slip boundary conditions. The definitions of these non-dimensional numbers and the results obtained in present simulations and results of \cite{stellmach2004cartesian} are shown in Table \ref{tab: validation}. Our results agree with those of the \cite{stellmach2004cartesian}. \\

 \begin{table}
  \begin{center}
  \def~{\hphantom{0}}
  \setlength{\tabcolsep}{9pt} 
  \renewcommand{\arraystretch}{1.2} 
  \begin{tabular}{lcccccccc}
               & $Re_0$   & $Nu_0$  & $Re_m$    & $Nu$  & $Re$  & $\Lambda$ & $ER$       \\[3pt]
    Present results  & 48.6   & 1.36  & 168.5    &  1.68  & 67.4  &  0.36 & 1.36    \\
    \cite{stellmach2004cartesian}  & 48.3   & 1.34  & 170.7 & 1.66  & 68.3  & 0.38  & 1.37 
  \end{tabular}
  \caption{Results from the present simulations and those of \cite{stellmach2004cartesian} for RC simulations with $Ra = {g \beta \Delta T d^3}/{\kappa \nu}=1.4\times10^7$, $E= {\nu}/{\Omega d^2}=5 \times 10^{-5}$, and $Pr = {\nu}/{\kappa}=1$, as well as for the DC simulations with $Pr_m={\nu}/{\mathcal{D}}=2.5$. Here, $d$ is the distance between parallel plates, $\Delta T$ is the temperature difference between the lower (hot) and the upper (cold) plate, and $\Omega$ is the angular velocity of rotation. Results are described by the following non-dimensional numbers: magnetic Reynolds number $Re_m=u_{rms} d/\mathcal{D}$, Reynolds number $Re=Re_m/Pr_m$, Nusselt number $Nu=1+\sqrt{Ra Pr} \langle u_3 \theta \rangle_\mathcal{V}$, Elsasser number $\Lambda=\sigma {B_{rms}}^2/\rho \Omega$, and magnetic to kinetic energy ratio $ER$, where $u_{rms}$ is the characteristic (root-mean-square ($rms$)) velocity, $\theta$ is the non-dimensional temperature, $\sigma$ is the electrical conductivity, $B_{rms}$ is the characteristics magnetic field strength, and $\langle \cdot \rangle_\mathcal{V}$ denotes average over the entire volume ($\mathcal{V}$). $Re_0$ and $Nu_0$ represents the results of non-magnetic RC simulation, whereas $Re_m$, $Re$, $\Lambda$, and $ER$ represents the results of DC simulation.}  \label{tab: validation}
  \end{center}
\end{table}







\end{document}